\documentclass[12pt]{article}
\usepackage{amsmath}
\usepackage{amssymb}
\usepackage[utf8]{inputenc}
\usepackage{graphicx}
\usepackage{subcaption}
\usepackage{mathtools}
\usepackage{empheq}
\usepackage{chngcntr}
\usepackage{esint}
\usepackage{colonequals}
\usepackage{mathrsfs}
\usepackage{extarrows}
\usepackage{bookmark}
\usepackage{outlines}
\usepackage{geometry}
\numberwithin{equation}{section}

\usepackage{tikz}
\usepackage[compat=1.1.0]{tikz-feynman}
\usetikzlibrary{calc,patterns,angles,quotes,intersections,arrows,shapes.misc}
\tikzset{cross/.style={cross out, draw=black, minimum size=2*(#1-\pgflinewidth), inner sep=0pt, outer sep=0pt},
cross/.default={1pt}}
\newcommand*\circled[1]{\tikz[baseline=(char.base)]{
            \node[shape=circle,draw,inner sep=2pt] (char) {#1};}}

\usepackage{xspace}
\newcommand{\vev}[1]{\langle #1 \rangle\xspace}

\newcommand{\im}[1]{\,\mathrm{Im}\,#1}

\newcommand{\NN}{\mathcal N}
\newcommand{\TT}{\mathcal T}
\newcommand{\OO}{\mathcal O}

\newcommand{\CC}{{\mathcal C}}
\newcommand{\D}{\Delta}

\let\a=\alpha \let\b=\beta \let\g=\gamma

\let\s=\sigma     
  \let\D=\Delta

\newcommand{\zb}{{\bar z}}

\newcommand{\rhob}{{\bar \rho}}

\newcommand{\bxx}[1]{\begin{#1}}
\newcommand{\be}{\bxx{equation}}
\newcommand{\ee}{\end{equation}}
\newcommand{\bea}{\bxx{equation}\bxx{aligned}}
\newcommand{\eea}{\end{aligned}\end{equation}}

\newcommand{\cO}{\mathcal O}

\newcommand{\Df}{{\Delta_{\cO}}}

\newcommand{\reef}[1]{(\ref{#1})}

\begin{document}

\begin{titlepage}
		\thispagestyle{empty}
		\setcounter{page}{0}
		\medskip
		\begin{center} 
			{\Large \bf Landau diagrams in AdS \vspace{10pt}\\and S-matrices from conformal correlators}

			\bigskip
			\bigskip
			\bigskip
			
			{\bf Shota Komatsu$^{1}$, Miguel F. Paulos$^2$, Balt C. van Rees$^{3}$ and Xiang Zhao$^{3}$\\ }
			\bigskip
			\bigskip
			${}^{1}$
			School of Natural Sciences, Institute for Advanced Study, \\1 Einstein Drive, Princeton, NJ 08540, USA
			\vskip 5mm
			${}^{2}$
			Laboratoire de Physique de l’Ecole Normale Sup\'erieure\\
			PSL University, CNRS, Sorbonne Universit\'es, UPMC Univ. Paris 06\\
			75231 Paris Cedex 05, France
			\vskip 5mm
			${}^{3}$
			Centre de Physique Th\'eorique (CPHT), Ecole Polytechnique\\91128 Palaiseau Cedex, France
			\vskip 5mm
			
			\texttt{~skomatsu@ias.edu,~miguel.paulos@ens.fr,\\
				~balt.van-rees@polytechnique.edu,~xiang.zhao@polytechnique.edu} \\
		\end{center}
		
		\bigskip
		\bigskip
		
		\begin{abstract}
			\noindent Quantum field theories in AdS generate conformal correlation functions on the boundary, and in the limit where AdS is nearly flat one should be able to extract an S-matrix from such correlators. We discuss a particularly simple position-space procedure to do so. It features a direct map from boundary positions to (on-shell) momenta and thereby relates cross ratios to Mandelstam invariants. This recipe succeeds in several examples, includes the momentum-conserving delta functions, and can be shown to imply the two proposals in \cite{QFTinAdS} based on Mellin space and on the OPE data. Interestingly the procedure does not always work: the Landau singularities of a Feynman diagram are shown to be part of larger regions, to be called `bad regions', where the flat-space limit of the Witten diagram diverges. To capture these divergences we introduce the notion of Landau diagrams in AdS. As in flat space, these describe on-shell particles propagating over large distances in a complexified space, with a form of momentum conservation holding at each bulk vertex.
			As an application we recover the anomalous threshold of the four-point triangle diagram at the boundary of a bad region.
		\end{abstract}
		
		\noindent

\end{titlepage}
	

\setcounter{tocdepth}{2}

\tableofcontents


\section{Introduction}
Consider a quantum field theory on a fixed $D$-dimensional Anti-de Sitter background. In this setup, take a correlation function of local operators and push its insertion points all the way to the conformal boundary, inserting scaling factors to obtain a finite answer. This LSZ-like limit gives rise to what we call \emph{boundary correlation functions}. If the AdS isometries are preserved then these obey all the useful axioms of usual CFT correlation functions: conformal invariance in $D-1$ dimensions, a large domain of analyticity and a convergent conformal block decomposition. All this is of course familar from AdS/CFT; the only difference is that there is no stress tensor in the boundary spectrum because the bulk metric is not dynamical.

Our main interest lies with the behavior of these boundary correlation functions in the \emph{flat-space limit}. (We will write it as $R\to \infty$ with $R$ the curvature radius of AdS.) Supposing that it exists, it is a natural expectation that the \emph{S-matrix} of the bulk theory is encoded in the flat-space limit of the boundary correlation functions. This idea has a long history, especially in the context of AdS/CFT (starting with \cite{Polchinski:1999ry,Giddings:1999jq,Gary:2009ae,Okuda:2010ym,joaomellin,Goncalves:2014ffa}) where one can try to extract string theory amplitudes from CFT correlators. Until recently comparatively little attention has been given to the setup where the bulk theory is gapped and does not contain gravity, but see \cite{QFTinAdS,Paulos:2016but,Paulos:2017fhb,Homrich:2019cbt,Aharony:2012jf,Aharony:2015zea,Carmi:2018qzm,Giombi:2020rmc,Giombi:2017cqn,Beccaria:2019stp} for works in  that direction. It is nevertheless an extremely interesting subject. This is because scattering amplitudes are rather mysterious objects with an interplay of analyticity and unitarity that appears to be at most partially understood. But via the QFT in AdS consturction we can obtain amplitudes as a limit of conformal correlation functions, and it is natural to expect that the well-established properties of the latter can clarify some of the mysteries surrounding the former.

As for the precise map from correlator to amplitude there exist several proposals. Two concrete proposals were written down in \cite{QFTinAdS}: one in Mellin space and a phase shift formula. In other work mention was made of a Fourier space algorithm \cite{Hijano:2019qmi}. In this work we propose a \emph{position-space} limit, which refines and generalizes the idea proposed in \cite{TTBar} for AdS$_2$. One might wonder why we need yet another formula given we already have concrete proposals. The main reason is because the existing proposals have their own shortcomings: for instance,  the phase shift formula has a drawback that it can only be defined in a physical kinematics and relies on averaging over the OPE data, which is sometimes difficult to perform in practice. On the other hand, the Mellin approach involves integral transforms of some correlators which make it hard to discuss their analytic properties at the nonperturbative level. In fact the existence of the Mellin-space representation of the correlators was established only quite recently in \cite{Penedones:2019tng} and yet, its analyticity is not fully understood. Another, more technical issue is that there are singularities of the flat-space amplitudes which are not well-understood in the Mellin approach. One representative example are anomalous thresholds, which come from on-shell propagations of particles in several different channels. Such singularities are hard to see from the Mellin approach since the poles of the Mellin representation of the correlator are normally associated with the operator product expansion in a single channel.

By contrast, our position-space approach has the distinguishing feature that it requires no OPE data manipulation or  integral transforms: instead the position-space correlator \emph{becomes} the S-matrix element. We propose, for example, that two-point functions $|x-y|^{-2\Delta}$ become single-particle norms, $\vev{\vec k|\vec p}\propto \delta^{(D-1)}(\vec k - \vec p)$, that contact diagrams in AdS become momentum-conserving delta functions, and more generally that
\be
\lim_{R \to \infty} \vev{\OO_1(x_1) \OO_2(x_2) \ldots \OO_n(x_n)} = \vev{ \vec k_1 \vec k_2 \ldots | \ldots \vec k_n }\,,
\ee
with a suitable normalization of the operators. To make the above formula work a map is needed from boundary positions to on-shell momenta, which indeed both have $D-1$ components. It turns out that this map is not without $i$'s, and for physical kinematics we need to move the $x_i$ to complex positions. This is maybe to be expected, since in real Lorentizan AdS massive particles cannot reach the boundary. The precise map is given in section \ref{subsec:conjectures}. For a four-point function of identical operators it implies a relation between cross-ratios and Mandelstam invariants as given in equation \eqref{rhorhobtomandelstam}.

Another distinguishing feature of our proposal is that it fails to work in certain kinematic regions. Starting with the exchange diagram, which is discussed in detail in section \ref{subsec:exchange}, we find that there are regions in the complex Mandelstam planes where the flat-space limit of the correlation function diverges, even after stripping off the momentum-conserving delta function, and therefore does not equal the scattering amplitude. As we explain qualitatively in section \ref{Section_Erasing the circle}, this is due to the possibility of exchanged particles going on-shell and propagating over distances of the order of the ever-growing AdS scale.\footnote{Landau diagrams in AdS were discussed also in \cite{Maldacena:2015iua}, but there are several important differences from our work. In \cite{Maldacena:2015iua}, the authors only considered trajectories of massless particles in Lorentzian AdS which interact at a single bulk point. Such diagrams give rise to singularities of the boundary correlation functions even before the flat-space limit is taken. On the other hand, in this paper we discuss Landau diagrams of massive particles in a complexified AdS space that interact at widely separated bulk points and which are responsible for singularities of the S-matrix in the flat-space limit.} Such a separation of the interaction vertices in a given diagram is of course reminisicent of flat-space Landau diagrams which can be used to deduce the location of potential singularities in flat-space scattering amplitudes. In AdS with finite $R$ the infrared is regulated and these Landau singularities do not exist, but that does not mean that they cannot spoil the flat-space limit. To understand them better we formulate in section \ref{sec:landaudiagrams} the general AdS Landau equations and compare them with their flat-space counterpart. We will argue that they are indicators of singularities in the flat-space amplitude, since, it appears that every flat-space Landau singularity is surrounded by a region in the Mandelstam planes where the flat-space limit does not work. This would imply that the AdS Landau equations can reproduce anomalous thresholds in the flat-space limit; to demonstrate this we include a numerical analysis of the triangle diagram in section \ref{subsec:triangle}.

In sections \ref{sec:mellin} and \ref{sec:blocks} we compare our proposal with the Mellin space and phase shift proposals of \cite{QFTinAdS}, respectively. We will find that the Mellin space proposal can be recovered from our proposal (for Mellin-representable correlation functions) via a saddle point analysis, and can understand the divergences from the AdS Landau singularities as originating from a contribution of Mellin poles that are picked up by moving the original integration contour to the steepest descent contour. (Conversely, it is natural to suspect that anomalous thresholds cannot appear if no poles are picked up.) Conformal blocks will really only enter our discussion in section \ref{sec:blocks}, where we will make contact with the phase shift formula of \cite{QFTinAdS} and formulate a condition on the OPE data such that the flat-space limit amplitude obeys unitary conditions. We will also see that in that context the singularities arise from divergent contributions of conformal blocks corresponding to ``bound states'' in the flat space limit. The results in this section should be viewed as a first exploration into the implications of the existence of an OPE for scattering amplitudes --- we hope to report more results in this direction in the near future.


\section{The flat-space limit in position space}
\label{Section_Erasing the circle}

In this section, we present our conjectural position-space recipe for obtaining flat-space scattering amplitudes from the conformal boundary correlation functions of a QFT in AdS. In order to motivate the conjectures, we first explain how the building blocks of AdS Witten diagrams, namely the bulk-boundary propagator and the bulk-bulk propagator, morph into their counterparts in flat space. Our main result will be that the bulk-boundary propagator becomes very simple in the flat-space limit and essentially reduces to a factor like $e^{i p x}$ with an on-shell momentum $p$ while the bulk-bulk propagator becomes the Feynman propagator $1/(p^2+m^2)$.

After presenting our position-space formulas, we discuss briefly its physical implications including a direct relation between the conformal cross ratios and the Mandelstam variables. We also give a heuristic argument on why such a formula may fail to work in certain kinematic regions. Understanding the details of why and how the formula fails is the main subject of the rest of this paper and that is what will lead us to propose the AdS analogue of Landau diagrams in section \ref{sec:landaudiagrams}. 

\subsection{Motivating the conjecture}
To motivate the conjecture, let us consider the flat-space limit of the bulk-boundary and bulk-bulk propagators. 

\subsubsection{Bulk-boundary propagator}
We follow the conventions of \cite{joaomellin}. This means that we will describe Euclidean AdS$_{d+1}$ using embedding space coordinates $X$ living in $d+2$ dimensional Minkowski space which obey:
\begin{align*}
-(X^0)^2+\sum_i (X^i)^2&=-R^2,\quad X^0>0
\end{align*}
and points on the conformal boundary of AdS are labeled by $d+2$ dimensional points $P$ on the projective null cone:
\be
-(P^0)^2+\sum_i (P^i)^2=0, \quad P \sim \lambda P \,\, (\lambda \in \mathbb R^*)
\ee
We can resolve the constraints and `gauge fix' as follows:
\begin{align*}
X&=\left(R \cosh \left(\frac{\rho}{R}\right),\,
R \sinh \left(\frac{\rho}{R}\right) 
 n_{X}\right),\\
P&=(1,\,  n_P),
\end{align*}
Where we introduced our choice of local coordinates for AdS$_{d+1}$: a radial coordinate $\rho$ and a $d+1$ dimensional unit norm vector $n^\mu$ which obeys $ n^\mu  n^\nu \delta_{\mu \nu} = 1$. The metric reads:
\be \label{adsmetric}
ds^2 = d \rho^2 + R^2 \sinh^2\left(\frac{\rho}{R} \right) d\Omega_d^2,
\ee
and therefore the flat-space limit in these coordinates is very simple: we just send $R \to \infty$ holding all of the coordinates fixed. The standard Euclidean coordinate $x$ is then:
\begin{align*}
x = \rho  n_X.
\end{align*}

Now consider the bulk-boundary propagator. It reads:
\begin{align}
G_{B\partial}(X,P)
&=\frac{\mathcal{C}_{\Delta}}{R^{(d-1)/2}(-2P\cdot X/R)^\Delta}
\nonumber\\
&=\frac{\mathcal{C}_{\Delta}}{2^\Delta R^{(d-1)/2}} e^{-\Delta\log(-P\cdot X/R)}
\end{align}
where
\begin{align*}
\mathcal{C}_{\Delta}=\frac{\Gamma(\Delta)}{2\pi^h \Gamma(\Delta-h+1)}, \qquad 
h=\frac{d}{2}, \qquad
\Delta(\Delta-d)=m^2 R^2.
\end{align*}
Substituting
\be
-P\cdot X/R= \cosh \left(\frac{\rho}{R}\right)
-\sinh \left(\frac{\rho}{R}\right)  n_P \cdot  n_X
\ee
straightforwardly yields that
\begin{align}
\boxed{
G_{B\partial}(X,P)\overset{R\to\infty}{\xrightarrow{\hspace*{0.6cm}}}
\frac{m^{h-1}}{2^{mR+1}\pi^h R^{1/2}}
e^{m  n_P \cdot x}.
}
\label{Eqn_FlatBulkBoundary}
\end{align}
This can be compared this with an external leg in a flat-space Feynman diagram, which would simply read
\be
\label{StandardSmatrixLeg}
\frac{1}{\sqrt{Z}} e^{i \eta_{\mu \nu} k^\mu y^\nu}.
\ee
with $k^\mu$ an on-shell Lorentzian momentum (so $k^2 = - m^2$), $y^\nu$ a Lorentzian position, $\eta_{\mu \nu} = \text{diag}(-++ \ldots +)$, and, because we work in conventions where all momenta are ingoing, $k^0 > 0$ or $k^0 < 0$ for an ingoing or outgoing momentum, respectively.

\begin{figure}[t]
\centering
\includegraphics[clip,height=4.5cm]{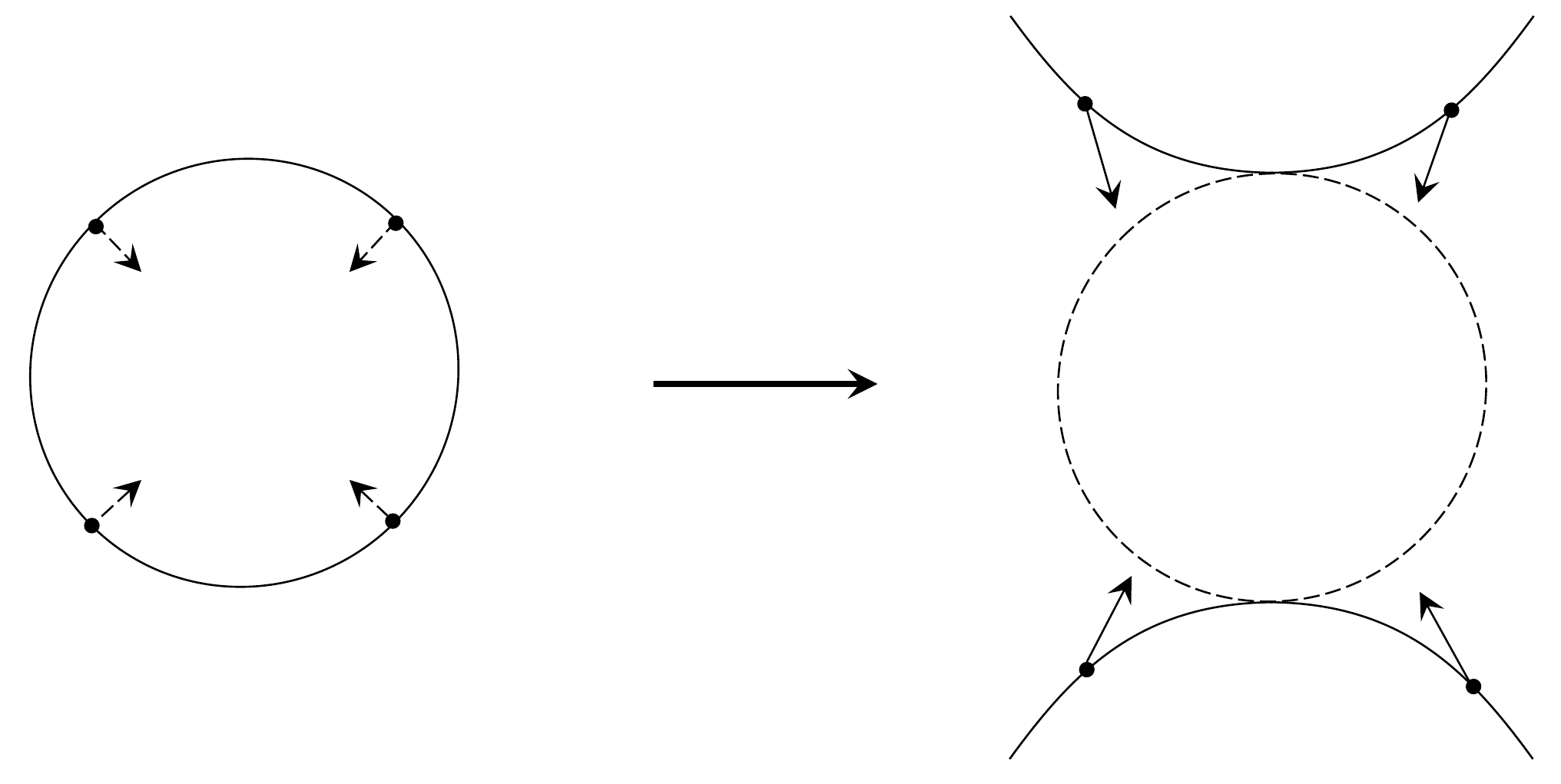}
\caption{Analytic continuation of the boundary points. We start from a CFT on $S^{d}$ and analytically continue the insertion points of operators to complex values in order to recover the flat-space S-matrix. Geometrically this corresponds to going from a sphere to a hyperboloid. \label{fig:hyperboloid}}
\end{figure}

Clearly we find the normalization factor
\begin{align}
\frac{1}{\sqrt{Z}} = \frac{\mathcal{C}_{\Delta}}{2^\Delta R^{(d-1)/2}} \overset{R\to\infty}{\xrightarrow{\hspace*{0.6cm}}}  \frac{m^{h-1}}{2^{mR+1}\pi^h R^{1/2}}\,,
\label{Eqn_normalization factor Z}
\end{align}
whereas the exponents are matched as follows. First we recognize that \eqref{Eqn_FlatBulkBoundary} was derived with a Euclidean signature bulk metric, so the contraction ${n}_P \cdot x$ is really equal to $\delta_{\mu \nu} n_P^\mu x^\nu$, and similarly $\delta_{\mu \nu} n_p^\mu n_p^\nu = 1$. On the other hand \eqref{StandardSmatrixLeg} requires Lorentzian signature, so if we write $y^\mu = (y^0, \underline{y})$ and $x^\mu = (x^0, \underline{x})$ then the standard \emph{bulk} analytic continuation dictates that\footnote{There is no real freedom here: the bulk point $x$ is integrated over and should be continued in accordance with the desired Wick rotation.}
\be \label{bulkcontinuation}
y^0 = - i x^0\,, \qquad \underline{y} = \underline{x}\,,
\ee
and therefore a match can be obtained if the \emph{boundary} points do something entirely different, namely we need to set
\be
n_p^0 = - k^0 / m  \qquad \underline{n}_p = i \underline{k} / m\,.
\ee
We conclude that physical S-matrix momenta correspond to complex boundary positions! More precisely, the above equation shows that if we start from real boundary coordinates in Euclidean signature (so real $n_p^\mu$), then we need to continue the spacelike components $\underline{n}_p$ to purely imaginary values, whereas the zero-component remains real but obeys $|n_p^0| > 1$ because $|k^0| > m$. Pictorially this corresponds to analytically continue the boundary sphere to a hyperboloid, see figure \ref{fig:hyperboloid}. Alternatively, supposing we start from real boundary coordinates in Lorentzian signature, which according to the bulk analytic continuation \eqref{bulkcontinuation} corresponds to real $\underline{n}_p$ but purely imaginary $n_p^0$, then we find that we need to continue \emph{all} the components to purely imaginary values. We should also note that these continuations always respect $1 = \delta_{\mu \nu} n_p^\mu n_p^\nu  = - \eta_{\mu \nu} k^\mu k^\nu / m^2$, so we are automatically on the mass shell. The analytic continuation will be discussed in more detail below.

\subsubsection{Bulk-bulk propagator}
\label{subsubsec:bulkbulknaive}
\begin{figure}[t]
\centering
\begin{minipage}{0.45\hsize}
\centering
\includegraphics[clip,height=3cm]{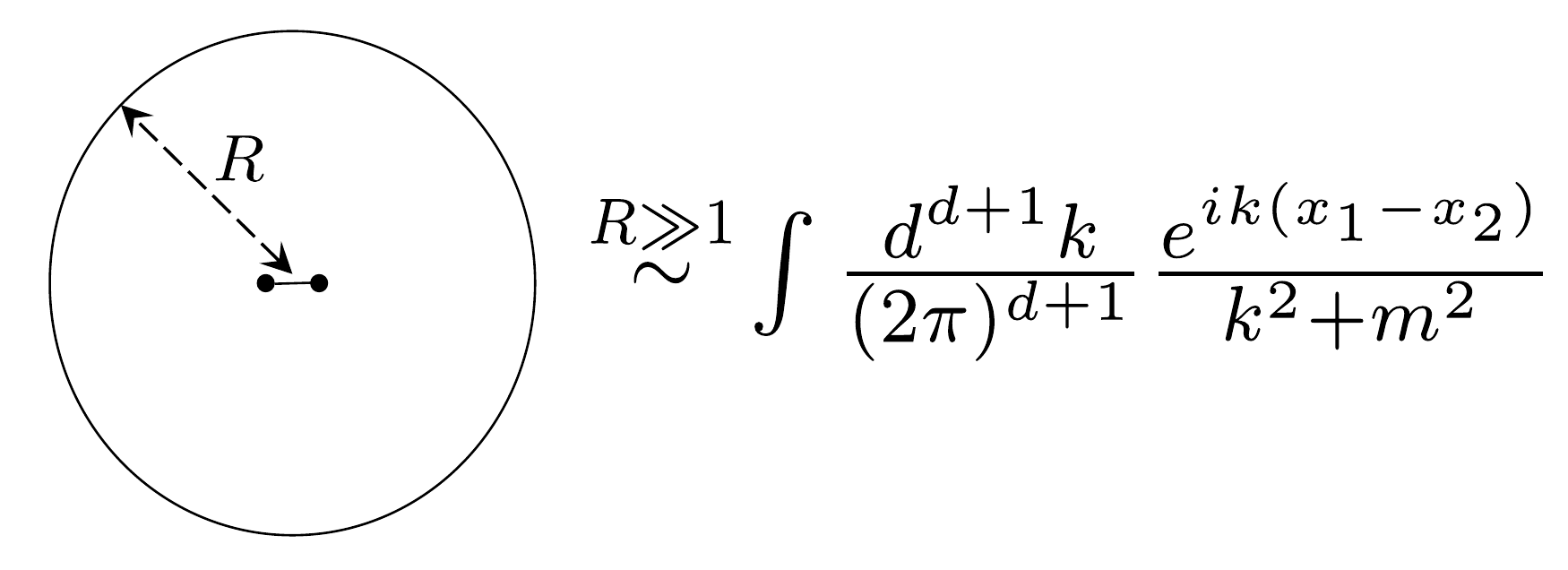}
\subcaption{$(X_1-X_2)^2\ll R^2$ \label{fig:propagator1}}
\end{minipage}\hspace{10pt}
\begin{minipage}{0.45\hsize}
\centering
\includegraphics[clip,height=3cm]{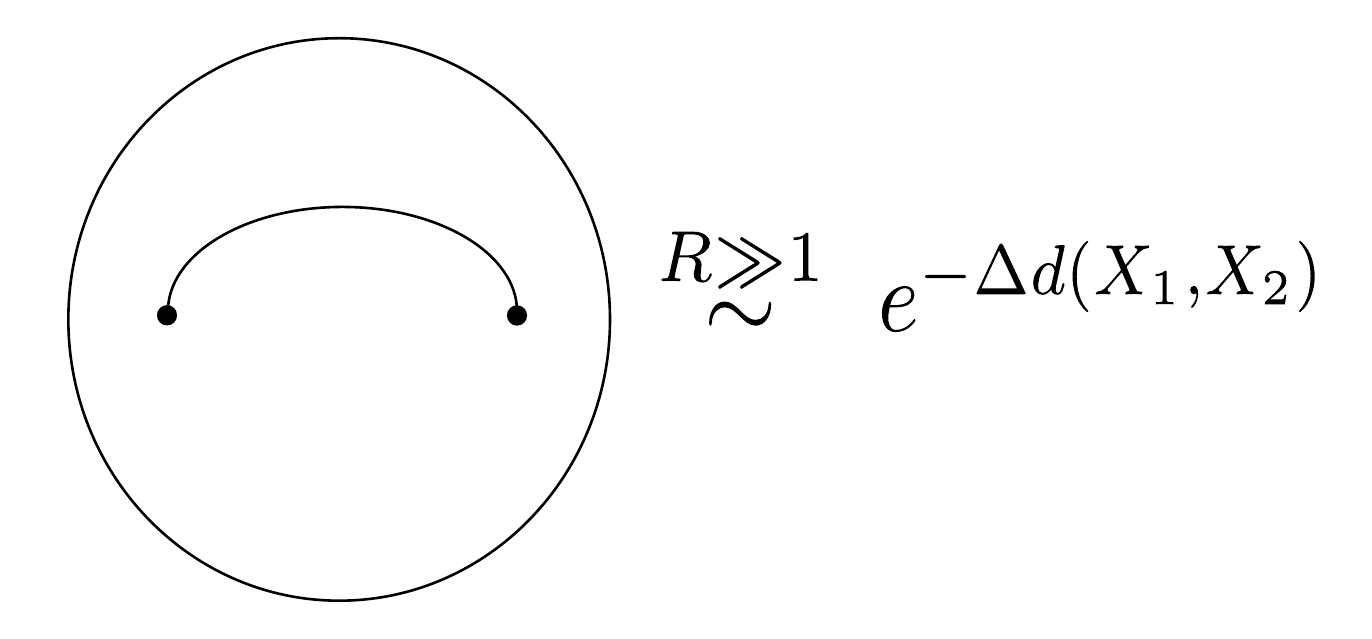}
\subcaption{$(X_1-X_2)^2\sim R^2$ \label{fig:propagator2}}
\end{minipage}
\caption{Two different limits of the bulk-bulk propagator. $(a)$ If the two bulk points are close to each other $(X_1-X_2)^2\ll R^2$, the large $R$ limit gives a propagator in flat space. $(b)$ If the two bulk points are kept apart, the limit is described by a geodesic in AdS which connects the two points. In this case, the propagator falls off exponentially $e^{-\Delta d(X_1,X_2)}$ where $d(X_1,X_2)$ is a geodesic distance between the two points. }
\end{figure}

Next we consider the bulk-bulk propagator $G_{BB}(X_1, X_2)$. Its defining equation reads
\be
\left( \square_g  - \Delta(\Delta-d) \right) G_{BB} (X_1, X_2) = \frac{1}{\sqrt{g}} \delta^{(d+1)}(X_1 - X_2)\,.
\ee
For the computations that are to follow it turns out that the most convenient solution is the split representation of \cite{joaomellin} where\footnote{The solutions can also be written as $G_{BB}(X_1,X_2) = \frac{R^{1-d}\CC_{\D}}{u^{\D}} {}_2 F_1 \left(\D, \D - d/2 + 1, 2 \D - d + 1; - 4 / u \right)$ with $u = (X_1 - X_2)^2 / R^2$. In this case the two limits discussed later in this section yield, respectively, the familiar Bessel function expression for the position-space Klein-Gordon propagator
and an expression which is familiar from the large $\Delta$ limit of a one-dimensional conformal block.}
\begin{align}
G_{BB}(X_1,X_2)&= 
\int_{-i\infty}^{i\infty}\frac{dc}{2\pi i}
\frac{2c^2}{c^2-(\Delta-h)^2} \int_{\partial AdS}dQ
\frac{R^{1-d} \, \mathcal{C}_{h+c} \, \mathcal{C}_{h-c}}{(-2Q\cdot X_1/R)^{h+c}(-2Q\cdot X_2/R)^{h-c}}
\end{align}
In the large $R$ limit we send $\Delta \to \infty$ but we have to give some thought to the scaling of $X_1$ and $X_2$. In the spherical AdS coordinates introduced above we have
\begin{align*}
X_1&=\left(R \cosh \left(\frac{\rho_1}{R}\right),\,
R \sinh \left(\frac{\rho_1}{R}\right) 
n_{X_1}\right),\\
X_2&=\left(R \cosh \left(\frac{\rho_2}{R}\right),\,
R \sinh \left(\frac{\rho_2}{R}\right)
n_{X_2}\right),\\
Q&=(1,\, n_Q).
\end{align*}
and the integration measure $dQ$ is the usual one on the $d$-dimensional sphere. In the flat-space limit we keep $\rho$ fixed as we send $R \to \infty$. The substitution
\begin{align}
c\equiv iKR,\qquad K\in\mathbb{R}
\end{align}
yields
\be
\frac{1}{(\Delta-h)^2-c^2}\qquad \to \qquad\frac{1}{R^2}\frac{1}{m^2+K^2}
\ee
with $\Delta(\Delta-d)=m^2R^2$ as before, and with the appropriate large $R$ limits of the other building blocks we find that
\begin{align}
G_{BB}(X_1,X_2)
&\to \int_{0}^{\infty} \frac{K^d dK}{(2\pi)^{d+1}} 
\int d\Omega_{d}
\frac{e^{iK\left(\rho_1 n_Q \cdot n_{X_1}-\rho_2 n_Q \cdot n_{X_2}\right)}}{m^2+K^2}\nonumber\\
&= \int \frac{d^{d+1}k}{(2\pi)^{d+1}} \,
\frac{e^{i k\cdot\left(x_1 - x_2\right)}}{m^2+k^2},
\end{align}
where the integral over the AdS boundary coordinate $Q$ simply becomes an integral over a $d$-dimensional unit sphere, and we have made the identification $K n_Q^\mu \to k^\mu$. Notice that we get the right answer on the nose: unlike the bulk-boundary propagator there are no relative factors of $i$ or normalization issues. See also figure \ref{fig:propagator1}.

In the above flat-space limit, we implicitly assumed that the two bulk points are close to each other, $(X_1 - X_2)^2 \ll R^2$. Although this would be appropriate for the flat-space limit, there also exists a pure \emph{large $\Delta$ limit} of the propagator where the bulk points are kept apart. To find the behavior in this limit we can close the $c$ contour in the appropriate right or left half plane to pick up the pole at $c = \pm (\D - h)$.
The $Q$ integral can then be done via a saddle point approximation (which is easy after choosing a specific frame) and results in
\be
G_{BB}(X_1,X_2) \to \exp( - \D \tilde \rho ) = \exp( - \D \, \text{arccosh}(- X_1 \cdot X_2 / R^2 ))
\ee
up to a prefactor and other non-exponential terms in $\Delta$ that will not matter below. Note that what appears in the exponent is a geodesic distance between the two points $X_1$ and $X_2$, so in this limit we recover a classical particle travelling along the geodesic between these two points (see figure \ref{fig:propagator2}).

In order for the flat-space limit to work all the interactions must take place at distances below the AdS scale. In the integrals over the bulk vertices $X_1$ and $X_2$ it is therefore essential that these large $\Delta$ limits are always suppressed for the correlation functions that we want to analyze. This is a nontrivial condition since the points $X_1$ and $X_2$ are integrated over in a Witten diagram and in principle both limits must be included. As we see in the subsequent sections, this large $\Delta$ limit is precisely what sometimes obstructs us from taking the flat-space limit in position space.  For now, we proceed to present our conjectures on the flat-space limit relegating detailed discussions about possible subtleties to subsection \ref{subsec:capsandsubleties}.

\subsection{S-matrix conjecture and amplitude conjecture}
\label{subsec:conjectures}
Any Witten diagram is a combination of bulk-bulk and bulk-boundary propagators which are connected at vertices to be integrated over all of $AdS_{d+1}$. In the preceding section we have seen that the flat-space limit of these building blocks (when holding the bulk coordinates $\rho$ and $n^\mu$ fixed) reduces them to the corresponding flat-space expression, and in particular bulk-boundary propagators reduce to the usual external leg factors for position-space Feynman diagrams after a suitable analytic continuation. 
It is then natural to formulate the
\be
\begin{aligned}
&\textbf{S-matrix conjecture}:\\
&\vev{\underline{\tilde k}_1\ldots \underline{\tilde k}_a|S| \underline{k}_1 \ldots \underline{k}_b} 
\stackrel{?}{=} 
\lim_{R \to \infty} 
\left(\sqrt{Z}\,\right)^{a + b} \left. \vev{\OO(\tilde n_1) \ldots \OO(\tilde n_a) \OO(n_1) \ldots \OO(n_b)} \right|_{\text{S-matrix}}
\end{aligned}
\label{Eqn_flatspaceSconj}
\ee
where the boundary correlator should be evaluated in the round metric on the boundary $S^d$ and analytically continued to the `S-matrix' configurations which in unit vector coordinates $\delta_{\mu \nu} n^\mu n_\nu = \delta_{\mu \nu} \tilde n^\mu \tilde n_\nu = 1$ correspond to the values
\be \label{momentaSmatrix}
\begin{split}
(n^0, \underline{n}) &= (-k^0, i \underline{k})/ m\,,\\
\end{split}
\ee
and similarly for the tilded variables, with $k^0 > 0$ for `in' and $\tilde k^0 < 0$ for `out' states. The normalization factor $\sqrt{Z}$ was given in \eqref{Eqn_normalization factor Z}.\footnote{This is the right normalization factor when operators are normalized as $\langle \OO |\OO\rangle = {\mathcal C}_{\Delta}$. For unit normalized operators one should replace $Z$ by $\tilde Z = {\mathcal C}_\Delta^2 Z$ in \eqref{Eqn_flatspaceSconj}. Notice also that in our conventions $\OO = \OO^{(\text{can})}/(2\D -d)$ where $\OO^{(\text{can})}$ would be the operator dual to a canonically normalized scalar field in AdS, a common normalization convention in the holographic renormalization literature \cite{Skenderis:2002wp}.} 

Notice that the object on the left-hand side of equation \eqref{Eqn_flatspaceSconj} is an S-matrix element and therefore includes possible disconnected components as well as an overall momentum-conserving delta function. Schematically we can write:
\begin{align}
\vev{\underline{\tilde k}_1\ldots \underline{\tilde k}_a|S| \underline{k}_1 \ldots \underline{k}_b} = \text{(disconnected)} + (2\pi)^{d+1} i \, \delta^{(d+1)}\left(\sum_{j=1}^a {\tilde k}_j + \sum_{i=1}^b k_i \right)\TT(\tilde k_1\ldots \tilde k_a;k_1\ldots k_b)
\label{MomentumConservation}
\end{align}
where the scattering amplitude $\TT(\ldots)$ normally has no further delta-function singularities. To obtain $\TT(\ldots)$ we can consider the \emph{connected} correlation function which we then divide by the contact diagram to get rid of the momentum-conserving delta function. This leads us to the
\be 
\label{flatspaceTconj}
\begin{aligned}
&\textbf{Amplitude conjecture}:\\
&\TT(\tilde k_1\ldots \tilde k_a;k_1\ldots k_b) \stackrel{?}{=} \lim_{R \to \infty} \left. \frac{\vev{\OO(\tilde n_1) \ldots \OO(\tilde n_a) \OO(n_1) \ldots \OO(n_b)}_{\text{conn}}}{D(\tilde n_1, \ldots, \tilde n_a, n_1, \ldots ,n_b)} \right|_{\text{S-matrix, cons}}
\end{aligned}
\ee
with $D(\tilde n_1, \ldots, \tilde n_a, n_1, \ldots ,n_b)$ denoting the contact diagram in AdS, which is most easily defined as the function that is a constant in Mellin space.\footnote{The masses of the external particles in the contact diagram should be taken to be the physical masses in the interacting theory.} Notice also that, as indicated by the subscript, we not only continue the momenta to the S-matrix configuration as in \eqref{momentaSmatrix} but we also evaluate it on the support of the momentum-conserving delta function in \eqref{MomentumConservation}.

\paragraph{Validity of the conjectures} We now make two important comments on our conjectures. First, precisely speaking these conjectures are valid only in certain kinematic regions. At the level of Witten diagrams, this is basically due to the large $\Delta$ limit of the bulk-bulk propagator, which we discussed at the end of the last subsection. We will give a heuristic explanation of why they can fail in subsection \ref{subsec:capsandsubleties} and discuss in more detail when the conjectures hold in the rest of this paper. Second, although here we motivated the conjectures by the analysis of perturbative Witten diagrams, one can arrive at the same conclusion from the conformal block expansion once one makes certain assumptions on the OPE coefficients. We will present a first exploration in this direction in section \ref{sec:blocks} while a more detailed analysis will be presented in an upcoming paper.

\paragraph{Comparison between the conjectures}Although the expressions look similar, there is an important difference between the S-matrix conjecture in \eqref{Eqn_flatspaceSconj} and the amplitude conjecture in \eqref{flatspaceTconj}. The former in its most general form only really makes sense for real (on-shell) momenta, because only in that case can we make sense of various delta functions. The latter has no such restriction and can be applied to complex values of the momenta. Because of this feature, one might think that the amplitude conjecture is more useful in practice. However we emphasize that being able to reproduce the momentum-conserving delta function is not merely of academic interest but is necessary in certain situations in order to capture the correct physics of scattering amplitudes. The best place to see this is the scattering amplitude in integrable field theories in two dimensions. Owing to the existence of higher conserved charges, the (higher-point) scattering amplitudes in integrable field theories come with extra factors of delta functions, one for each pair of incoming and outgoing momenta, $\propto \prod_{j}\delta^{(2)}(\tilde{k}_j+k_j)$. As we see in the next section, the momentum-conserving delta function in general come from a certain exponentially growing piece of the boundary CFT correlator. This suggests that the higher-point correlation functions in integrable field theories in $AdS_2$ grow much faster than the corresponding counterparts in non-integrable field theories when we take the flat-space limit. This feature is arguably what distinguishes integrable field theories from non-integrable field theories in $AdS_2$. It would be interesting to make this precise and check it in explicit examples\footnote{See recent works \cite{Giombi:2017cqn,Beccaria:2019dws,Beccaria:2019stp,Beccaria:2019ibr,Beccaria:2019mev,Beccaria:2019dju,Beccaria:2020qtk} on integrable (or solvable) theories on $AdS_2$.}.

\paragraph{Connections to previous results}
The S-matrix conjecture \eqref{Eqn_flatspaceSconj} would lead to an elegant way to obtain scattering amplitudes directly from the correlation function in position space. A similar conjecture was published for AdS$_2$ in \cite{TTBar}, where it was claimed that the Euclidean amplitude could be obtained as a limit of the position-space expression. The derivation in that paper however required a more involved wave-packet analysis, and the momentum-conserving delta function was left implicit. In unrelated work, the paper \cite{QFTinAdS} presented both a Mellin space formula and a phase shift formula that could be used in certain cases to extract a flat-space scattering amplitude from CFT data. A detialed comparison with these two prescriptions will be presented in section \ref{sec:mellin} and \ref{sec:blocks}, respectively. Finally our conjectures are rather closely related to a recent proposal in \cite{Hijano:2019qmi}. Although we do not see the need to perform any Fourier transforms as was proposed in that work, the underlying picture is quite appealing --- both to explain the complexification of the boundary positions and to highlight potential issues with the conjectures.

\subsection{Potential subtleties}
\label{subsec:capsandsubleties}
\begin{figure}[t]
\centering
\includegraphics[clip,height=5.5cm]{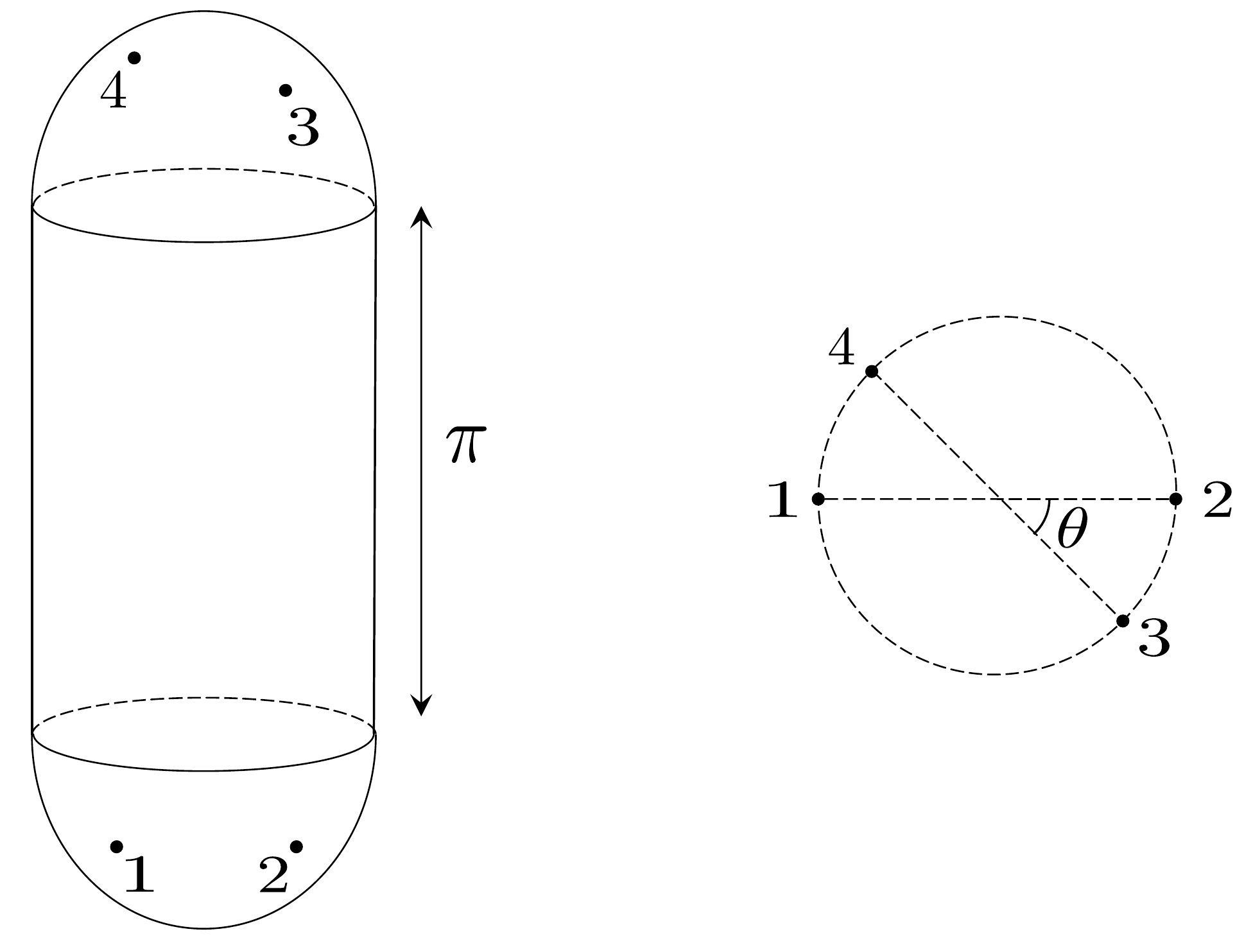}
\caption{A `cylinder with caps' configuration discussed in \cite{Hijano:2019qmi}. We consider two Euclidean hemispheres and connect them by a Lorentzian cylinder of length $\pi$. We then insert two operators on the upper cap and the remaining two operator on the lower cap. The right figure shows a configuration of operator when viewed from the bottom of the lower cap. The angle $\theta$ depicted in the figure becomes a scattering angle in the flat-space limit.\label{fig:cylinderconfig}}
\end{figure}
We now present a heuristic explanation on why the conjectures can fail in certain kinematic regions. For this purpose it is useful to connect our conjectures to a `cylinder with caps' picture put forward in \cite{Hijano:2019qmi} (see figure \ref{fig:cylinderconfig})~\footnote{We thank Jo\~ao Penedones for pointing out the relevance of this picture for our formulae.}. This picture involves the complexification of the boundary positions and naturally ties the `real-time AdS/CFT' prescriptions of \cite{Skenderis:2008dh,Skenderis:2008dg} to the extraction of scattering amplitudes from conformal correlation functions. To see this, introduce new coordinates as:
\be
X = (R \cosh(r/R) \cosh(\tau), R \cosh(r/R)\sinh(\tau), R \sinh(r/R) \underline n')
\ee
with $\underline n'$ a new unit norm vector. Next we set $\tau = i t$ so we are in Lorentzian signature and the metric becomes:
\be
ds^2 = dr^2 - R^2 \cosh^2(r/R) dt^2 + R^2 \sinh(r/R) \,d\Omega_{d-1}^2
\ee
These are the standard global coordinates for Euclidean AdS. The map between old boundary coordinates $n^\mu$ and the new boundary coordinates $(\tau, \underline n')$ is easily found, and \eqref{momentaSmatrix} then implies that we need to set:
\be
\tanh(\tau) = - \frac{k^0}{m}, \qquad \frac{\underline n'}{\cosh(\tau)} = i \frac{\underline k}{m}
\ee
to obtain an S-matrix element with external momentum $k^\mu$. In terms of the Lorentzian coordinate $t$ this means that we can take:
\be \label{capvariables}
\begin{split}
\text{in state with $k^0 > 0$:} \qquad & t = - \pi / 2 + i\, \text{arccoth}(\phantom | k^0 \phantom |/m), \qquad \underline n' = - \frac{\underline k}{|\underline k|}\\
\text{out state with $k^0 < 0$:} \qquad &t = + \pi / 2 - i\, \text{arccoth}(|k^0|/m), \qquad \underline n' = - \frac{\underline k}{|\underline k|}
\end{split}
\ee
Notice that with our in-going conventions it is entirely reasonable that $\underline n'$ points in the opposite direction of $\underline k$. What is more interesting is the behavior of $t$: when tracing it in the complex time plane we see that we arrive exactly at the complex time contour sketched already in figures 1 and 2 in \cite{Skenderis:2008dh}, the essential bits of which we reproduced in figure \ref{fig:cylinderconfig}. The idea is that a Lorentzian segment, now with a length in global time of exactly $\pi$, is sandwiched between two Euclidean `caps' that are responsible, via operator insertions on their conformal boundary, for the initial and final state of the scattering event. Modulo Fourier transforms, this is exactly the same picture as transpired from \cite{Hijano:2019qmi}.

We can now use the `caps' picture to explain potential subtleties of our conjectures. To simplify the discussion we will consider a four-point function with two operators in the upper cap and the remaining two in the lower cap as in figure \ref{fig:cylinderconfig}.

\begin{figure}[t]
\centering
\begin{minipage}{0.32\hsize}
\centering
\includegraphics[clip,height=5.5cm]{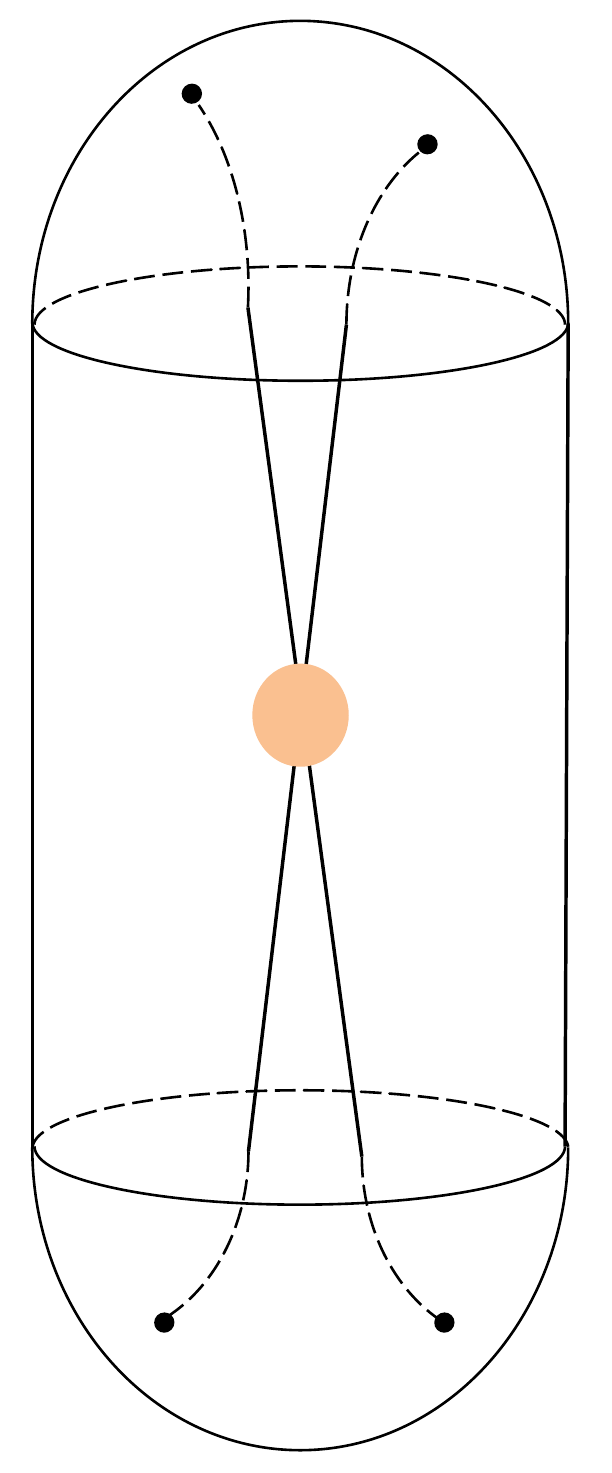}
\subcaption{ \label{fig:cylinder1}}
\end{minipage}
\begin{minipage}{0.32\hsize}
\centering
\includegraphics[clip,height=5.5cm]{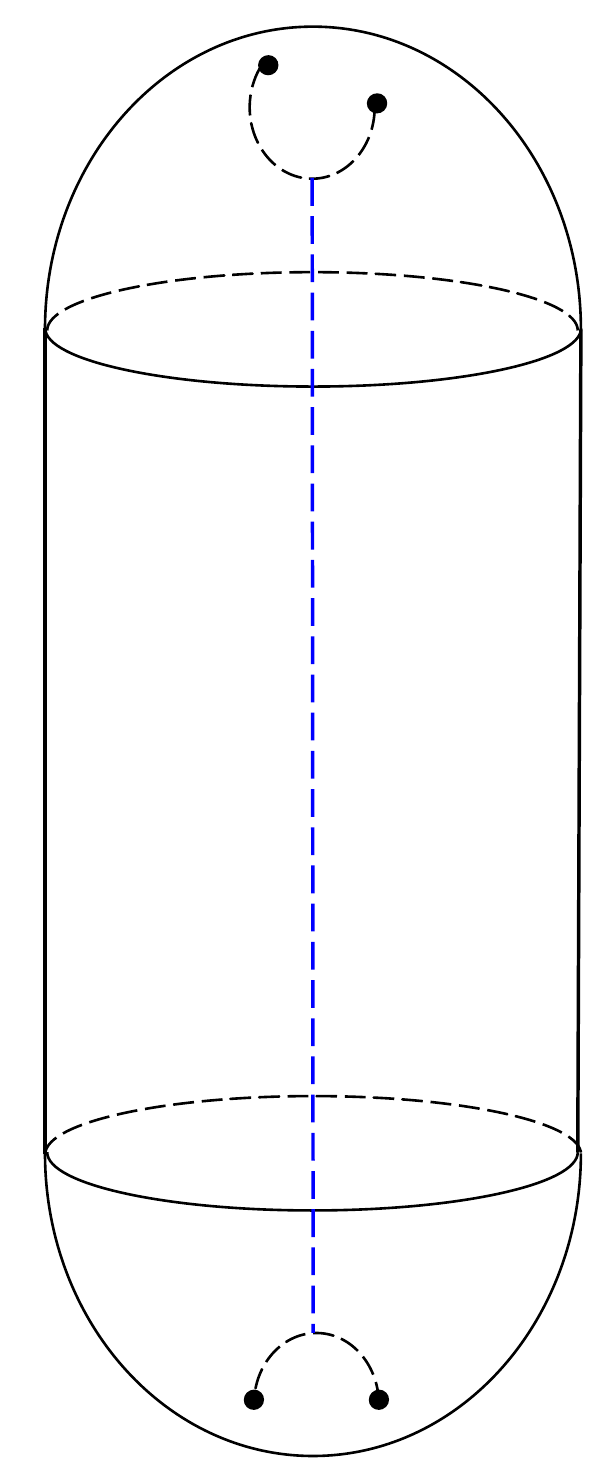}
\subcaption{ \label{fig:cylinder2}}
\end{minipage}
\begin{minipage}{0.32\hsize}
\centering
\includegraphics[clip,height=5.5cm]{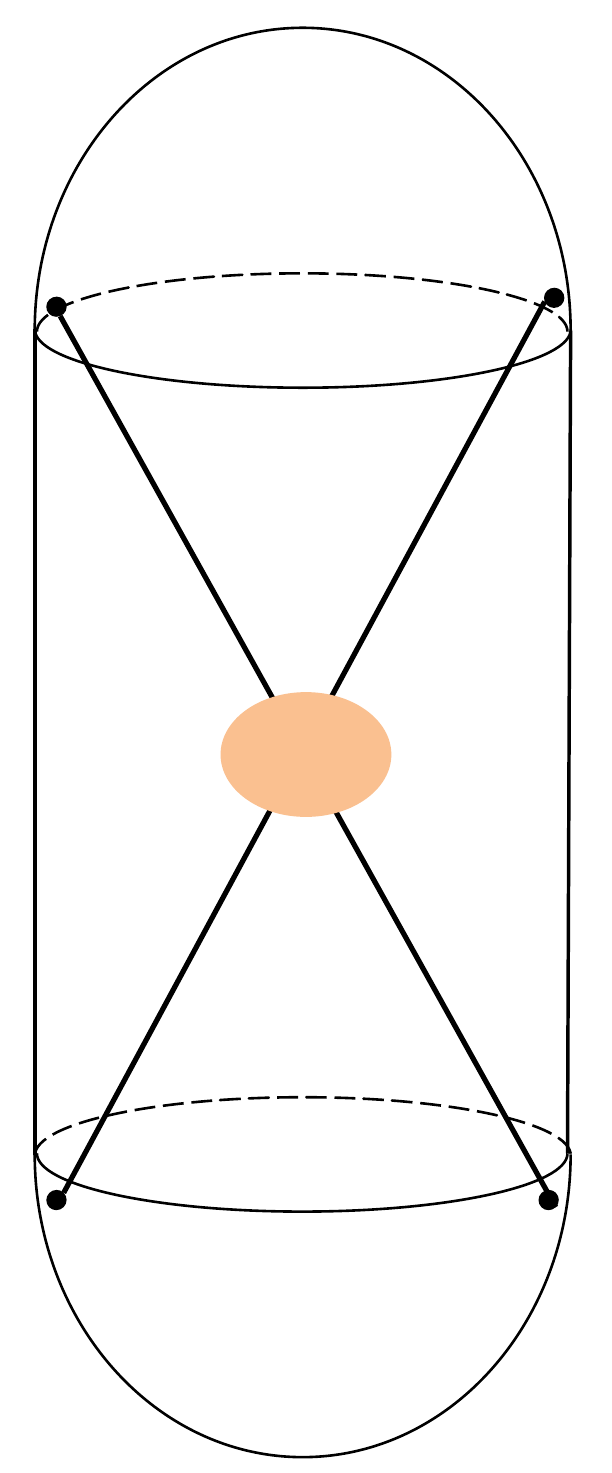}
\subcaption{ \label{fig:cylinder3}}
\end{minipage}
\caption{Geodesic configurations and flat-space scattering. (a) The geodesics that describe the flat-space scattering in the large $R$ limit. Two particles emitted from the operators in the lower Euclidean cap first tunnel to the Lorentzian cylinder, then scatter at the center of AdS and tunnel back to the upper Euclidean cap. (b) The geodesics that give a dominant contribution when two operators are close to each other. Two operators in the lower cap are directly connected by a Euclidean geodesic, and so are the operators in the upper cap. These two geodesics can also be connected by an exchange of some light particle (denoted by a blue dashed line). (c) The geodesics relevant for the bulk point limit which corresponds to $|k^0| \to \infty$ in our setup. When two operators are close to the edge of the lower cap, the dominant contribution is given by geodesics which connect four operators in the Lorentzian cylinder.}
\end{figure}

\paragraph{AdS as a particle accelerator}As discussed previously, particles dual to CFT operators become classical in the flat space limit and travel along geodesics inside AdS. In the present case, we have to find a geodesic in a mixed-signature spacetime since the bulk geometry consists of the Euclidean part and the Lorentzian part. To understand such a geodesic, let us start with the two operators inserted in the lower cap, see figure \ref{fig:cylinder1}. The particles emitted from these operators first need to `tunnel' to the Lorentzian cylinder following the Euclidean geodesics. In order to smoothly connect them to the Lorentzian geodesics, these two particles must have zero velocities when they emerge into the Lorentzian cylinder from below. Once they appear at the bottom of the cylinder, they then start to accelerate and approach each other owing to an attractive potential coming from the AdS curvature. In this sense, the AdS spacetime acts like a {\it particle accelerator}. Eventually, these particles collide at the center of AdS and scatter. Since the Compton wavelengths of these particles ($\propto 1/\Delta$) are much smaller than the AdS radius, the scattering process must be described by the flat-space S-matrix. The energy of the collision is detemined by how far apart the particles were at the bottom of the cylinder; the farther they were, more they get accelerated. After the collision, the particles move away from each other and eventually reach the top of the cylinder and then tunnel into the upper Euclidean cap.

\paragraph{Other geodesics}The discussion so far seems to support our conjectures on the flat-space limit. There is however one important subtlety: {\it the geodesic configuration described above is not the only one that contributes to the four-point function}. To understand this, let us consider a limiting case in which the two operators in the lower cap are very close to each other. In this case, we should take into account a Euclidean geodesic which directly connects these two operators (see figure \ref{fig:cylinder2}). This latter geodesic has a smaller Euclidean action than the one described above and therefore gives a dominant contribution. 

On the other hand, if we separate the two operators in the Euclidean cap, this latter geodesic tends to have a larger Euclidean action and therefore can be neglected. In particular, if we consider the so-called bulk-point limit \cite{Maldacena:2015iua} which in this picture corresponds to inserting the two operators at the edge of the cap, the former geodesic becomes entirely Lorentzian while the latter geodesic is Euclidean and is therefore suppressed. See figure \ref{fig:cylinder3}. 

These considerations suggest that the validity of our conjectures depends on the kinematics. Of course, it is hard to tell just from this heuristic argument when precisely they work. The purpose of the rest of this paper is to perform a more careful analysis and delineate the kinematic region in which they are supposed to hold.

\subsection{Conformal Mandelstam variables and kinematics}
In this subsection we discuss some important aspects of the kinematical relation \eqref{momentaSmatrix} between real Lorentzian momenta and complexified boundary positions. More details and technical derivations can be found in appendix \ref{Analytic continuation subsec}.

\subsubsection{Conformal Mandelstam variables}
Let us first say a few words about cross ratios. In terms of the spherical coordinates on the boundary of AdS, we have
\be\label{eq:PiPjmimj}
- P_i \cdot P_j = 1 - n_i \cdot n_j=1+\frac{k_i\cdot\eta\cdot k_j}{m_i m_j}
\ee
This equation immediately implies the following relation between conformal cross ratios and momenta:
\be\label{eq:crossratiorelationMandelstam}
\frac{(P_i\cdot P_j)(P_k\cdot P_l)}{(P_i\cdot P_k)(P_j\cdot P_l)}=\frac{(m_im_j+k_i\cdot \eta \cdot k_j)(m_km_l+k_k\cdot \eta \cdot k_l)}{(m_im_k+k_i\cdot \eta \cdot k_k)(m_jm_l+k_j\cdot \eta \cdot k_l)}\,.
\ee
By further imposing the momentum conservation\footnote{Interestingly, there always exists a conformal transformation that places the external points $P_i$ such that $\sum_i k_i = 0$ and therefore $s + t + \tilde u = 4m^2$ in terms of Mandelstam invariants. This is why, in the equations below, there are never three independent Mandelstam invariants which would be too many to match against the two independent cross ratios.} 
\be
\sum_i k_i = 0\,,
\ee
one can rewrite the right hand side of \eqref{eq:crossratiorelationMandelstam} in terms of Mandelstam invariants. 

It is instructive to work out the relation explicitly for four-point functions of identical scalar operators. The familiar cross ratios are then:
\be
u\colonequals \frac{P_{12} P_{34}}{P_{13}P_{24}}\,,\qquad v\colonequals \frac{P_{14}P_{23}}{P_{13}P_{24}}\,,\qquad \big(P_{ij}\colonequals -2 P_i\cdot P_j\big)\,,
\ee
but it will sometimes be better to use either the Dolan-Osborn variables $(z,\bar{z})$ \cite{Dolan_Osborn,Dolan:2003hv} or the radial coordinates $(\rho,\bar{\rho})$ \cite{Hogervorst:2013sma}:
\be\label{eq:CRdef}
u=z\bar{z}\,,\quad v=(1-z)(1-\bar{z})\,,\quad z=\frac{4\rho}{(1+\rho)^2}\,,\quad \bar{z}=\frac{4\bar{\rho}}{(1+\bar{\rho})^2}\,.
\ee
We then get the relation\footnote{While preparing this paper, \cite{Aprile:2020luw} appeared in arXiv
in which a similar relation between the cross ratios and the Mandelstam variables was discussed. It also
discusses the relation between the momentum conservation and the saddle-point equation, which we explain
in section 3.2. As acknowledged in that paper, the results in this paper were obtained prioir to the publication
of \cite{Aprile:2020luw}.}
\be
\begin{aligned}
&\\
&\hspace{-4cm}\textbf{Conformal Mandelstam variables}:\\
&s\colonequals -(k_1+k_2)^2=4m^2 \left(\frac{1-\sqrt{\rho\bar{\rho}}}{1+\sqrt{\rho \bar{\rho}}}\right)^2\,,\\
&t\colonequals -(k_1+k_4)^2=4m^2 \left(\frac{\sqrt{\rho}+\sqrt{\bar{\rho}}}{1+\sqrt{\rho \bar{\rho}}}\right)^2\,,\\
&\tilde{u}\colonequals -(k_1+k_3)^2=-4m^2 \left(\frac{\sqrt{\rho}-\sqrt{\bar{\rho}}}{1+\sqrt{\rho \bar{\rho}}}\right)^2\,.
\end{aligned} \label{eq:conformalMandelstam}
\ee
(Here we used $\tilde{u}$ instead of the conventional notation $u$ in order to distinguish it from the cross ratio $u$.) These equations \eqref{eq:conformalMandelstam} are a new parametrization of the conformal cross ratios of the boundary CFT correlators, chosen precisely such that they become the Mandelstam variables of the scattering amplitudes in flat space. For AdS$_2$ this relation was derived previously in \cite{TTBar} and the result here generalizes it to arbitrary dimensions.

\begin{figure}[t]
\centering
\includegraphics[clip,height=4cm]{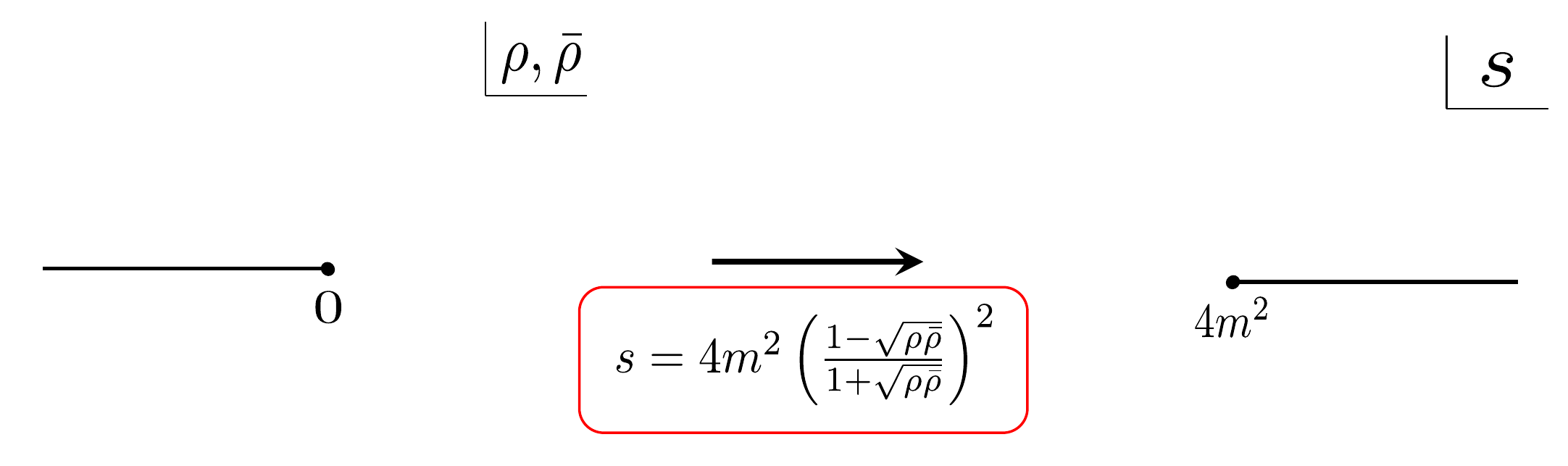}
\caption{The analytic structure of the four-point function in CFT and the flat-space S-matrix. The conformal Mandelstam variables map the branch-cut singularities of the four-point function to a two-particle threshold of the S-matrix. \label{fig:analyticmap}}
\end{figure}

The above equations map the Euclidean CFT kinematics where $\rho$ and $\bar \rho$ are complex conjugates 
to the \emph{Euclidean region} where $0 < s, t, \tilde u < 4m^2$, which is the orange triangle in the center of figure \ref{fig:mandelstamplanemain} shown below. 
To reach physical kinematics of scattering amplitudes, or indeed any other region, some careful analytic continuations are needed and we will discuss those below. One interesting initial observation is that the expected two-particle branch cut at $s = 4m^2$ is built in from the beginning: according to equation \eqref{eq:conformalMandelstam} we just inherit it from the branch cuts at $\rho, \bar \rho = 0$ that exist in any correlation function, even before taking the flat-space limit.\footnote{This simply follows from the fact that the operator product expansion generally gives terms like $(\rho\bar{\rho})^{\Delta}$ with $\Delta$ being noninteger. Therefore if we view the correlator as a function of two independent parameters $\rho$ and $\bar{\rho}$, it has a branch cut at $\rho,\bar{\rho}=0$.} We illustrated this in figure \ref{fig:analyticmap}. This behavior should perhaps be contrasted with the Mellin space prescriptions: in Mellin space the idea is that infinite sequences of poles condense into cuts and it is for example impossible to explore other Riemann sheets before taking the flat-space limit. On the other hand, whereas our prescription nicely yields the two-particle threshold there is no sign of any further cuts or poles (at least on the first sheet) because conformal correlation functions are always perfectly analytic in the Euclidean region. In section \ref{sec:examples} and beyond we will see that this is very much related to the subtleties already discussed in section \ref{subsec:capsandsubleties}.

\subsubsection{Analytic continuations}
By conformal invariance, the $n$-point boundary correlation functions in the flat Euclidean boundary metric depend only on the combinations $- P_i \cdot P_j$. Contact or light-cone singularities arise when there are $i$, $j$ such that $- P_i \cdot P_j = 0$. These singularities correspond to the end points of branch cuts of position-space CFT correlators, which extend to infinity along the negative real axis in the complex plane of $- P_i \cdot P_j$. From the last expression in equation \eqref{eq:PiPjmimj} we find that
\begin{align}
&\text{$i$ in, $j$ out or vice versa: } & 
 - P_i \cdot P_j|_{\text{S-matrix}} \geq 2 \nonumber \\
&\text{$i$, $j$ both in or both out: } &
 - P_i \cdot P_j|_{\text{S-matrix}} \leq 0
\label{continuation_Pi_Pj}
\end{align}
with the inequalities holding by virtue of the fact that all the $k^\mu_i/m_i$ are unit norm timelike vectors with $|k^0_i| \geq m_i$. The second continuation precisely lands us on a branch cut. To see how we should approach the branch cut, recall that in flat space one requires the corresponding Mandelstam invariants like $s_{ij} = - (k_i + k_j) \cdot \eta \cdot (k_i + k_j)$, to have a small \emph{positive} imaginary part. In terms of such variables $ - 2 m_i m_j P_i \cdot P_j = (m_i+m_j)^2 - s_{ij}$, so we will need to give a small \emph{negative} imaginary part to $- P_i \cdot P_j$ when $i$ and $j$ are either both `in' or both `out'.

Let us return to the cross ratios for the four-point functions of identical operators. With the $i \epsilon$ prescription understood, it is not hard to deduce (see appendix \ref{Analytic continuation subsec}) that reaching physical kinematics in the $s$-channel means that we should set the $(\rho,\bar \rho)$ variables to:
\be
\label{rhorhobtomandelstam}
\rho = \frac{\sqrt{s}-2 m}{\sqrt{s} + 2 m}e^{i (\theta - 2 \pi )},\qquad \rhob = \frac{\sqrt{s}-2 m}{\sqrt{s} + 2m}e^{-i \theta}
\ee
where $\theta$ is the scattering angle defined through
\be
t = \frac{1}{2} (4m^2 - s) (1 - \cos(\theta)),\qquad \tilde u = \frac{1}{2} (4m^2 - s)(1 + \cos(\theta))\,.
\ee
The factor $2 \pi i$ indicates that $\rho$ should be evaluated on the second sheet obtained by circling around zero in a clockwise fashion whereas $\bar \rho$ remains on the first sheet.

\begin{figure}[ht]
\begin{center}
\includegraphics[width=12cm]{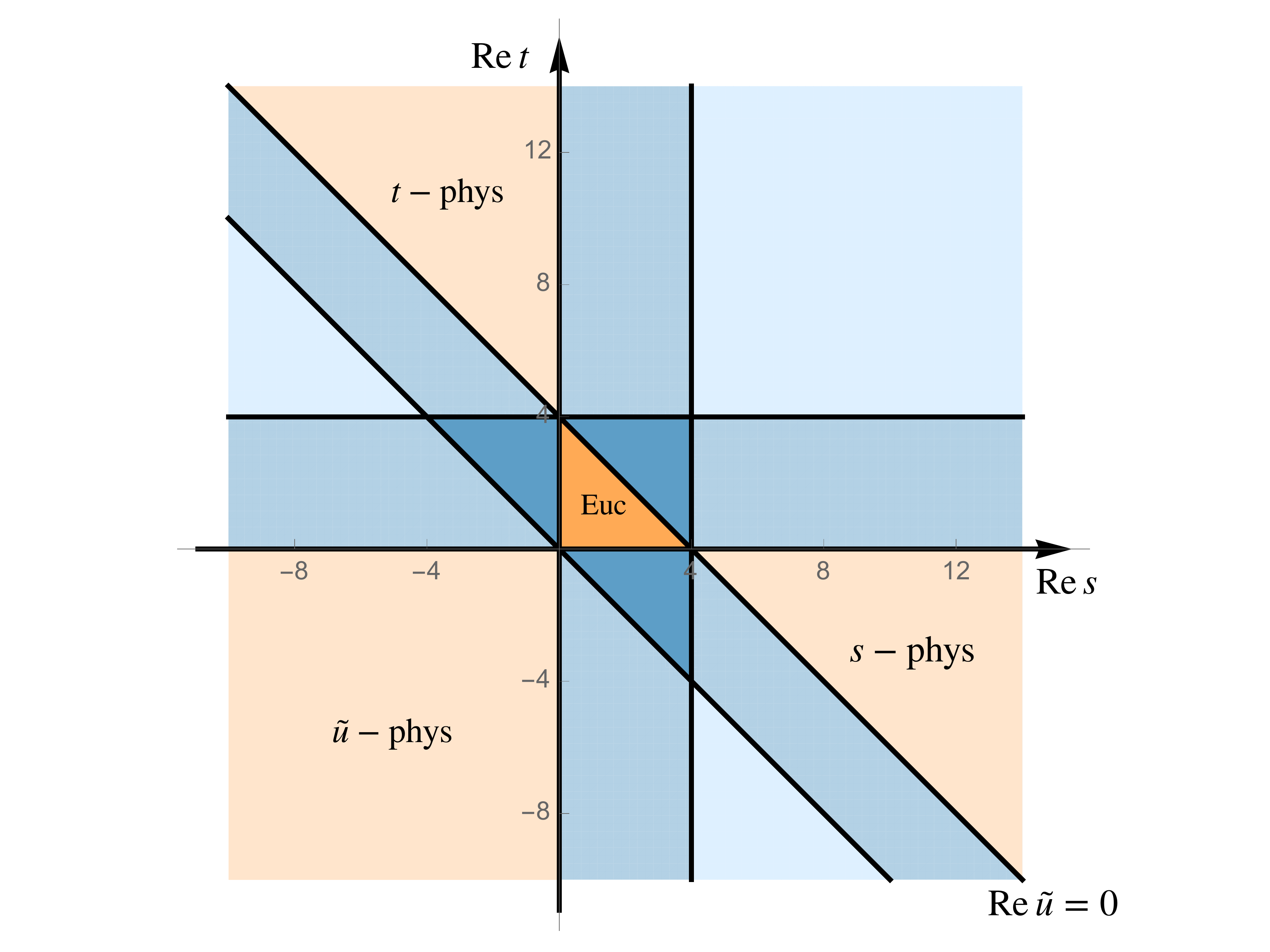}
\caption{\label{fig:mandelstamplanemain}The Mandelstam $(s,t)$ plane, colored according to the analytic continuation necessary to reach each region. The main distinction is orange versus blue: in the former one should take $\rho$ and $\bar \rho$ to be complex conjugates and in the latter they are real and independent. A refinement is the dark versus light shading. In the darkest central triangles $\rho$ and $\bar \rho$ should be taken to live on the first sheet, and passing to lighter shades corresponds to one or two analytic continuations of either $\rho$ or $\bar \rho$ around either $0$ or $1$. Exactly which sheet corresponds to which region is detailed in appendix \ref{Analytic continuation subsec}.}
\end{center}
\end{figure}

We can also consider the Mandelstam plane more generally and to find the analytic continuations in the cross ratios that are necessary to reach all its different regions. This is done in detail in appendix \ref{Analytic continuation subsec} and we summarize a few essential fact in figure \ref{fig:mandelstamplanemain}.

\subsubsection{Kinematic limits}
\label{sec:kinematic limits}
The conformal Mandelstam variables allow us to relate kinematic limits of the CFT correlator to those of the scattering amplitude. Here we summarize the correspondence relegating more details including the derivations to appendix \ref{Analytic continuation subsec}.

\paragraph{OPE limit} Let us first analyze the $s$-channel OPE limit in which both $\rho$ and $\bar{\rho}$ approach zero: $\rho,\bar{\rho}\to 0$. Using the conformal Mandelstam variables \eqref{eq:conformalMandelstam}, we can immediately see that this limit corresponds to the low-energy limit, or more precisely the near-threshold limit
\be
s\sim 4m^2\,,\qquad t,\tilde{u}\sim 0\,.
\ee
\paragraph{Regge limit} We next consider the Regge limit of the CFT correlator discussed in \cite{Cornalba:2006xk,Cornalba:2007fs,Maldacena:2015iua,Costa:2012cb}. Following \cite{Costa:2012cb}, we write the cross ratios as
\be
u=\sigma^2\,,\qquad v=(1-\sigma e^{\rho})(1-\sigma e^{-\rho})\simeq 1-2\sigma \cosh\rho\,.
\ee
Then, to get to the Regge limit, we first analytically continue $v$ as
\be
v\to e^{2\pi i }v \,, 
\ee
and send $\sigma\to 0$ while keeping $\rho$ fixed. In this limit, the conformal Mandelstam variables scale as
\be
s\sim 4m^2\left(1-\frac{1}{\cosh^2 \frac{\sigma}{2}}\right)\,,\qquad t\sim -\tilde{u} \sim \frac{4m^2}{\sigma}\frac{1}{\cosh^2 \frac{\sigma}{2}}\gg 1\,.
\ee
This in fact corresponds to the Regge limit of scattering amplitudes (in the $t$-channel). This is of course expected from various results in the literature but the virtue of the conformal Mandelstam variables is that it makes the relation transparent. 
\paragraph{Bulk-point limit} Another interesting limit is the so-called bulk-point limit studied in \cite{Maldacena:2015iua}. This limit corresponds to the following analytic continuation\footnote{Here we are following the definition of the bulk-point limit in \cite{Maldacena:2015iua}. To relate it to the analytic continuation to physical scattering kinematics discussed above \eqref{rhorhobtomandelstam}, we need to perform a further Euclidean rotation $\rho \to e^{-\pi i} \rho$ and $\bar{\rho}\to e^{\pi i}\bar{\rho}$.} of the radial coordinates $\rho$,
\be
\rho =e^{-i\pi -\epsilon}e^{i\varphi}\,,\qquad \bar{\rho}=e^{-i\pi -\epsilon}e^{-i\varphi}\,.
\ee
Here $\epsilon$ is the regularization parameter, which will be sent to $0$ in the bulk-point limit. In this limit, the conformal Mandelstam variables scale as
\be
s\to \infty\,,\qquad t\to -\infty\,,\qquad  \tilde{u}\to -\infty\,,
\ee
while the scattering angle $\theta$ is finite. This is a fixed-angle high energy scattering limit, which was studied by  Gross and Mende \cite{Gross:1987ar} in string theory.
\paragraph{Massless limit} Although this is not a kinematic limit, it is interesting to discuss the massless limit $m\to 0$. If we naively take this limit, the conformal Mandelstam variables \eqref{eq:conformalMandelstam} all vanish. In order to have finite Mandelstam variables, we need to approach the bulk-point limit as we send $m$ to $0$. This is consistent with the results in the literature on the flat-space limit of the massless scattering, all of which involve taking the bulk-point limit. It would be interesting to clarify the precise relation between our proposal and those results, in particular the position-space approach to the massless scattering discussed in \cite{Joao_string_scattering}. We leave it for future investigations.
\paragraph{Double lightcone limit} Finally let us briefly mention the double lightcone limit, which corresponds to
\be
u=\epsilon(1-\eta)\,,\qquad v=(1-\epsilon)\eta\,,
\ee
with $\epsilon\ll \eta \ll 1$. In this limit, we obtain
\be
s\sim 4m^2(1-2\sqrt{\epsilon})\,,\quad t\sim 4m^2 (1-2\sqrt{\eta})\,,\quad \tilde{u}\sim -4m^2 (1-2(\sqrt{\epsilon}+\sqrt{\eta}))\,. 
\ee
To our knowledge, this limit does not correspond to a well-studied limit of scattering amplitudes. However, given the role the double lightcone limit played in the development of the analytic conformal bootstrap, it might be worth studying this limit in the flat-space scattering.


\section{Illustrative examples}
\label{sec:examples}
We now test our conjectures in several simple and illustrative examples: two-point functions, contact diagrams and four-point exchange diagrams. The goal of this section is threefold. First, we explain the details of how the formula works in simple cases. Second, we point out that the saddle-point equations for the geodesic networks in AdS can be interpreted as the momentum conservation at each bulk vertex. We also see a natural connection with the flat-space limit of the Mellin amplitude, which we explore more in section \ref{sec:mellin}. Third, we discuss how and when our formula stops working using the exchange diagram as an illustrative example. In section \ref{sec:landaudiagrams} below, we combine the latter two observations and propose the AdS analogue of Landau diagrams, which delineate the kinematic regions in which the position-space recipe for the flat-space limit gives a divergent answer.

\subsection{Two-point functions}
\label{subsec:twopt}
The easiest example for which we can test \eqref{Eqn_flatspaceSconj} is the two-point function. In our normalization
\be
\vev{\OO(n_1)\OO(n_2)} = \frac{\CC_\Delta 2^{-\Delta} }{(1 - n_1 \cdot n_2)^{\Delta} }
\ee
We multiple by the factor $Z$ and use the continuation in \eqref{momentaSmatrix} to move particle 1 to the `in' position, so $(n_1^0, \underline n_1) = (-q^0_1,i\underline{q}_1)/m$ and particle 2 to the `out' position, which we can write as $(n_2^0, \underline n_2) = (q_2^0, -i {\underline q}_2)/m$, with $q^0 \geq 0$ and $q^2 = - m^2$ in both cases. This yields
\be
Z \vev{\OO(n_1)\OO(n_2)}|_{\text{S-matrix}} = \frac{Z \,\CC_\Delta 2^{-\D} }{ (1 + q_1^0 q_2^0/m^2 - \underline q_1 \cdot \underline q_2 / m^2)^\Delta} = \frac{2^{\Delta} \CC_\Delta^{-1} R^{d-1}}{(- \eta_{\mu \nu} (q_1 + q_2)^\mu (q_1 + q_2)^\nu / m^2)^\Delta}
\ee
This expression is best understood by going to a frame where $q_1 = (m,\underline 0)$. We get
\be
\frac{2^{\Delta} \CC_\Delta^{-1}R^{d-1} }{\left(1 + \sqrt{1 + \underline q_2^2/m^2}\right)^{\Delta}}
\ee
and we observe that for large $\Delta$ this function starts to look like a delta function singularity, in the sense that it becomes a positive `bump' with support contracting to the point where $\underline q_2 = 0$. To check that all the factors come out right we can integrate:
\be
\int d^d {\underline q} \frac{2^{\Delta} \CC_\Delta^{-1}R^{d-1} }{\left(1 + \sqrt{1 + \underline q^2/m^2}\right)^{\Delta}} = 2 ( 2 \pi m R)^d \frac{\D\Gamma(\D - d)}{\Gamma(\D)} \overset{R\to\infty}{\xrightarrow{\hspace*{0.6cm}}} 2 m (2 \pi)^d 
\ee
which demonstrates that, more generally,
\be
Z \vev{\OO(n_1)\OO(n_2)}|_{\text{S-matrix}} \overset{R \to \infty}{\longrightarrow} 2 E_1 (2 \pi)^d \delta^{(d)}(\underline q_1 - \underline q_2)
\ee
thus proving our general formula \eqref{Eqn_flatspaceSconj} for single-particle states.

\subsection{Contact diagram and momentum conservation}
\label{subsec:contact}
For our next example we consider $n$-point contact diagrams, which according to our conjectures should give rise to the momentum-conserving delta function in the flat-space limit. The diagram can be written as
\be
G_c(P_i) = \int d X \prod_{i = 1}^n G_{B\partial}(X,P_i) =  \frac{1}{R^{n(d-1)/2}} \int d X \prod_{i = 1}^n \mathcal{C}_{\Delta_i}2^{-\Delta_i} e^{-\Delta_i\log(-P_i\cdot X/R)}
\ee
\subsubsection{Vertex momenta and vertex Mandelstam invariants}
\begin{figure}[t]
\centering
\includegraphics[clip,height=5cm]{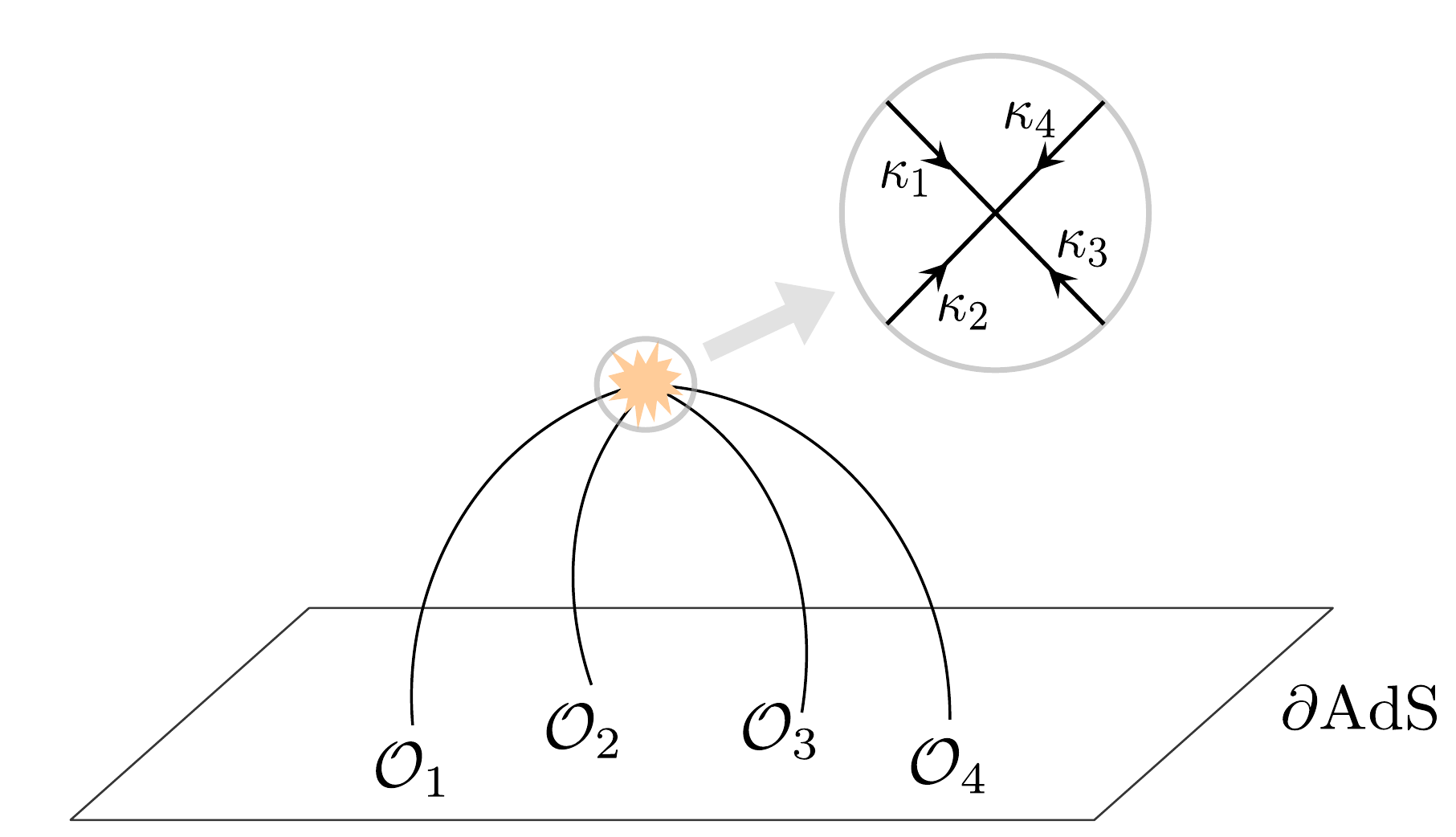}
\caption{Vertex momenta and their conservation. The saddle-point equation for the contact diagram can be interpreted as the momentum conservation of `vertex momenta' (denoted by $\kappa_j$'s), which are momenta of particles measured at a point where all particles meet. Since the particles follow curved trajectories, these momenta in general do not coincide with the boundary momenta introduced in section \ref{Section_Erasing the circle}.\label{fig:contactmomenta}}
\end{figure}

We will analyze the Euclidean correlator for now, which means that the integral over $X$ is over the hyperboloid $X^2 = - R^2$ and $X^0 > 0$. In the flat-space limit all the scaling dimensions become large and we can use a saddle point approximation for the integral. The relevant function to extremize is then
\be
f_c(X) = - \sum_i \Delta_i \log(-P_i \cdot X/R) + \lambda (X^2 + R^2)
\ee
with $\lambda$ a Lagrange multiplier. In more detail, we define the integrals as
\be
\int_{\text{AdS}} dX =  2 R \int_{-\infty}^{\infty}\! d^{d+2} X \, \theta(X^0) \delta(X^2 + R^2) = 2 R \int_{-i\infty}^{i \infty} \frac{d\lambda}{2 \pi i} \int_{-\infty}^{\infty} \!d^{d+2} X \, \theta(X^0) e^{\lambda(X^2 + R^2)}
\ee
with the factor $2R$ inserted so the volume element agrees with the one given by the metric in equation \eqref{adsmetric}. The saddle point equation becomes:
\be
\sum_i \Delta_i \frac{P_i}{P_i \cdot X} - 2 \lambda X  = 0
\ee
which we can contract with $X$ to yield 
\be
 \lambda = - \frac{1}{2 R^2} \sum_i \Delta_i\,.
\ee
and substituting this back we obtain that
\be \label{contactsaddle}
\sum_i \Delta_i \left(\frac{P_i}{P_i \cdot X / R} + X/R \right)= 0\,.
\ee
Now comes a crucial observation: We can interpret \eqref{contactsaddle} as a momentum-conservation condition at the interaction vertex in AdS. To see this, introduce $\kappa_i$ defined as
\be
\kappa_i \colonequals \frac{\Delta_i}{R} \left(\frac{P_i}{P_i \cdot X /R} + X/R \right)\,.
\ee
In terms of these variables, the saddle-point equation \eqref{contactsaddle} indeed takes the form of the momentum conservation
\be
\sum_i \kappa_i = 0\,.
\ee
In addition, they are on-shell (in the Euclidean sense) and tangent vectors to AdS, i.e.,
\be
\kappa_i^2 = m_i^2, \qquad \qquad X \cdot \kappa_i = 0\,.
\ee
For these reasons, we call these variables `vertex momenta'. Geometrically these vectors measure the momenta of particles at the position of the interaction vertex in AdS (see figure \ref{fig:contactmomenta}). The fact that the saddle-point equation for the geodesics coincides with the momentum conservation was first pointed out in \cite{Minahan:2012fh} for three-point functions. Our analysis provides a simple generalization of that statement to higher-point functions.  

It is instructive to introduce also the `vertex Mandelstam invariants'. If we set
\be
\sigma_{ij} \colonequals - \frac{\D_i \D_j}{\sum_k \D_k} \frac{P_i \cdot P_j}{(P_i \cdot X / R)(P_j\cdot X / R)} = \frac{\D_i \D_j}{\sum_k \D_k} \left( 1 - \frac{\kappa_i \cdot \kappa_j}{m_i m_j} \right) 
\ee
then the saddle point equations \eqref{contactsaddle}, contracted with $P_j$, can be written as:
\be\label{eq:sumsigmas}
\sum_{j} \sigma_{ij} = \Delta_i\,.
\ee
Note furthermore that for $n \geq 4$ the relation between the $\sigma_{ij}$ and the $P_{i}$ is constrained via their cross ratios,
\be\label{eq:additionalcond}
\frac{\sigma_{ij} \sigma_{kl}}{\sigma_{ik} \sigma_{jl}} = \frac{(P_i \cdot P_j)(P_k \cdot P_l)}{(P_i \cdot P_k)(P_j \cdot P_l)}\,.
\ee
Namely the cross ratios of the vertex Mandelstam invariants coincide with the conformal cross ratios of CFT.
The previous two equations give precisely enough constraints to completely determine the $\sigma_{ij}$. The beauty of using the vertex Mandelstam variables is that they turn the saddle-point equations, which are originally constraints on $(d+2)$-component vectors, into simple algebraic equations \eqref{eq:sumsigmas} and \eqref{eq:additionalcond} for $\sigma_{ij}$. Once they are determined, one can compute $X$ from 
\be \label{XviaPi}
P_i \cdot X/R = -  \frac{\D_i}{\sqrt{\sum_k \D_k}} \sqrt{ - \frac{\s_{jk}}{\s_{ij} \s_{ik}} \frac{(P_i \cdot P_j)(P_i \cdot P_k)}{P_j \cdot P_k} }\,.
\ee

Readers familiar with the Mellin space description of a correlator \cite{Mack:2009mi,joaomellin} would immediately notice an interesting similarity: if one replaces $\sigma_{ij}$ in \eqref{eq:sumsigmas} with the Mellin variables $\gamma_{ij}$, equation \eqref{eq:sumsigmas} coincides with the familiar constraint on $\gamma_{ij}$. There is indeed a more precise connection. In section \ref{sec:mellin} we will explain that the Mellin representation in the flat-space limit can be evaluated via the saddle point method, and at the saddle point the $\gamma_{ij}$ become equal to the $\sigma_{ij}$ and therefore in particular obey equation \eqref{eq:additionalcond}. In the remainder of this section we will however stick to the position-space description and discuss the explicit solutions to the saddle-point equations and their physical implications.

\subsubsection{Boundary momenta vs. vertex momenta}
So far, we have introduced two different notions of momenta for the CFT correlators. First there are the boundary momenta $k_i$ that we used in section \ref{Section_Erasing the circle} to state our conjecture. In Euclidean kinematics it is better to momentarily forget about the $i$'s in equation \eqref{momentaSmatrix} and set:
\be
k^\mu_j = - \frac{\Delta_j}{R} n_{P_j}^\mu, \qquad \text{where } P_j = (1, n_{P_j})\,.
\label{Eqn_P_i - k_i relation}
\ee
These momenta are on-shell, $k_j^2 = m_j^2$, but it is not at all necessary for them to be conserved since we are free to choose arbitrary values of the $n_P$. Physically the $k_j^\mu$ measure the momenta of particles at the boundary of AdS, and the relative minus sign means that they are ingoing.

A second set of momenta are the vertex momentum $\kappa_i$ which measure the momenta of particles at the position of the interaction vertex in AdS and were introduced in the previous subsection. Like the boundary momenta these are also on-shell, but unlike the boundary momenta they always satisfy the momentum conservation condition. This indicates that the vertex and boundary momenta do not agree in general.

The discrepancy arises, of course, because the hyperbolic space is not flat and particles move along curved trajectories.\footnote{The $k_i$ and $\kappa_i$ are normalized tangent vectors to the geodesic, so the contraction with any Killing vector field is conserved along the trajectories. But this is not relevant for the component-wise comparison in these paragraphs.} For given $P_i$ the saddle point equations select the particular bulk point $X$ such that the particles interact at the vertex in a \emph{locally} momentum-conserving fashion. The relation between $P_i$ and $X$ is, especially for higher-point functions, quite complicated, and it is therefore not always easy to determine the vertex momenta. Nevertheless, there is a simple and beautiful relation between these two momenta if we are in a special kinematics in which the {\it boundary momenta} $k_i$ are also conserved. To see this, let us take
a closer look at the momentum conservation condition for the boundary momenta $\sum_i k_i = 0$. Using \eqref{Eqn_P_i - k_i relation}, we can rewrite it into
\be
\frac{1}{\sum_k \Delta_k} \sum_{i} \Delta_i P_i = C \colonequals (1, 0, 0, \ldots)\,.
\ee
Now, for this particular choice of the boundary points, the saddle-point equation for the vertex momenta \eqref{contactsaddle} becomes trivial to solve: we find that $X = RC$ does the job since $P_i \cdot C = - 1$. It immediately follows that $\kappa_i = (0, k_i)$, so bulk and vertex momenta agree, and therefore the Mandelstam invariants for the boundary momenta also coincide with the vertex Mandelstam invariants $\sigma_{ij}$.

Restricting the $P_i$ to the support of the momentum conserving delta function would be sufficient to extract the amplitude $\mathcal T(\ldots)$ as follows from our amplitude conjecture. That said, we should remember that important information is lost if we impose the boundary momentum conservation from the outset: our S-matrix conjecture states that the contact diagram, when suitably continued to Lorentzian signature, becomes a momentum-conserving delta function. To verify this, we need to start with a more general configuration in which the boundary momenta are {\it not} conserved and carefully analyze what happens if we approach the support of the momentum-conserving delta function. This analysis turns out to be quite complicated in general. So we will consider only $n = 3$ and $n = 4$ in what follows.

\subsubsection{General momenta, $n = 3$}
As a warm up, let us consider the three-point function, $n=3$. In this case, we can solve the saddle point equation \eqref{contactsaddle} even in the absence of the conservation of the boundary momenta. Specifically we try an ansatz of the form
\be \label{Xansatz}
X/R = \sum_j c_j P_j
\ee
and the saddle point equations then determine the coefficients
\be
c_1 = \sqrt{- \frac{P_2 \cdot P_3}{2(P_1\cdot P_2)(P_1 \cdot P_3)} \frac{\D_{12|3} \D_{13|2}}{\D_{23|1} \sum_k \D_k}}
\ee
with $\D_{12|3} = \D_1 + \D_2 - \D_3$, and cyclic permutations thereof. The vertex momenta obey
\be\label{eq:vertexman3}
\sigma_{12} = \frac{1}{2} \Delta_{12|3}\,.
\ee
Note that, for the three-point function, \eqref{eq:vertexman3} immediately follows from the saddle-point equation written in terms of the vertex Mandelstam variables, \eqref{eq:additionalcond}.

The analysis of the three-point function is certainly a simple and instructive exercise but unfortunately there is not much more we can say, since there are no physical three-point scattering processes and we cannot really see the emergence of the delta-function. Thus we will not work out the details of the flat-space limit any further. 

\begin{figure}[t]
\centering
\begin{minipage}{0.45\hsize}
\centering
\includegraphics[clip,height=4cm]{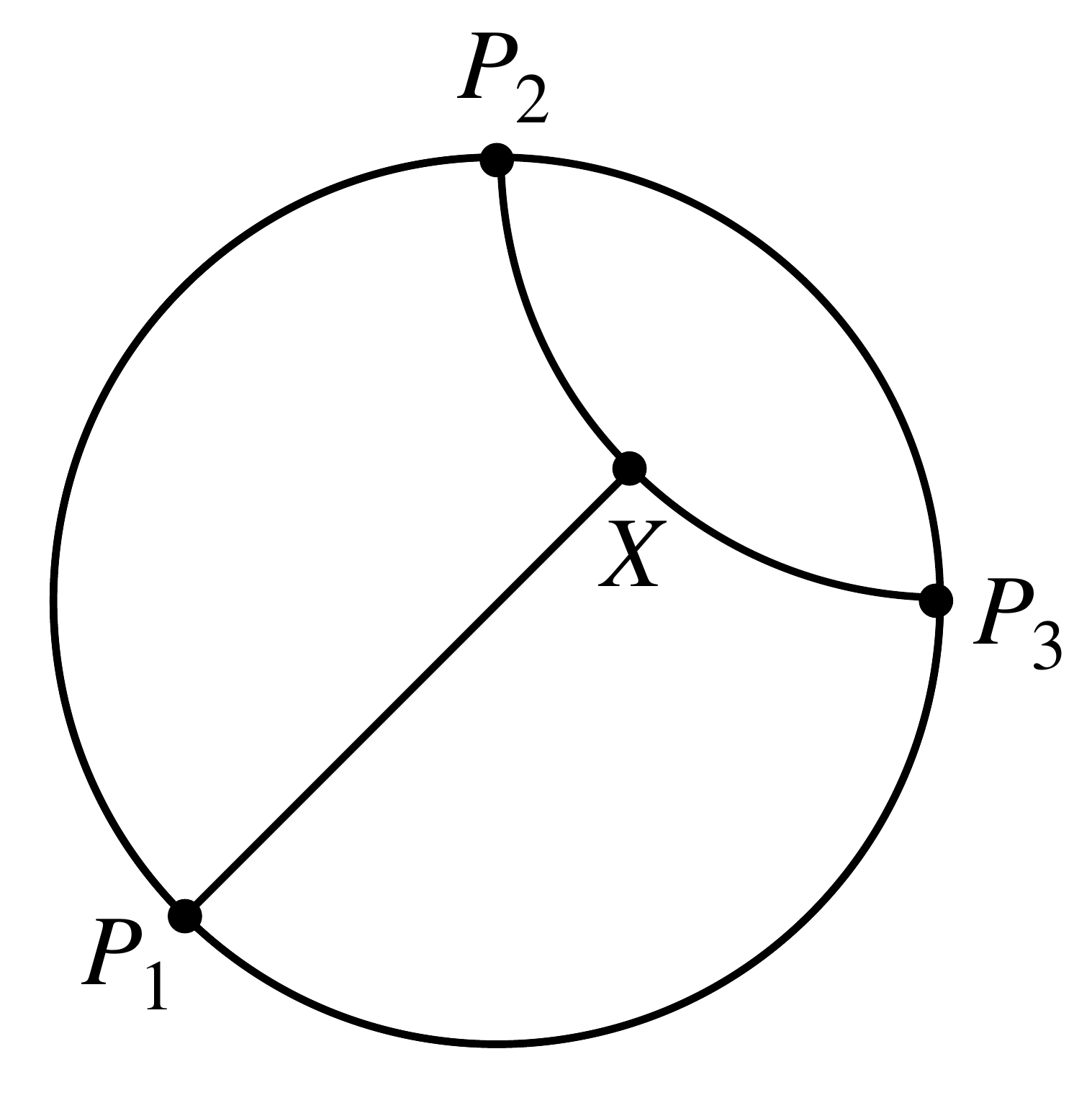}
\subcaption{$\D_1\to0$ \label{Fig_3pt_diagram_1}}
\end{minipage}\hspace{10pt}
\begin{minipage}{0.45\hsize}
\centering
\includegraphics[clip,height=4cm]{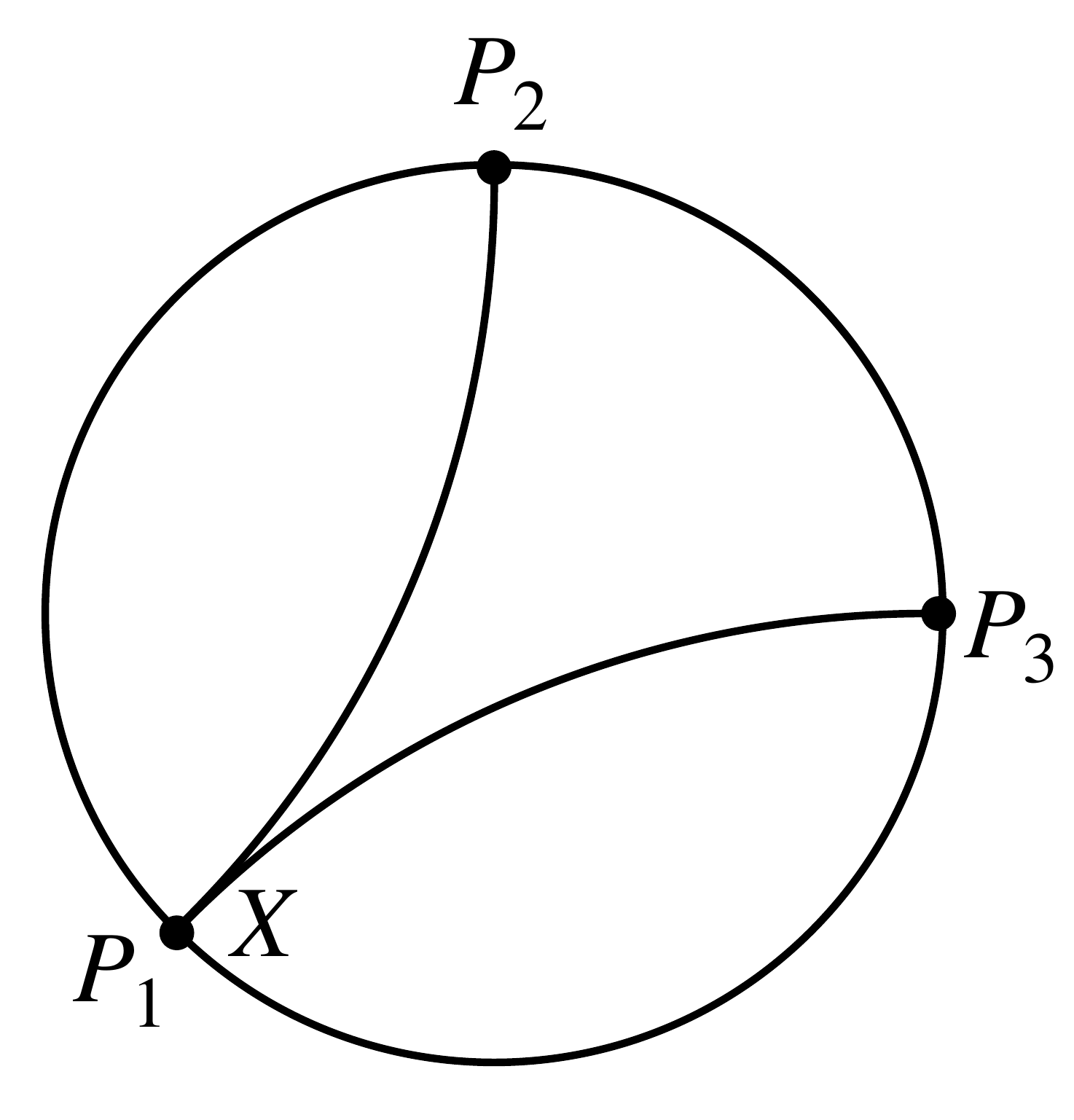}
\subcaption{$\D_{23|1}\to0$ \label{Fig_3pt_diagram_2}}
\end{minipage}
\caption{Two different limits of the three-point diagram with $\D_2=\D_3$.}
\label{Fig_3pt_diagram}
\end{figure}

Instead let us briefly mention two specific limits of the $\Delta_i$ variables for later reference. First we can take the limit where $\Delta_2 = \Delta_3$ and send $\Delta_1$ to 0. In that case $c_1$ goes to zero, so $X$ becomes a linear combination of $P_2$ and $P_3$ which means that $X$ lies on the geodesic between $P_2$ and $P_3$ in AdS. Another possible limit is the `decay' limit where we send, say $\D_{23|1}$  to zero so particle $1$ can (almost) decay into particles 2 and 3. In that case $c_1$ blows up whereas $c_2$ and $c_3$ go to zero, which means that $X$ approaches $P_1$. These are drawn in figure \ref{Fig_3pt_diagram}.

\subsubsection{General momenta, $n = 4$}
Let us now consider a more physically interesting example, the contact diagram for four identical particles. We will do a detailed analysis and show that the momentum-conserving delta function appears from the saddle point value of the diagram, in accordance with our S-matrix conjecture. 
Let us first determine the $\sigma_{ij}$ by solving the algebraic equations \eqref{eq:sumsigmas} and \eqref{eq:additionalcond}. The result is given purely in terms of the conformal cross ratios \eqref{eq:CRdef} as
\be\label{eq:sigmasconformal} 
\begin{aligned}
&\sigma_{12}=\sigma_{34}=\Delta \frac{\sqrt{u}}{1+\sqrt{u}+\sqrt{v}}\,,\\
&\sigma_{13}=\sigma_{24}=\Delta \frac{1}{1+\sqrt{u}+\sqrt{v}}\,,\\
&\sigma_{14}=\sigma_{23}=\Delta \frac{\sqrt{v}}{1+\sqrt{u}+\sqrt{v}}\,.
\end{aligned}
\ee
One can then define corresponding vertex Mandelstam invariants via $s\colonequals 4m^2 -\frac{8}{\Delta}\sigma_{12}$, $t \colonequals 4m^2 -\frac{8}{\Delta}\sigma_{13}$ and $u \colonequals  4m^2 -\frac{8}{\Delta}\sigma_{14}$. In equation \eqref{eq:conformalMandelstam} we wrote an expression for the conformal (or boundary) Mandelstam invariants which was valid on the support of the momentum-conserving delta function; one can verify that it agrees precisely with equation \eqref{eq:sigmasconformal}.

With the $\sigma_{ij}$ in hand we can determine $X$ via equation \eqref{XviaPi}. The bulk point is most easily described as a linear combination of the boundary points as in equation \eqref{Xansatz}. With a little work we find that \eqref{XviaPi} reduces to:
\be
P_i \cdot X = - \frac{R}{4 c_i}\,.
\ee
and that
\be\label{eq:c1exp}
c_1 = \frac{\alpha}{(P_{12}P_{13}P_{14})^{1/2}}
\ee
with others obtained through cyclic permutations of the indices. The common factor $\alpha$ is given by
\be
\alpha = \frac{ \left( P_{12} P_{13} P_{14} P_{23} P_{24} P_{34} \right)^{1/4}}{\sqrt{2} \left( \sqrt{P_{12} P_{34}} + \sqrt{P_{13}P_{24}} + \sqrt{P_{14}P_{23}} \right)^{1/2}}\,.
\ee
The on-shell value of $f_c(X)$ reads:
\be
f_c(X) =  \D \sum_{i = 1}^4 \log(4 c_i)
\ee
To compute the full saddle-point approximation we need to compute the determinant of the Hessian. The second derivatives are given by:
\be
\begin{split}
\frac{\partial^2 f_c(X)}{\partial X_\mu \partial X_\nu} &= \sum_{i=1}^4 \Delta \frac{P_i^\mu P_i^\nu}{(P_i \cdot X)^2} + 2 \lambda \eta^{\mu \nu}  = \frac{4\Delta}{R^2} \left( \sum_{i=1}^4 4 c_i c_i P_i^\mu P_i^\nu  - \eta^{\mu \nu}\right)\\
\frac{\partial^2 f_c(X)}{\partial X_\mu \partial \lambda} &= 2 X^\mu = \sum_{i=1}^4 2 c_i P_i^\mu\\
\frac{\partial^2 f_c(X)}{\partial \lambda^2} &= 0
\end{split}	
\ee
After introducing the vectors
\be
\tilde P_i = 2 c_i P_i
\ee
one uses the matrix determinant lemma for
\be
\begin{split}
\begin{vmatrix}
- \eta + \sum_i \tilde P_i (\tilde P_i)^T & \sum_i \tilde P_i\\
\sum_i (\tilde P_i)^T & 0
\end{vmatrix}
&= (-1)^{d} 4\, \text{det}_{ij} \left[\delta_{ij} - (\tilde P_i)^T \tilde P_j - \frac{1}{4} e_{ij}\right] \\
&= (-1)^{d} \frac{32 \sqrt{P_{12}P_{13}P_{14}P_{23}P_{24}P_{34}}}{\left( \sqrt{P_{12} P_{34}} + \sqrt{P_{13}P_{24}} + \sqrt{P_{14}P_{23}} \right)^3}
\end{split}
\ee
where $e_{ij} = 1$ for all $i$ and $j$. Mopping up all the other factors ultimately gives
\be
G_c(P_i) \overset{R\to\infty}{\xrightarrow{\hspace*{0.6cm}}} R^{-d + 3} 2^{2 \D - d/2 - 6} \pi^{-3d/2 + 1/2} \Delta^{3d/2 - 9/2}\frac{\left( \sqrt{P_{12} P_{34}} + \sqrt{P_{13}P_{24}} + \sqrt{P_{14}P_{23}} \right)^{-2\D + 3/2}}{\left(P_{12}P_{13}P_{14}P_{23}P_{24}P_{34}\right)^{1/4}}
\label{Eqn_flat space limit G_c}
\ee
As expected, this is a manifestly crossing symmetric function of the positions that also obeys the right conformal transformation properties for a four-point function of identical operators.

Now, according to the prescription dictated by the `S-matrix' conjecture \eqref{Eqn_flatspaceSconj}, we should obtain a momentum conserving delta function if we multiply the contact diagram $G_c(P_i)$ by the normalization factor $\sqrt{Z}$, analytically continue to the `S-matrix' configuration and then take the flat-space limit. In equations this means that
\begin{align}
Z^2 G_c(P_i)|_{\text{S-matrix}}
\overset{R\to\infty}{\xrightarrow{\hspace*{0.6cm}}}
i(2\pi)^{d+1} \delta^{(d+1)}( k_1 + k_2 + k_3 + k_4 )
\label{Eqn_delta function claim}
\end{align}
should hold, with the momenta $k_i$ related to the boundary points $P_i$ through \eqref{Eqn_P_i - k_i relation}. In appendix \ref{app:delta function} we prove that this is indeed the case.\footnote{In particular, the seemingly random and certainly lengthy prefactor in equation \eqref{Eqn_flat space limit G_c} is essential to reproduce the correct normalization in equation \eqref{Eqn_delta function claim}.}

\subsection{Exchange diagram and geodesic networks}
\label{subsec:exchange}
We now discuss the next-to-simplest Witten diagram for the four-point functions, namely the exchange diagram. By analyzing its flat-space limit, we encounter an interesting obstruction against the position-space recipe for the flat-space limit. This naturally leads us to propose the notion of Landau diagrams in AdS, which will be the subject of the next section.

The exchange diagram is given by
\be
G_e(P_i) = \int dX dY G_{B\partial}(X,P_1) G_{B\partial}(X,P_2) G_{BB}(X,Y) G_{B\partial}(Y,P_3) G_{B\partial}(Y,P_4)  
\ee
We will set all the external dimensions equal $\D_1 = \D_2 = \D_3 = \D_4 = \D$ and the dimension of the exchanged particle equal to $\D_b$ for simplicity. Using the split representation for the bulk-bulk propagator this becomes
\begin{align}\label{eq:contourintegralGe}
\begin{aligned}
G_e(P_i)=&\int_{-i\infty}^{i\infty}\frac{dc}{2\pi i}
\frac{2c^2}{c^2-(\Delta_b-h)^2} \int dQ\int dX dY
\frac{R^{3-3d} \, \mathcal{C}_{h+c} \, \mathcal{C}_{h-c}}{(-2Q\cdot X/R)^{h+c}(-2Q\cdot Y/R)^{h-c}}\\
&\times \frac{(\mathcal{C}_{\Delta})^4}{(-2P_1\cdot X/R)^{\Delta}(-2P_2\cdot X/R)^{\Delta}(-2P_3\cdot Y/R)^{\Delta}(-2P_4\cdot Y/R)^{\Delta}}\,.
\end{aligned}
\end{align}
\subsubsection{Contribution from the saddle}
In the flat space limit, we expect this integral to be dominated by the saddle point.  After introducing the Lagrange multipliers $\lambda_{Q,X,Y}$ and $\theta_Q$, the function to extremize becomes
\be\label{eq:feextremize}
\begin{aligned}
f_{e} (X,Y,Q,c)=&-c\left[\log (- Q\cdot X/R)-\log (- Q\cdot Y/R)\right]\\
&-\Delta\left[\sum_{j=1,2}\log (- P_j\cdot X/R)+\sum_{j=3,4}\log (- P_j\cdot Y/R)\right]\,\\
&+\lambda_Q Q^2 + \theta_Q (Q^0 - 1) + \lambda_X (X^2+R^2)+\lambda_Y (Y^2+R^2)\,.
\end{aligned}
\ee
Imposing
\be
\frac{\partial f_e}{\partial c}=\frac{\partial f_e}{\partial Q}=\frac{\partial f_e}{\partial \lambda_Q}=\frac{\partial f_e}{\partial \theta_Q}=\frac{\partial f_e}{\partial \lambda_X}=\frac{\partial f_{e}}{\partial \lambda_Y}=0\,,
\ee
we find that the two bulk points must coincide at the saddle point; namely $X=Y$ (and $\lambda_Q=\theta_Q=0$). The remaining saddle point equations reduce to
\be\label{eq:cQXp}
\begin{aligned}
c\frac{Q}{Q\cdot X}+\Delta\sum_{j=1,2}  \frac{P_j}{P_j \cdot X} - 2 \lambda_{X} X  &= 0\,,\\
-c\frac{Q}{Q\cdot X}+\Delta\sum_{j=3,4}  \frac{P_j}{P_j \cdot X} - 2 \lambda_{Y} X  &= 0\,.
\end{aligned}
\ee
Contracting these equations with $X$, we get
\be\label{eq:saddlelamXY}
\lambda_X=-\frac{1}{2R^2}(2\Delta+c)\,,\qquad \lambda_Y=-\frac{1}{2R^2}(2\Delta-c)\,.
\ee
Now, to understand the physical meaning of the saddle-point equations, it is again useful to use the vertex momenta
\be
\kappa_j\colonequals \frac{\Delta_j}{R} \left(\frac{P_j}{P_j \cdot X /R} + X/R \right)\,,\qquad \chi \colonequals \frac{c}{R} \left(\frac{Q}{Q \cdot X /R} + X/R \right)\,.
\ee
Here $\chi$ is a vertex momentum associated with the exchanged particle. Unlike the external vertex momenta, it is {\it off-shell}, meaning that $\chi^2 \neq m_b^2$ with $m_b \colonequals \Delta_b/R$. In terms of these variables, the saddle-point equation again takes the form of the momentum conservation
\be\label{eq:conservexchange}
\kappa_1+\kappa_2 +\chi=0\,,\qquad \kappa_3+\kappa_4-\chi=0\,.
\ee
This in particular means that the vertex momenta of the external particles are conserved, $\sum_j \kappa_j=0$. Note that \eqref{eq:conservexchange} matches our expectation in the flat-space limit: the momentum conservation holds at each vertex but the internal particle is off-shell.

To determine the saddle-point values of $c$, $Q$, and $X(=Y)$ we can again consider the vertex Mandelstam variables. Owing to the momentum conservation of the external particles $\sum_{j}\kappa_j=0$, $\sigma_{ij}$'s are given by the same expressions as the contact diagram \eqref{eq:sigmasconformal} and so is $X$. On the other hand, if we use the momentum conservation at each interaction vertex \eqref{eq:conservexchange}, we obtain alternative expressions
\be
c^2 =4 \Delta^2-8\Delta\sigma_{12} = 4 \Delta^2-8\Delta\sigma_{34}
\ee
With \eqref{eq:sigmasconformal} this yields\footnote{The sign of $c$ is arbitrary, since the equations are invariant under $c \to - c$ and $Q \to X + Q/(2 Q\cdot X)$.}
\be\label{eq:csaddle}
c^2= 4\Delta^2 \frac{- \sqrt{P_{12}P_{34}} + \sqrt{P_{13}P_{24}} + \sqrt{P_{14}P_{23}}}{\phantom{-}\sqrt{P_{12}P_{34}} + \sqrt{P_{13}P_{24}} + \sqrt{P_{14}P_{23}}}
\ee
In terms of the vertex momenta, or in terms of the boundary momenta on the support of the momentum-conserving delta function, this expression is actually much simpler: 
\be\label{eq:csaddle22}
c^2 =R^2s\,,
\ee 
with $s$ the conformal Mandelstam variable \eqref{eq:conformalMandelstam}. This is simply a manifestation of the fact that $c^2$ measures the energy of the exchanged particle. We can then determine $Q$ solving \eqref{eq:cQXp}. The result reads
\be
Q\propto\lambda_Y\sum_{j=1,2}c_j P_j -\lambda_X\sum_{k=3,4}c_k P_k\,,
\ee
where $c_j$'s are given by \eqref{eq:c1exp} and $\lambda_{X,Y}$ given by \eqref{eq:saddlelamXY}. Not written is an unimportant proportionality factor that fixes the gauge $Q^0 = 1$.

With $X=Y$ it is immediate that the on-shell value of $f_e$ coincides with that of the contact diagram, so we find again that
\be\label{eq:saddlesaddlevalue}
f_e=\Delta \sum_{i=1}^{4}\log (4c_i)\,.
\ee
As is the case with the contact diagram, we also need the one-loop fluctuation around the saddle point to reproduce the correct flat-space limit. It turns out that the computation is most efficiently done if we first perform integration of $X$, $Y$ and $Q$ exactly and then compute the fluctuation around the saddle point of $c$ \eqref{eq:csaddle}. Relegating the details to Appendix \ref{ap:exchange}, here we display the final result
\be
G_{e}(P_i)\overset{R\to\infty}{\xrightarrow{\hspace*{0.6cm}}} \left.G_{c}(P_i)\right|_{R\to \infty} \times \frac{R^2}{\Delta_b^2-c^2}\,,
\ee
where the first factor $\left.G_{c}(P_i)\right|_{R\to \infty}$ is the flat-space limit of the contact diagram \eqref{Eqn_flat space limit G_c}. Thus, once we multiply $G_{e}(P_i)$ with the normalization factors $\sqrt{Z}$ and perform the analytic continuation, we recover the result for the exchange diagram in the flat-space limit including the momentum conserving delta function:
\be
\left.Z^2 G_{e}(P_i)\right|_{\text{S-matrix}}\overset{R\to\infty}{\xrightarrow{\hspace*{0.6cm}}} i (2\pi)^{d+1}\delta^{(d+1)}(k_1+k_2+k_3+k_4)\frac{1}{(k_1+k_2)^2+m^2}\,.
\ee
\subsubsection{Contribution from the pole}
\begin{figure}[t]
\begin{center}
\includegraphics[clip,width=12cm]{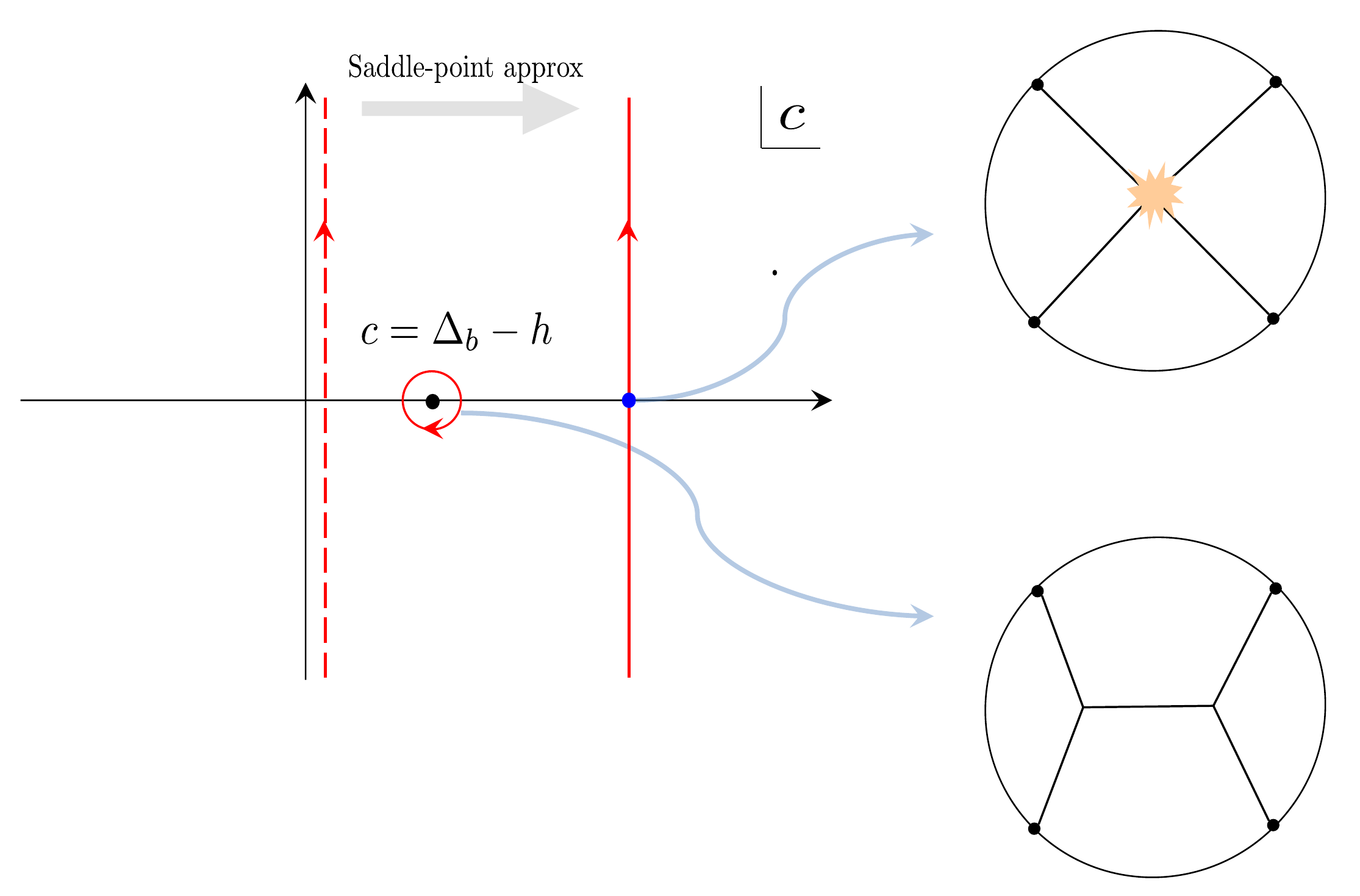}
\caption{\label{fig:saddleandpole}The contributions from the saddle point and the pole for the exchange diagram. To evaluate the exchange diagram using the saddle-point approximation of the $c$-integral, we need to deform the original contour (the red dashed line) to a steepest descent contour (the solid red line) that goes through the saddle point (the blue dot). Upon doing so, the contour sometimes needs to cross the poles of the integrand and this produces an additional contribution. Physically, the contribution from the saddle-point corresponds to a scattering process in which the four-particles meet at a point while the contribution from the pole corresponds to a geodesic network. The former is related to a flat-space S-matrix while the latter is not.}
\end{center}
\end{figure}

We have seen above that the contribution from the saddle point beautifully reproduces the flat-space limit of the exchange diagram. There is however one subtlety in the argument above: initially the contour of integration of $c$ is placed along the imaginary axis. In order to evaluate the integral using the saddle-point approximation, we need to shift the contour so that it goes through the saddle point given by \eqref{eq:csaddle22}. Upon doing so, the contour sometimes crosses poles in the integrand of \eqref{eq:contourintegralGe}, namely the poles at $c=\pm (\Delta_b-h)$. When this happens, the full contribution in the large $R$ limit is given by a sum of two terms, the saddle-point contribution determined above, and the contribution from the residue of the pole (see figure \ref{fig:saddleandpole}). 

\begin{figure}[t]
\centering
\includegraphics[clip,height=5cm]{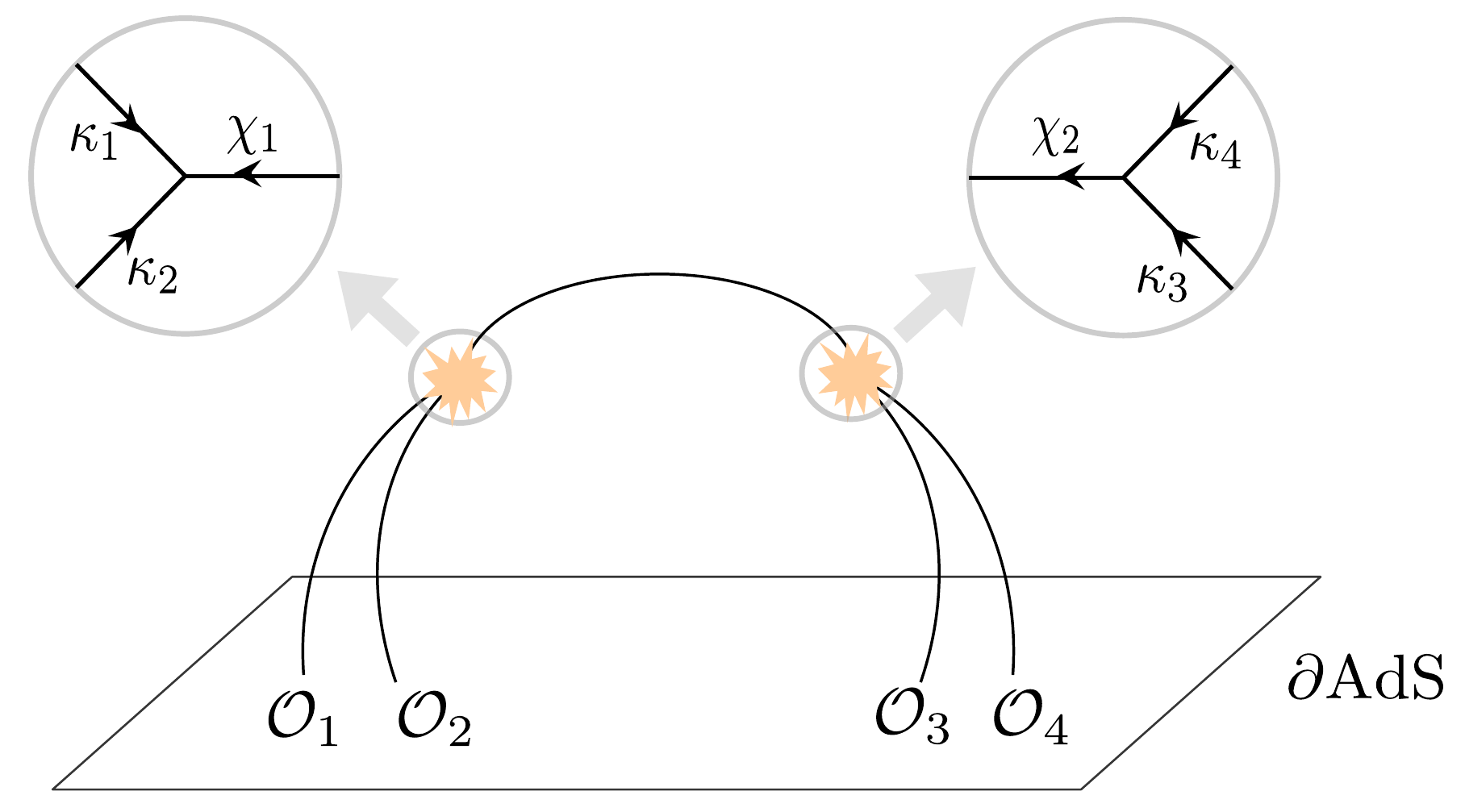}
\caption{Geodesic network. The contribution from the pole $c=\Delta_b$ corresponds to a network of geodesics in which the conservation of momenta holds at each vertex. Here $\chi_{1,2}$ are internal vertex momenta while $\kappa_j$'s are external vertex momenta.\label{fig:exchangemomentacon}}
\end{figure}

Let us for now discuss the contribution from the pole in the right half plane, $c=\Delta_b-h$. Evaluating the integral \eqref{eq:contourintegralGe} at the pole we get
\begin{align}\label{eq:poleintegral}
\begin{aligned}
\left.G_e(P_i)\right|_{\text{pole}}=&(h-\Delta_b) \int dQ\int dX dY
\frac{R^{3-3d} \, \mathcal{C}_{\Delta_b} \, \mathcal{C}_{d-\Delta_b}}{(-2Q\cdot X/R)^{\Delta_b}(-2Q\cdot Y/R)^{d-\Delta_b}}\\
&\times \frac{(\mathcal{C}_{\Delta})^4}{(-2P_1\cdot X/R)^{\Delta}(-2P_2\cdot X/R)^{\Delta}(-2P_3\cdot Y/R)^{\Delta}(-2P_4\cdot Y/R)^{\Delta}}\,.
\end{aligned}
\end{align}
To analyze the rest of the integral, we can again use the saddle-point approximation. This is a safe manipulation since the integrand  is not singular (for generic $P_j$'s). Now the function to extremize is
\be
\begin{aligned}
f_{e,{\rm pole}} (X,Y,Q)=&-\Delta_b\left[\log (- Q\cdot X/R)-\log (- Q\cdot Y/R)\right]\\
&-\Delta\left[\sum_{j=1,2}\log (- P_j\cdot X/R)+\sum_{j=3,4}\log (- P_j\cdot Y/R)\right]\,\\
&+\lambda_Q Q^2 + \theta_Q (Q^0-1) + \lambda_X (X^2+R^2)+\lambda_Y (Y^2+R^2)\,.
\end{aligned}
\ee
Since the integration variable $c$ in \eqref{eq:feextremize} is replaced by a fixed number $\Delta_b$, the saddle-point equations do not set the two bulk points to be coincident, so generically $X\neq Y$. Instead we obtain
\begin{align}
&\Delta_b\frac{Q}{Q\cdot X}+\Delta\sum_{j=1,2}  \frac{P_j}{P_j \cdot X} - 2 \lambda_{X} X  = 0\,,\qquad
-\Delta_b\frac{Q}{Q\cdot Y}+\Delta\sum_{j=3,4}  \frac{P_j}{P_j \cdot Y} - 2 \lambda_{Y} Y  = 0\,,\nonumber\\
&-\Delta_b \frac{X}{Q\cdot X}+ \Delta_b \frac{Y}{Q\cdot Y} + 2 \lambda_Q Q =0\,,\label{eq:saddlepointpole}
\end{align}
where $\lambda_{X,Y}$ are given by
\be
\lambda_X =-\frac{1}{2R^2}\left(2\Delta+\Delta_b\right)\,,\qquad \lambda_Y =-\frac{1}{2R^2}\left(2\Delta-\Delta_b\right)\,.
\ee
The first two equations can be recast into the momentum conservations at the two bulk vertices $X$ and $Y$. To see this, we introduce internal vertex momenta
\be
\chi_1\colonequals \frac{\Delta_b}{R} \left(\frac{Q}{Q \cdot X /R} + X/R \right)\,,\qquad \chi_2\colonequals \frac{\Delta_b}{R} \left(\frac{Q}{Q \cdot Y /R} + Y/R \right)\,.
\ee
Then the first two equations in \eqref{eq:saddlepointpole} can be rewritten as
\be\label{eq:momentumconservationpole}
\kappa_1+\kappa_2+\chi_1=0\,,\qquad \kappa_3+\kappa_4-\chi_2=0\,.
\ee
There are two important differences compared to equation \eqref{eq:conservexchange}: first the two internal momenta $\chi_{1,2}$ are in general different, and second, unlike $\chi$ in \eqref{eq:conservexchange} the internal momenta $\chi_{1,2}$ are on-shell, i.e.~$\chi_{1}^2 =\chi_2^2 =m_b^2$. Geometrically, these features reflect the fact that the contribution from the pole describes a network of geodesics in which two interaction points are macroscopically separated in AdS.\footnote{Unlike the `geodesic Witten diagram' introduced in \cite{GeodesicWittenDiagram}, here the bulk vertices do not lie on the geodesic connecting the boundary points (as long as $m_b > 0$). Both constructions do reduce to a conformal block at large $\Delta$, albeit with different normalizations.} This is also consistent with the analysis in section \ref{Section_Erasing the circle}, which showed that the pole contribution to the bulk-to-bulk propagator corresponds to a geodesic connecting two bulk points (see also figure \ref{fig:exchangemomentacon}). This is reminiscent of Landau diagrams in flat space, which correspond to trajectories of on-shell particles in complexified Minkowski space. In section \ref{sec:landaudiagrams}, we will use this observation to propose the AdS version of Landau diagrams.

There are two different ways to evaluate the saddle-point value of $f_{e,{\rm pole}}$. The first approach is to explicitly determine the saddle point by solving all the equations \eqref{eq:saddlepointpole} and then to evaluate $f_{e,{\rm pole}}$ on that saddle point. The second approach is to perform the integrals of $X$, $Y$ and $Q$ in \eqref{eq:poleintegral} exactly and use the asymptotic form of the conformal block. As explained in appendix \ref{ap:exchange}, the second approach turns out to be simpler and the result reads
\be\label{eq:polesaddlevalue}
f_{e,{\rm pole}}=-\Delta \log \left(\frac{P_{12}P_{34}}{16}\right)+g\left(\Delta_b\right)\,.
\ee
with
\be
g(x)\colonequals -4\Delta \log (\Delta) +(2\Delta+x)\log (\Delta+\tfrac{x}{2})+(2\Delta-x)\log (\Delta-\tfrac{x}{2})+x\log \left(\frac{2m-\sqrt{s}}{2m+\sqrt{s}}\right)\,,
\ee
where $s$ is the Mandelstam variable.

\subsubsection{Exchange of dominance}
We have seen that the exchange diagram receives two different contributions, the one associated with the saddle point of $c$ and the other associated with the pole of $c$. As discussed above, the first contribution correctly reproduces the flat-space limit while the second contribution corresponds to a network of geodesics in which the two bulk points are macroscopically separated in AdS. Therefore the large $R$ limit of the exchange diagram gives the flat-space result if and only if the second contribution can be neglected. 

\begin{figure}[t]
\begin{center}
\includegraphics[width=10cm]{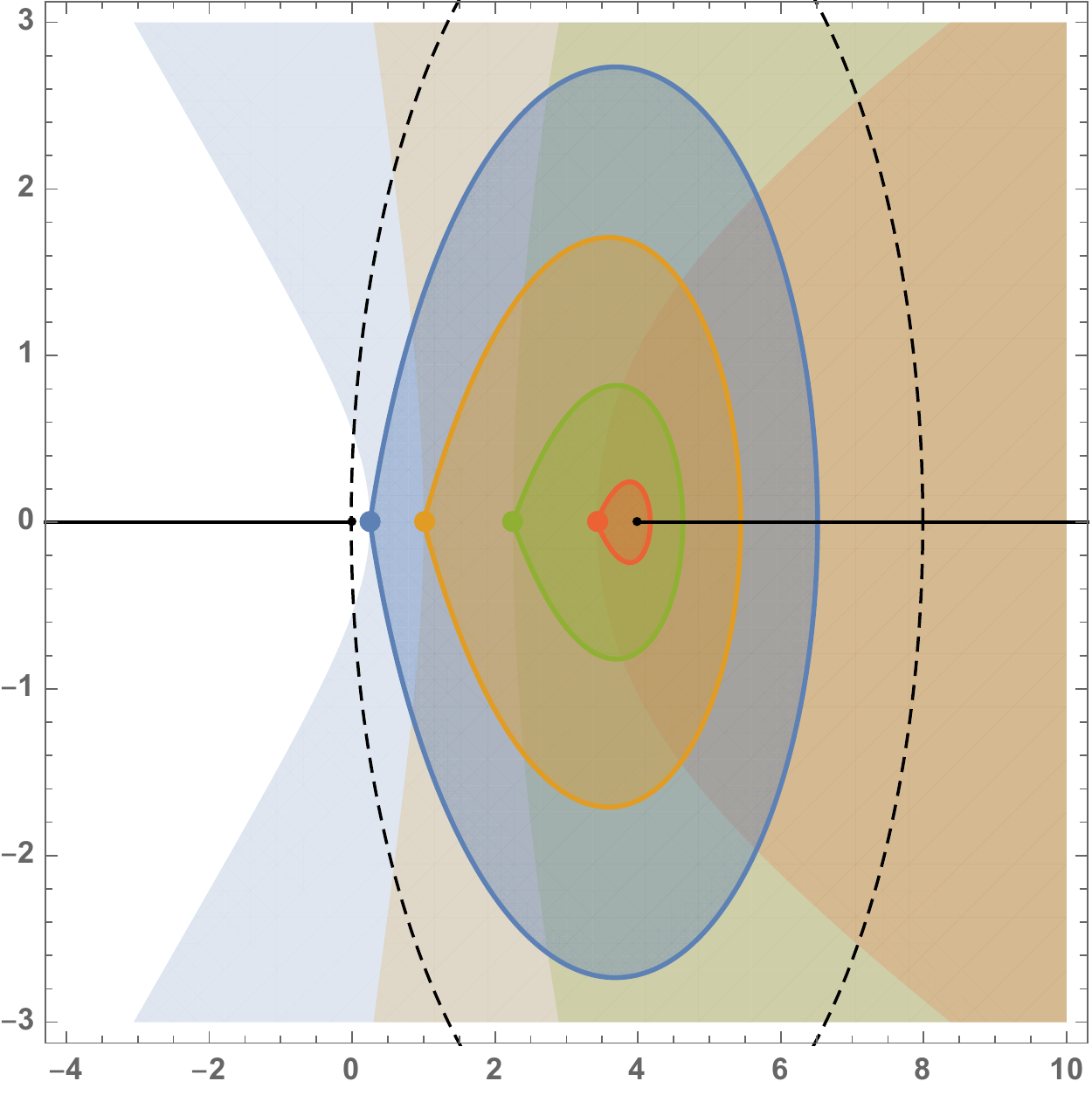}
\caption{\label{fig_regions3}Regions in the complex $s$ plane where the pole at $c = \Delta_b$ is picked up (lighter shaded) and the smaller subregion where the flat-space limit diverges (darker shaded). The different colors correspond to $m_b = 0.5, 1, 1.5, 1.85$ and we have set $m = 1$. The problematic region for the exchange diagram always lies within the disk given by $|s - 4 m^2| < 4m^2$. We assumed that the external momenta are chosen such that momentum conservation holds.}
\end{center}
\end{figure}

Clearly the most interesting configuration is when we are on the support of the momentum-conserving delta function so the external momenta obey $\sum_{i=1}^4 k_i = 0$. In that case the algorithm is the following (see figure \ref{fig:saddleandpole}).
\begin{enumerate}
\item Compare the position of the steepest descent contour through the saddle point at $c=R \sqrt{s}$ and the position of the pole at $c=\Delta_b$. If the steepest descent contour is to the left of the pole, the flat space limit gives the correct answer.
\item If the steepest descent contour is to the right of the pole, compare the real parts of the exponents \eqref{eq:saddlesaddlevalue} and \eqref{eq:polesaddlevalue}. In fact, if $\sum_{i=1}^4 k_i = 0$ then the value of $f_e$ at the saddle point value vanishes and so we simply need to check the sign of the real part of \eqref{eq:polesaddlevalue}: if it is positive then the flat-space limit diverges, and if it is negative then the flat-space limit gives the correct answer.
\end{enumerate}

In figure \ref{fig_regions3}, we plotted the region in which the flat-space limit works/fails in the complex $s$-plane. As one can see there, the bound-state pole $s=m_b^2$ is at the edge of a larger `blob' where the flat-space limit gives a divergent answer. The size of the region grows as we decrease the mass of the bound state, and when it becomes massless $m_b\to 0$, the region is given by a disk of radius $4m^2$ centered at the two-particle threshold $s=4m^2$. Therefore, if we are agnostic about $m_b$ then the flat-space limit of the exchange diagram is only guaranteed to be finite for $|s - 4m^2| > 4m^2$. We furthermore observe that the bad region vanishes entirely as $m_b \to 2m$, and that for any $0 \leq m_b < 2m$ it always includes at least a little bit of the physical line $s > 4 m^2$.


\section{Landau diagrams in AdS}
\label{sec:landaudiagrams}
In section \ref{Section_Erasing the circle} we discussed two large $\Delta_b$ limits of the bulk-bulk propagator $G_{BB}(X,Y)$: one where the distance between $X$ and $Y$ becomes much smaller than the AdS radius $R$, which reproduced the flat-space Klein-Gordon propagator, and one where this distance is kept finite in units of $R$, which reproduced the simple exponential $G_{BB}(X,Y) \sim \exp( - \Delta_b d(X,Y))$ with $d(X,Y) = \text{arccosh}(- X \cdot Y/R^2)$ the geodesic distance between $X$ and $Y$ (in units of the AdS radius).

In the flat-space analysis of Witten diagrams the bulk points $X$ and $Y$ are integrated over and in the large $\Delta$ analysis their locations are dynamically determined by the saddle point equations. It is therefore not entirely surprising that both behaviors of the bulk-bulk propagator can play a role. As was exemplified by the exchange diagram of the previous subsection, for a propagator in a generic Witten diagram the saddle point equations read:
\begin{align} \label{twosaddlesbulkbulk}
\frac{\partial f}{\partial c} &= 0  & &\implies & \log(- Q\cdot X) - \log( - Q\cdot Y) &= 0\\
\frac{\partial f}{\partial Q} &= 0 & &\implies & - c \left( \frac{X}{Q \cdot X} - \frac{Y}{Q \cdot Y} \right) + 2 \lambda_Q Q &= 0
\end{align}
Together they yield $Y = X$, and with all the bulk vertices close together we reproduce the flat space result because all the interactions take place at a distance much smaller than the AdS radius. The only potential hiccup in this procedure are the poles in the complex $c$ plane: if the steepest descent contour through the saddle point in the $c$ plane passes on the wrong side of one of these poles then its residue needs to be taken into account, resulting in unwanted additional contributions that can spoil the extraction of an amplitude from the position-space correlator.

An \emph{AdS Landau diagram} can be defined as a network of geodesics in AdS such that vertex momentum is conserved at every interaction point. We recall that the vertex momentum for each external leg is
\be
\kappa_i = \frac{\Delta_i}{R} \left(\frac{P_i}{P_i \cdot X /R} + X/R \right)\,,
\ee
and for each internal leg it is
\be \label{internalvertexmomentum}
\chi_i = \frac{\Delta_i}{R} \left(\frac{Q_i}{Q_i \cdot X /R} + X/R \right)\,,
\ee
where for every bulk-bulk propagator $G_{BB}(X,Y)$ the value of $Q$ is determined through momentum conservation and
\be \label{Qsaddleeqn}
2 \lambda_Q Q = \Delta_b \left( \frac{X}{Q \cdot X} - \frac{Y}{Q \cdot Y} \right) 
\ee
and $Q^2 = 0$ and $Q^0 = 1$. Notice that the first equation in \eqref{twosaddlesbulkbulk} no longer needs to be obeyed because $c$ is fixed to one of its poles; for definiteness we chose the pole at $c = \Delta_b$ but we will discuss this further below. Clearly the AdS Landau diagrams extremize the `action'
\be \label{adslandauaction}
\begin{split}
f = &- \sum_{\vev{ik}\, \in \text{ Ext}} \Delta_i \log( - P_i \cdot X_k) - \sum_{\vev{kl}\, \in \text{ Int}} \Delta_{\vev{kl}} \left( \log( - Q_{\vev{kl}}  \cdot X_k) - \log(- Q_{\vev{kl}} \cdot X_l)\right) \\
&+ \sum_k \lambda_{k} (X_k^2 + R^2) + \sum_{\vev{kl}\, \in \text{ Int}} \left( \lambda_{\vev{kl}} Q_{\vev{kl}}^2 + \theta_{\vev{kl}} (Q_{\vev{kl}}^0 - 1)\right)
\end{split}
\ee
where $\vev{ik}$ runs over the set of external legs between the boundary points $P_i$ and bulk points $X_k$ connected by a bulk-boundary propagator, and $\vev{kl}$ labels all pairs of internal legs, so legs connected by a bulk-bulk propagator. We will call the saddle point equations the \emph{AdS Landau equations} and the on-shell value of this action then gives the contribution of the AdS Landau diagram to the flat-space limit of the Witten diagram.

Notice that an alternative action can be obtained by eliminating $Q$ and simply using the large $\Delta$ expression for the bulk-bulk propagator:
\be \label{Eqn_AdS landau action simpler}
\tilde f = - \sum_{\vev{ik}\, \in \text{ Ext}} \Delta_i \log( - P_i \cdot X_k) - \sum_{\vev{kl}\, \in \text{ Int}} \Delta_{\vev{kl}} \, d(X_k,X_l) + \sum_k \lambda_{k} (X_k^2 + R^2) 
\ee
with $d(X_k,X_l) = \text{arccosh}(- X_k \cdot X_l/R^2)$ as above. In this sense an AdS Landau diagram is a sort of `minimal distance' diagram: the $\Delta$'s provide a `spring constant' that determines how much the action decreases if we pull points further apart, and the external leg factors provide a `renormalized distance' between bulk and boundary points.

Returning to our conjectures we see that we can divide the configuration space of all values of the Mandelstam invariants into different regions as follows. We first look at the original saddle point equations and determine the integration contours for the $c$ variables in the correct flat space limit. If poles have been crossed in deforming the original integration contour to this steepest descent contour then the corresponding leg is `freed' and the bulk points are allowed to separate. For each region we can construct the AdS Landau diagram with the corresponding free internal legs, demanding that all the non-free internal legs are contracted to a point. In regions where the number of free legs is not zero, our conjectures have a chance of working only if the on-shell value of the action is subleading compared to the contact diagram.\footnote{In fact, for the contact diagram on the support of the momentum conserving delta function the on-shell action is zero. Therefore the condition for the conjectures to hold becomes $\text{Re}(f) < 0$.}

\subsection{Comparison with flat space Landau diagrams}
Much like flat space Landau diagrams, our AdS Landau diagrams correspond to classical on-shell particles propagating over large distances with momentum conservation holding at the vertices. Let us compare the equations in a bit more detail.

In flat space the Landau conditions can be formulated as follows \cite{Landau:1959fi,Coleman:1965xm}.\footnote{A priori all the positions and momenta here are to be understood in Minkowski space with a metric with mostly plus signature. More interesting singularities can be obtained by complexification. In particular, to obtain singularities in the `Euclidean domain' where $s, t, u$ are all positive (and below threshold), we can analytically continue the spacelike components of $k^\mu$ and $x^\mu$ to purely imaginary values (for example, in the center of mass frame $s = 4m^2 + 4 \vec p^2$ so $s < 4 m^2$ requires $\vec p^2 < 0$). Absorbing the signs in a redefinition of the metric, this is commonly described as a configuration with all minus signature metric and real momenta. However our conjectures carry additional factors of $i$. More precisely, equation \eqref{momentaSmatrix} informs us that imaginary $\vec p$ corresponds to real $n^\mu$, which means that standard Euclidean AdS is appropriate for the Euclidean domain in the Mandelstam invariants.} Suppose the external momenta are $q_i^\mu$. One then associates a momentum $k^\mu_r$ and a parameter $\alpha_r$ to each internal leg $r$ and a position $x^\mu_s$ to each vertex $s$. Then for the internal leg between position $x^\mu$ and $y^\mu$ we impose that any non-zero propagation is physical, so
\be
x^\mu - y^\mu = \alpha k^\mu
\ee
Now either $\alpha$ is zero, the leg is contracted and the diagram said to be \emph{reduced}, or the propagation needs to be on-shell. In equations, for every leg we need that:
\be
\alpha( k^2 + m^2) = 0
\ee
The last condition is momentum conservation for each vertex. If we ignore signs corresponding to in- or outgoing momenta then this can be schematically written as:
\be
\sum_r k_r^\mu  + \sum_i q_i^\mu = 0
\ee
with the sum running over all legs, both external and internal (and both contracted and not contracted), that end on the given vertex. The parameter $\alpha$ is interesting here: for a large range of values of the Mandelstam invariants (in particular, all the physical values as well as the Euclidean region) the only possible singularities have $\alpha \geq 0$. Singularities with other values of $\alpha$ can appear on other sheets. More extensive reviews of the Landau equations can be found, for example, in \cite{bjorken1965relativistic,eden2002analytic,sterman1993introduction}.

From the preceding discussions one can distill a nearly perfect analogy with the AdS equations: the conservation of the on-shell momenta in flat space becomes simply the conservation of the on-shell vertex momenta in AdS, and equation \eqref{Qsaddleeqn} fixes the direction of the `momentum' $Q^\mu$ to be a linear combination of $X$ and $Y$ such that the relevant vertex momenta at $X$ and $Y$ are tangential to the geodesic between $X$ and $Y$. This latter condition is precisely the expected AdS analogue of flat-space propagation with a fixed momentum.

An final subtlety is the parameter $\alpha$ in the flat-space Landau equations. Its AdS analogue is $\lambda_Q$ since that is the natural relative parameter between the momentum through a leg and its displacement. In particular, $\lambda_Q = 0$ if the leg is contracted which corresponds to $\alpha = 0$ in flat space. But in flat space we can furthermore deduce that $\alpha \geq 0$ for singularities corresponding to physical or Euclidean kinematics, and it is not immediately clear that the sign of $\lambda_Q$ is similarly important. To see this we will consider the defining equation for $\lambda_Q$, which is
\be \label{lambdaQeqn}
\lambda_Q = \frac{\D_b}{4} \left( \frac{1}{(Q \cdot Y)^2} -\frac{1}{(Q\cdot X)^2}\right)
\ee

To see the relevance of the sign of $\lambda_Q$ we first have to discuss the irrelevance of the sign of $c$. Consider then a solution of the AdS Landau equations for a given value of $c$. We claim that there must exist a solution at the opposite value $- c$ with $Q$ pointing in the opposite direction. To see this, notice that we can always choose a frame where we only need to consider the $\mathbb{R}^2$ spanned by $X$ and $Y$, implying that we can take $X = (\cosh(\rho_X),\sinh(\rho_X))$, $Y = (\cosh(\rho_Y),\sinh(\rho_Y))$. Since $Q$ is a linear combination of $X$ and $Y$ which obeys $Q^2 = 0$ and $Q^0 = 1$, it must be that $Q = (1, \pm 1) \equalscolon Q_\pm$ and that correspondingly
\be
\lambda_Q^\pm = \frac{c}{4} \left( e^{\pm 2 \rho_Y} - e^{\pm 2\rho_X}\right)
\ee
The choice between the `$+$' and the `$-$' sign is not fixed in our partial analysis, but it will be fixed by the vertex momentum conservation equations which given $c$ provide a definite direction to $Q$. However, we also notice that the momentum conservation equations are invariant under sending $c \to - c$ and exchanging $Q^+$ and $Q^-$. Doing so sends\footnote{The gauge constraint $Q^0 = 1$ introduces some non-covariance in the expression for $\lambda_Q$. This is why $\lambda_Q$ is not completely invariant.}
\be
\lambda_Q^\pm \to - \frac{c}{4} \left( e^{\mp 2 \rho_Y} - e^{\mp 2\rho_X} \right) = e^{\mp 2 (\rho_X + \rho_Y)} \lambda_Q^{\pm}\,.
\ee
So, no matter whether the original solution had $Q_+$ or $Q_-$ at $+c$, if $\lambda_Q$ was positive at $+c$ then it is also positive at $-c$, and vice versa. In summary: the sign of $\lambda_Q$ does not depend on the choice between picking up the pole at $+\Delta_b$ or $-\Delta_b$. Suppose then that we set $c = \Delta_b$. To see that positive $\lambda_Q$ is the `correct' direction, note that from equations \eqref{internalvertexmomentum} and \eqref{Qsaddleeqn} it follows that the corresponding vertex momentum points inward at $X$ and outward at $Y$. This is exactly in agreement with the relative minus signs in vertex momentum conservation equations like equation \eqref{eq:momentumconservationpole}, where all the other vertex momenta are defined to point inward.

\begin{figure}[t]
\centering
\begin{minipage}{0.45\hsize}
\centering
\includegraphics[clip,height=3.5cm]{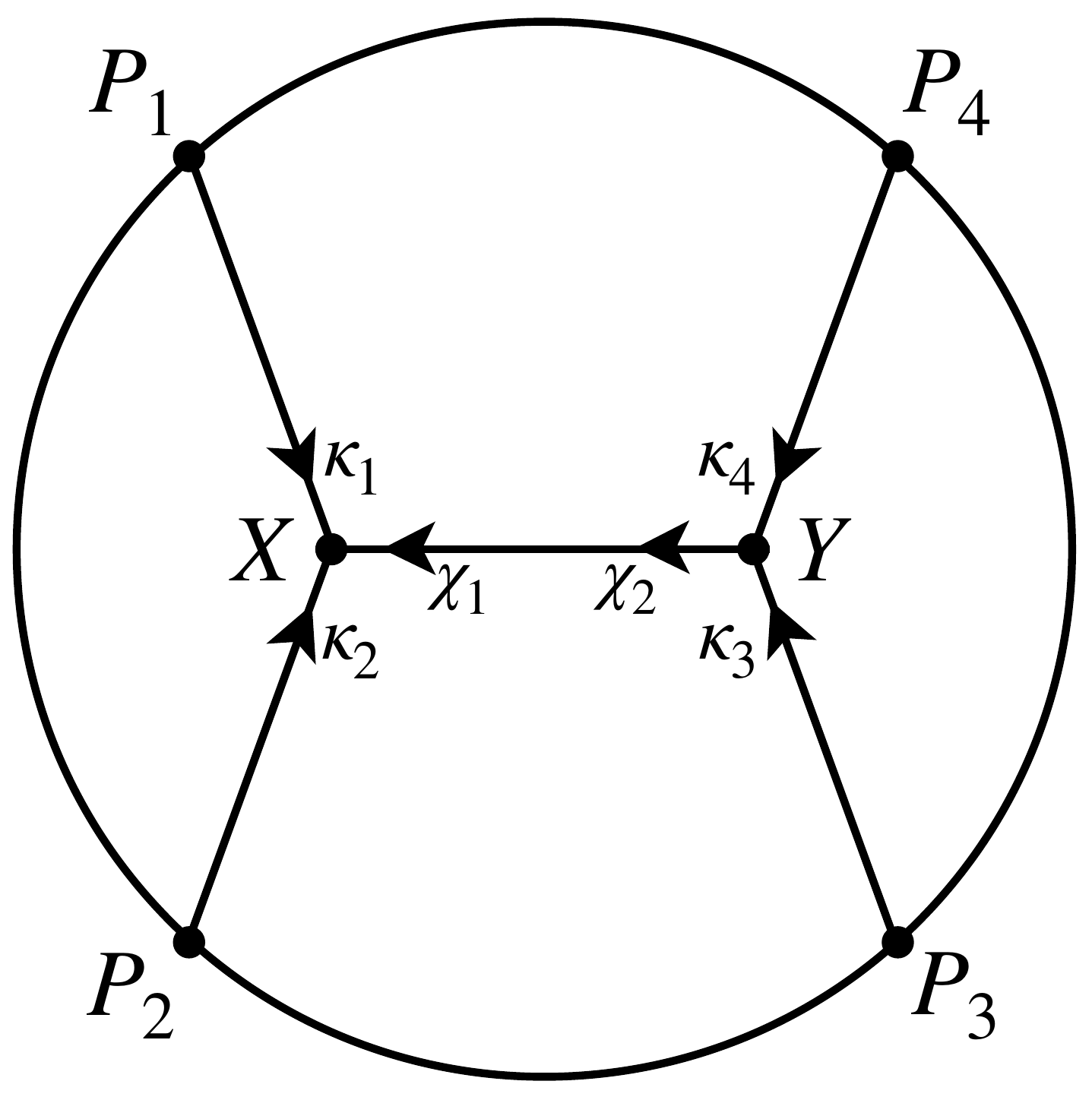}
\subcaption{Normal diagram with $\lambda_Q>0$ \label{Fig_normal_exchange_diagram}}
\end{minipage}\hspace{10pt}
\begin{minipage}{0.45\hsize}
\centering
\includegraphics[clip,height=3.5cm]{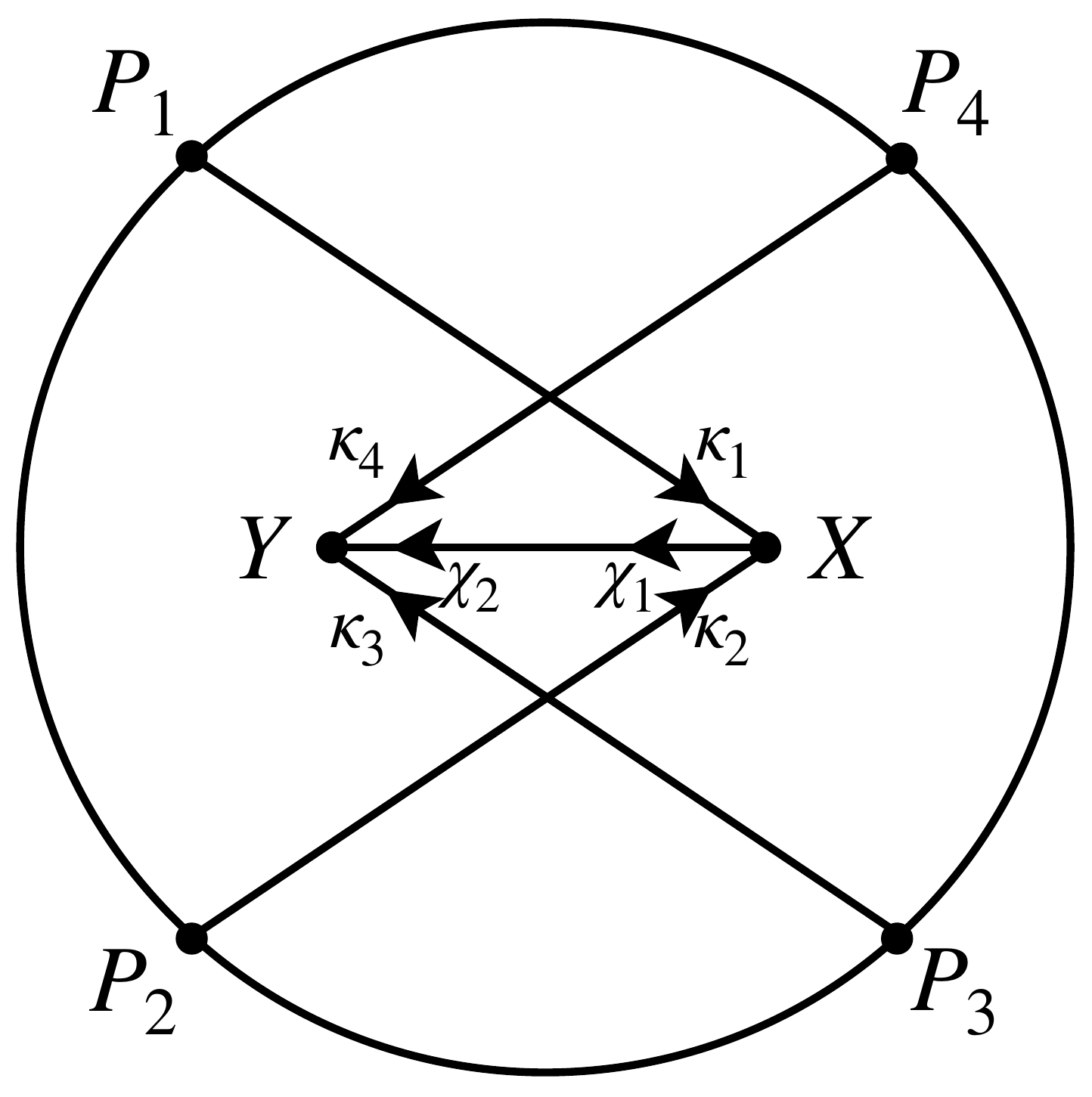}
\subcaption{`Flipped' diagram with $\lambda_Q<0$ \label{Fig_reverted_exchange_diagram}}
\end{minipage}
\caption{Normal and `flipped' scalar exchange diagram.}
\label{Fig_exchange_diagrams}
\end{figure}

To conclude the analogy we just need to deduce that saddle points with $\lambda_Q < 0$ are unimportant in Euclidean configuration. Clearly something is amiss with them, since they would correspond to `flipped' solutions where particles propagate in the direction opposite to their vertex momenta. This is illustrated in figure \ref{Fig_exchange_diagrams} using the scalar exchange diagram as an example. In equations what happens is the following. If $\lambda_Q < 0$ and $c = + \Delta_b$ then $(-Q \cdot Y) > (- Q \cdot X)$, implying that the action can be reduced by decreasing $c$. We take this as an indication that the saddle point in the $c$ plane lies to the \emph{left} of the pole, much like in the unshaded region in figure \ref{fig_regions3} for the exchange diagram. This means that the pole is not picked up and indeed the solution with $\lambda_Q < 0$ is unimportant.

Let us stress that this argument was restricted to Euclidean kinematics. Note that in flat space Landau singularities with non-positive $\alpha$ can become important when considering more involved analytic continuations in the Mandelstam variables, for example by passing onto other sheets. In AdS the same observation holds: such continuations may force a deformation of the $c$ integration contour which forces one to take into account a contribution of the pole independently of the location of the saddle point. It would be interesting to see if a criterion similar to the positivity of $\lambda_Q$ can be formulated for general complex values of the Mandelstam invariants. 

Finally let us stress the most striking difference between the flat space and AdS Landau equations appears to be that the latter can be solved much more generally than the former. In fact, if we do not require the $\lambda_Q$ variables to be positive then it appears that the AdS Landau equations have a solution for \emph{any} values of the external momenta. This is not because the number of equations has changed but rather because of the more permissive nature of the AdS Landau equations. Most importantly, conservation of the vertex momenta can be achieved by moving the bulk point $X$ to a suitable location -- something which is impossible in flat space because of translation invariance.

Of course the AdS Landau diagrams are only important if (1) the poles are picked up, and (2) the on-shell value of the action \eqref{adslandauaction} has positive real part.  As we have seen above, this leads to `blobs' where our conjectures are not valid because the flat-space limit diverges. It is our expectation that there is always a large region where the conjectures work and the AdS Landau diagrams do not dominate, and it would be interesting to find such a region. For four-point functions we can use the conformal block decomposition and some initial steps in this direction are discussed in section \ref{sec:blocks}. It would also be worthwhile to further investigate the natural conjecture that the support for flat-space Landau diagrams lies within the closure of the divergent blobs for AdS Landau diagrams.

\subsection{Anomalous thresholds and the triangle diagram}
\label{subsec:triangle}
\begin{figure}
\begin{center}
\begin{tikzpicture}
\coordinate (center) at (0,0);
  \def\radius{2.5cm}
  \draw [thick] (center) circle[radius=\radius];
  \fill[black] (center) ++(135:\radius) coordinate (P1) circle[radius=2pt] node[above left] {\(P_1\)};
  \fill[black] (center) ++(225:\radius) coordinate (P2) circle[radius=2pt] node[below left] {\(P_2\)};
  \fill[black] (center) ++(35:\radius) coordinate (P4) circle[radius=2pt] node[above right] {\(P_4\)};
  \fill[black] (center) ++(325:\radius) coordinate (P3) circle[radius=2pt] node[below right] {\(P_3\)};
  \draw [fill] (-0.6,0.8) coordinate (X) circle[radius=2pt] node[above] {\(X\)};
  \draw [fill] (-0.6,-0.8) coordinate (Y) circle[radius=2pt] node[below] {\(Y\)};
  \draw [fill] (0.8,0) coordinate (Z) circle[radius=2pt] node[right] {\(Z\)};
  \draw [thick] (P3) -- node [right] {\(\Delta\)}(Z);
  \draw [thick] (P4) -- node [right] {\(\Delta\)} (Z);
  \draw [thick] (P1) -- node [above] {\(\Delta\)} (X);
  \draw [thick] (P2) -- node [below] {\(\Delta\)} (Y);
  \draw [dashed, thick] (X) -- node[above] {\(\Delta_b\)} (Z);
  \draw [dashed, thick] (Y) -- node [below] {\(\Delta_b\)} (Z);
  \draw [dashed, thick] (X) -- node [left] {\(\Delta_b\)} (Y);
\end{tikzpicture}
\end{center}
\caption{\label{fig:trianglediagram}The triangle diagram in AdS.}
\end{figure}
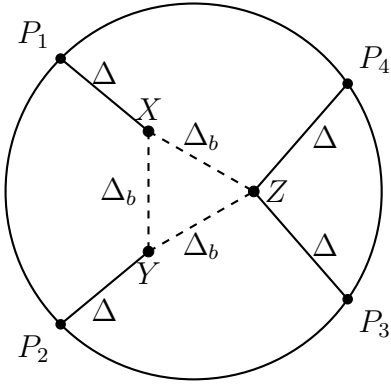

It is worthwhile to illustrate the general discussion of AdS Landau diagrams with the example of the triangle diagram, which is the simplest diagram exhibiting an anomalous threshold, i.e.~a singularity in the Mandelstam $s$ plane not directly attributable to an intermediate physical state. In more detail, we will consider an all-scalar triangle diagram with equal external masses $m$, equal internal masses $\mu$, and a momentum configuration as in figure \eqref{fig:trianglediagram}. In flat space the loop integral is easily written down and one finds that for
\begin{align}
\frac{1}{2}m<\mu<\frac{\sqrt{2}}{2}m,\qquad 0<s<2m^2,
\end{align}
there is a cut in the Mandelstam $s$ variable on the physical sheet starting at
\begin{align}
s^{\text{anom}}=4m^2-\frac{m^4}{\mu^2}.
\end{align}
Since this is less than the natural physical threshold $4 \mu^2$, the singularity is said to be anomalous. Our aim in this section is to reproduce this anomalous threshold from the corresponding Witten diagram.

In AdS the triangle Witten diagram reads
\begin{align}
G_{t}(P_i)=
&\int dX dY dZ G_{B\partial}(X,P_1)G_{B\partial}(Y,P_2)G_{B\partial}(Z,P_3)G_{B\partial}(Z,P_4)\nonumber\\
&\times G_{BB}(X,Y)G_{BB}(Y,Z)G_{BB}(X,Z).
\end{align}
with scaling dimensions $\D \sim m R$ and $\D_b \sim \mu R$ for the propagators as given in figure \ref{fig:trianglediagram}. As per the previous discussion, the action of the AdS Landau diagram to extremize is:
\begin{align}
\tilde{f}_t(X,Y,Z)=
&-\D \log(-P_1 \cdot X/R)-\D \log(-P_2\cdot Y/R)\nonumber\\
&-\D \log(-P_3\cdot Z/R)-\D \log(-P4\cdot Z/R)\nonumber\\
&-\D_b d(X, Y)-\D_b d(Y, Z)-\D_b d(X, Z),
\end{align}
with $d(X,Y) = \text{arccosh} (-X \cdot Y / R^2)$. We will again assume that we are on the support of the momentum conserving delta function and will also pass to the center of mass frame. Then the symmetries of the problem dictate the following ansatz:
\begin{align}
P_1 &=\left(1,\cos(\theta_P),\sin(\theta_P\right),\qquad 
P_2 =\left(1,\cos(\theta_P),-\sin(\theta_P\right),\nonumber\\[4pt]
P_3 &=\left(1,-\cos(\theta_P),\sin(\theta_P\right),\quad 
P_4 =\left(1,-\cos(\theta_P),-\sin(\theta_P\right),\nonumber\\[4pt]
X&=\left(
R\cosh\left(\frac{\rho_X}{R}\right),R\sinh\left(\frac{\rho_X}{R}\right)\cos(\theta_X),R\sinh\left(\frac{\rho_X}{R}\right)\sin(\theta_X)
\right),\\[4pt]
Y&=\left(
R\cosh\left(\frac{\rho_X}{R}\right),R\sinh\left(\frac{\rho_X}{R}\right)\cos(\theta_X),-R\sinh\left(\frac{\rho_X}{R}\right)\sin(\theta_X)
\right),\nonumber\\[4pt]
Z&=\left(
R\cosh\left(\frac{\rho_Z}{R}\right),R\sinh\left(\frac{\rho_Z}{R}\right),0
\right)\nonumber.
\end{align}
Although we wrote equations for AdS$_2$, the diagram does not depend on Mandelstam $t$ so our result is actually valid for any spacetime dimension. The only independent parameters are $\rho_X$, $\rho_Z$ and $\theta_X$ and the remaining AdS Landau equations can be found from:
\begin{align}
\begin{gathered}
\frac{\partial \tilde{f}_t}{\partial\rho_{X}} =
\frac{\partial \tilde{f}_t}{\partial\rho_{Z}} =
\frac{\partial \tilde{f}_t}{\partial\theta_{X}} =0\,,
\end{gathered}
\label{Eqn_triangle_saddle_eqn}
\end{align}
whereas the Mandelstam $s$ variable is related to $\theta_P$ as:
\be
\theta_P=-\frac{1}{2}\arccos\left(\frac{s-2m^2}{2m^2}\right)+\pi\,.
\ee
The precise branch can be fixed by requiring $\theta_P\in[\pi/2,\pi]$ for $s\in[0,4]$. 

\begin{figure}[t!]
\centering
\hspace{-2cm}
\begin{subfigure}{.5\textwidth}
\centering
\includegraphics[width=10cm]{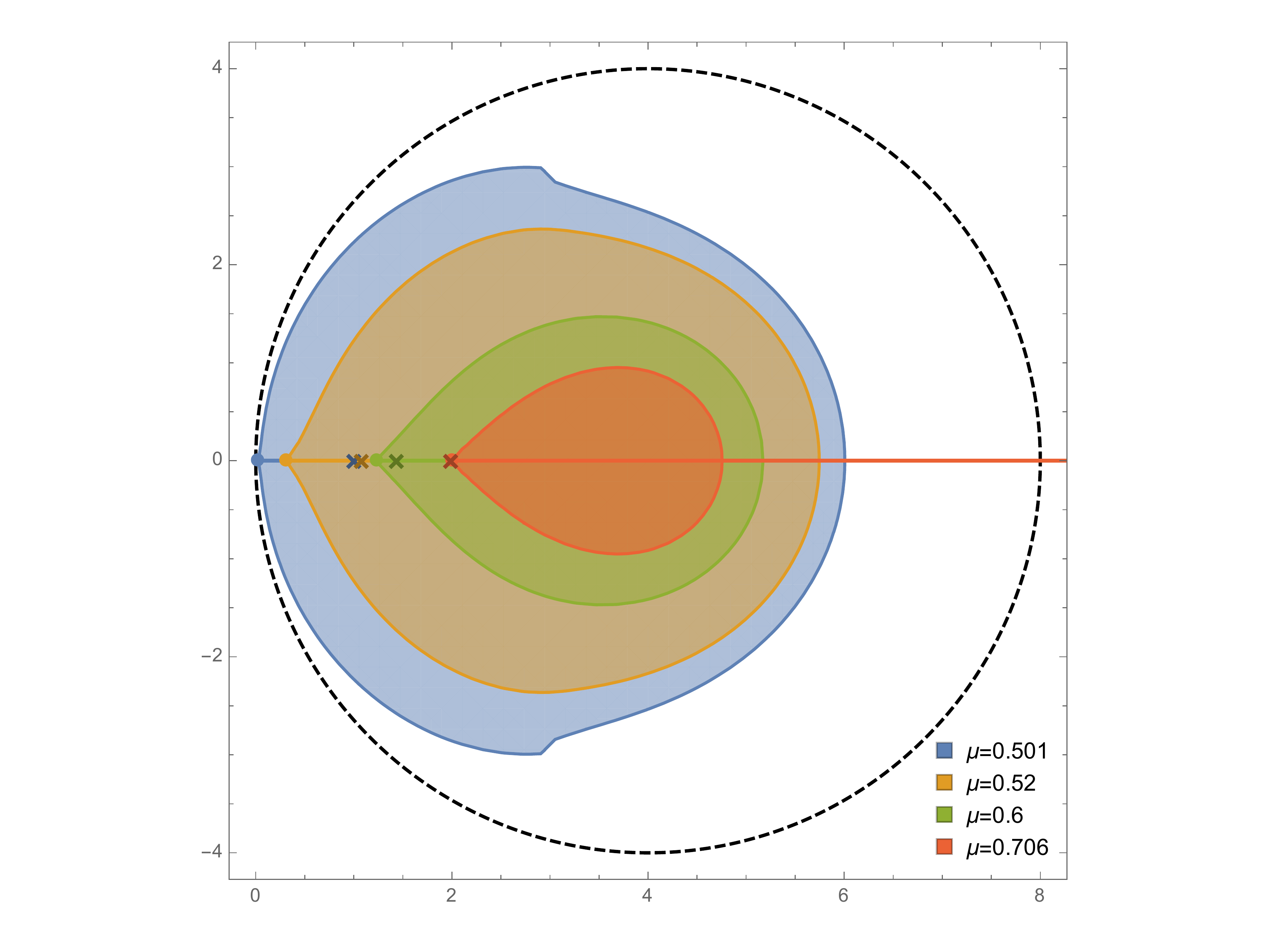}
\end{subfigure}%
\hfill
\begin{subfigure}{.5\textwidth}
\centering
\includegraphics[width=10cm]{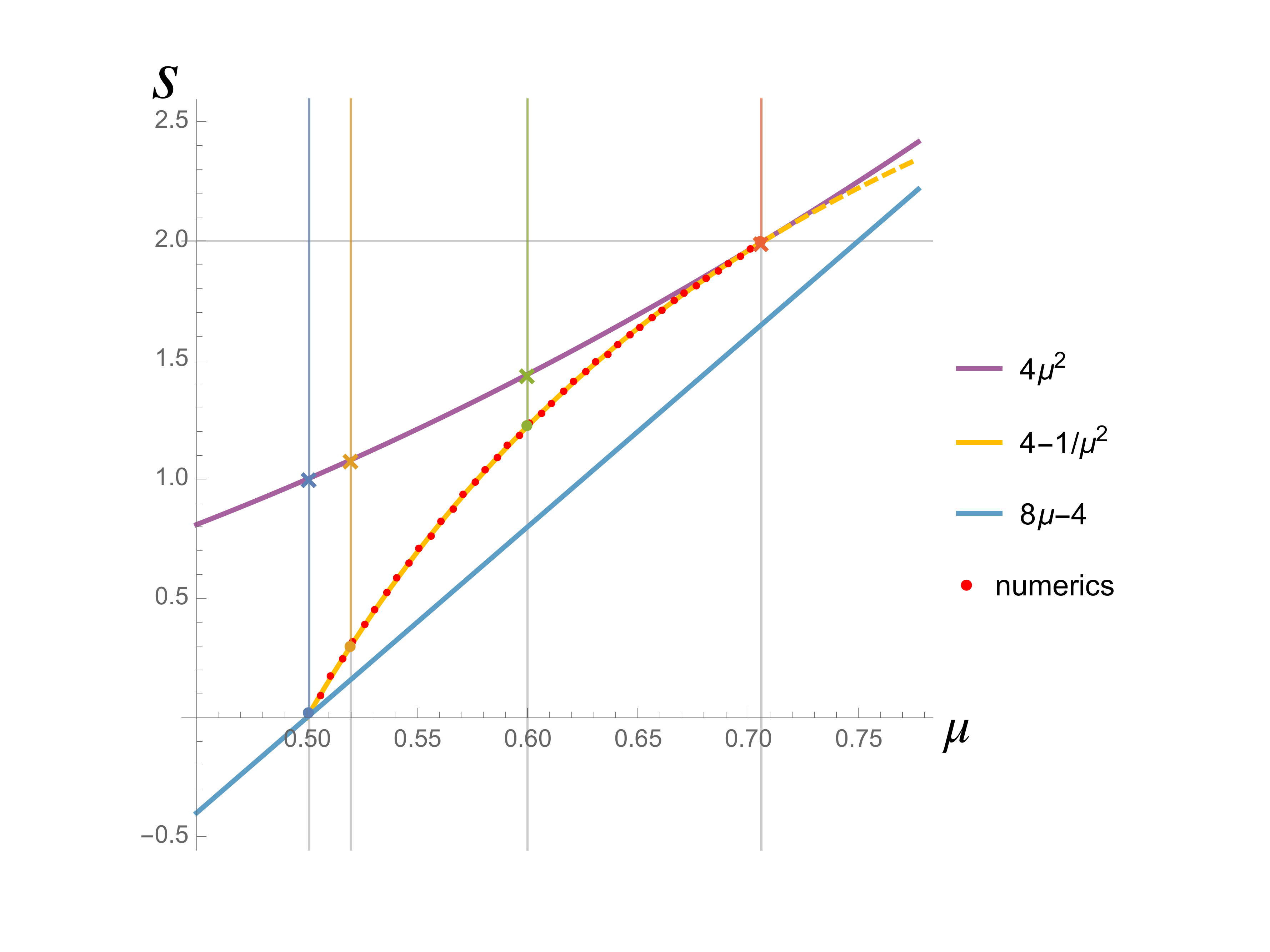}
\end{subfigure}
\caption{Left: Regions in the complex $s$ plane (shaded) where the AdS Landau diagram dominates over the flat-space saddle point. We have set $m=1$. The dots, lines and crosses are flat-space data: they are respectively the start of the anomalous threshold at $4m^2-m^4/\mu^2$, the branch cuts emanating from it, and the physical threshold $4\mu^2$ which is a little further along the cut. (The anomalous and physical threshold coincide when $\mu/m=\sqrt{2}/2$.) The shaded regions are the `bad' regions obtained numerically. Note that they always lie within the disk given by $|s-4m^2|<4m^2$ whose boundary is the black dashed circle. Right: the red dots indicate, for a given $\mu\in\left(1/2, 1/\sqrt{2}\right)$, the smallest real $s$ for which the AdS Landau diagram dominates. They are in perfect agreement with the yellow curve which is the flat-space anomalous threshold. The physical threshold is indicated by the purple curve. The blue curve will be useful later for discussion in section \ref{sec:mellin}. The vertical lines correspond to the branch cuts for different $\mu$ as in the left figure.}
\label{Fig_Triangle_numerics}
\end{figure}

Unfortunately the equations in \eqref{Eqn_triangle_saddle_eqn} are still somewhat difficult to solve analytically and so we will proceed numerically. In agreement with the general discussion earlier in this section, but in marked contrast with the flat-space equations, we managed to find solutions to the AdS Landau equations everywhere we looked in the complex $s$ plane.\footnote{The actual algorithm was fairly delicate. We used \texttt{FindRoot} in \texttt{Mathematica} to solve \eqref{Eqn_triangle_saddle_eqn} for different values of $s$. Unfortunately this method would often fail without a nearly perfect initial guess of the values $\{\rho_X,\rho_Z,\theta_X\}$. We therefore proceeded iteratively, starting from an `easy' point like $s = 4$ and then searching radially outward in small steps. We adjusted the step size dynamically, reducing it if no solution could be found. Additional radial searches were applied to check for non-convexities in the domain. Additional care needs to be taken because of the branch cuts in the square roots in the saddle point equations.} For each value we computed the on-shell action and checked the sign of $\text{Re}(\tilde{f}_t)$. The region where $\text{Re} (\tilde{f}_t)> 0$ is the problematic region because the Landau diagram dominates over the correct saddle point. In figure \ref{Fig_Triangle_numerics} we show the result for $\mu/m = 0.501,0.52,0.6,0.706(\approx\sqrt{2}/2)$ in detail on the left. We again uncover a blob-like region, entirely contained in the region $|s - 4 m^2| < 4 m^2$, where the flat-space limit gives a divergent answer and the conjectures of section \ref{Section_Erasing the circle} do not hold. It is satisfying to see that the anomalous threshold is correctly reproduced: on the right we show that for general $\mu$ there is a perfect agreement between the flat-space and AdS Landau diagram thresholds.

\section{Mellin space}
\label{sec:mellin}

In this section we will compare our conjectures with the flat-space prescription of \cite{QFTinAdS} that was based on Mellin space \cite{Mack:2009mi,joaomellin}. Recall that the Mellin space expression for a Witten diagram with $n$ points $G(P_{ij})$ is
\be
G(P_{ij}) 
= \int [d \gamma_{ij}] M(\gamma_{ij}) \prod_{1 \leq i < j \leq n} \Gamma(\gamma_{ij}) P_{ij}^{-\gamma_{ij}}
\label{mellindefinition}
\ee
where 
\be
P_{ij} \colonequals - 2 P_i \cdot P_j
\ee
and the Mellin space (integration) variables $\gamma_{ij}$ satisfy
\begin{align}\label{gammaijconstraints}
\sum_{j \neq i}^n \gamma_{ij} = \D_i, \qquad
\gamma_{ij}=\gamma_{ji}.
\end{align}
The integration measure $[d\gamma]$ is shorthand for a contour integral over the $n(n-3)/2$ independent Mellin variables and includes a factor of $1/2\pi i$ for each variable. The integration contour is of Mellin-Barnes type: it runs from $- i \infty$ to $+ i \infty$ and separates the poles in the Gamma functions and the Mellin amplitude in the usual way. The Witten diagram is then encoded in the meromorphic \emph{Mellin amplitude} $M(\gamma_{ij})$. For example, for an $n$-point contact Witten diagram the Mellin amplitude is just a constant:
\be\label{mellincontact}
\text{contact diagram:} \qquad M(\gamma_{ij}) = {\mathcal N} / R^{n(d-1)/2 - d - 1} 
\ee
with the canonical normalization constant
\begin{align*} 
\NN = \frac{1}{2} \pi^h \Gamma\left(\frac{\sum_i \D_i -d}{2}\right) \prod_{i=1}^n \frac{\CC_{\D_i}}{\Gamma(\D_i)}.
\end{align*}
 
Below we will investigate what happens to the Mellin representation in the flat-space limit in order to see what our conjecture \eqref{Eqn_flatspaceSconj} becomes for Mellin-representable functions.\footnote{Although the Mellin space representation works very well for Witten diagrams \cite{Fitzpatrick:2011ia,Paulos:2011ie}, it is more general and according to recent work \cite{Penedones:2019tng} can also be used for certain non-perturbative CFT correlation functions. It would be extremely interesting to further explore the applicability of Mellin space because of the immediate implication on our conjectures as outlined below.} This will also allow us to tie our conjectures to one of the two conjectures in \cite{QFTinAdS}, which we recall claimed that the scattering amplitude can be obtained directly in terms of the Mellin amplitude via:
\begin{align}
\label{flatspaceMellin}
(m_1)^{n(d-1)/2-d-1} &\TT(k_1 \ldots k_n) = \lim_{\D_i \to \infty}
\frac{(\D_1)^{n(d-1)/2-d-1}}{\NN} M \left(
\gamma_{ij}=\frac{\D_i \D_j}{\sum_{k} \D_k}\left(1+\frac{k_i \cdot k_j}{m_i m_j}\right)
\right),
\end{align}
Notice that all momenta in this equation are again taken to be ingoing (so $k^0 < 0$ for momenta corresponding to `out' states). As a zeroth order check, the prescription \eqref{flatspaceMellin} clearly works for the contact diagram given in \eqref{mellincontact}. Below we will essentially recover this conjecture, but in the process we will also be able to explain the appearance of the momentum-conserving delta function and find several subtleties that will explain the anomalous behavior discussed in sections \ref{sec:examples} and \ref{sec:landaudiagrams}.

\subsection{The saddle point}
In the large $R, \Delta$ limit it is natural to rescale $\gamma_{ij} = R \sigma_{ij}$ with $\sigma_{ij}$ fixed. We then find that
\be
G(P_{ij})\overset{\mathrm{R \to \infty}}{\longrightarrow}  R^{n(n-3)/2} e^{\sum_{i} \D_i \log(R) / 2} \int [d\sigma_{ij}] M(R \sigma_{ij}) \exp\left(R\, F(\sigma_{ij}; P_{ij})\right)
\ee
with
\be
F(\sigma_{ij}; P_{kl}) \colonequals \sum_{1 \leq i < j \leq n} \left( \sigma_{ij} \log(\sigma_{ij}) - \sigma_{ij} - \sigma_{ij} \log(P_{ij}) \right)
\ee
If the Mellin amplitude does not scale exponentially with $R$ then it is natural to assume that the integral can be approximated using a saddle point point analysis. Since the $\sigma_{ij}$ variables obey linear constraints the saddle point equations are actually a little bit involved because one needs to pull back the partial derivatives $\partial F / \partial \sigma_{ij}$ to the constraint surface. When the dust settles one finds that
\be
0 = \frac{\partial F}{\partial \sigma_{ij}} + \frac{2}{(n-1)(n-2)} \sum_{1 \leq k < l \leq n} \frac{\partial F}{\partial \sigma_{kl}} - \frac{1}{n-2}\left( \sum_{k \neq i} \frac{\partial F}{\partial \sigma_{ik}} + \sum_{k \neq j} \frac{\partial F}{\partial \sigma_{kj}}\right) 
\ee
A simple check of these equations is that they are trivially obeyed for $n = 3$, in agreement with the fact that there are no independent integration variables left in the original Mellin integral. Notice that these saddle point equations have to be solved simultaneously with the constraints $\sum_{j \neq i} \sigma_{ij} = \Delta_i / R$. For the given $F(\sigma_{ij};P_{kl})$ we find that
\be
\frac{\partial F}{\partial \sigma_{ij}} = \log(\sigma_{ij}) - \log(P_{ij})
\ee
The case $n = 4$ is illustrative. We find that
\be
\frac{\partial F}{\partial \sigma_{12}} + \frac{\partial F}{\partial \s_{34}} = \frac{\partial F}{\partial \sigma_{13}} + \frac{\partial F}{\partial \s_{24}} = \frac{\partial F}{\partial \sigma_{14}} + \frac{\partial F}{\partial \s_{23}}
\ee
If all the $\sigma_{ij}$ and $P_{ij}$ are real and positive then this gives:
\be
\frac{\sigma_{12} \sigma_{34}}{\sigma_{13} \sigma_{24}} = \frac{P_{12}P_{34}}{P_{13} P_{24}} \qquad \frac{\sigma_{14} \sigma_{23}}{\sigma_{13} \sigma_{24}} = \frac{P_{14}P_{23}}{P_{13} P_{24}}
\ee
so the cross-ratios in the Mellin variables should equal the cross-ratios in position space! Similar-looking expressions arise for $n > 4$.

Instead of solving these equations in full generality, let us consider the obvious attempt for the solution $\sigma_{ij}^*$ which is
\be
\sigma_{ij}^* = \frac{\D_i \D_j}{2 R \sum_k \D_k} P_{ij} = \frac{\D_i \D_j}{R \sum_k \D_k} \left(1 + \frac{k_i \cdot \eta \cdot k_j}{m_i m_j}\right)
\ee
which are just the values for the Mellin space prescription \eqref{flatspaceMellin}. This attempt solves the saddle point equations but the constraint equations now read:
\be
 \sum_{j \neq i}k_i \cdot \eta \cdot k_j = m_i^2
\ee
up to unimportant subleading terms in $1/R$. But this is now a constraint on the $k_i$ (so on the $P_i$) which is solved whenever
\be
\sum_{j} k_j^\mu = 0
\ee 
so precisely when we position the boundary operators on top of the support of the momentum-conserving delta function. Positioning the operators in this configuration is therefore part of the amplitude conjecture as discussed below equation \eqref{flatspaceTconj}. Since the Mellin amplitude $M(R\sigma_{ij})$ did not affect the location of the saddle point it is just an overall factor, and we can efficiently write the value at the saddle point as:
\be
G(P_{ij})\overset{\mathrm{R \to \infty}}{\longrightarrow}  G_c(P_{ij}) R^{n(d-1)/2 - d - 1}  M(R \sigma^*_{ij}) / {\mathcal N}
\ee
where $G_c(P_{ij})$ is the contact diagram in the same large $R$ limit. It then immediately follows that our amplitude conjecture reduces precisely to the Mellin space prescription conjecture \eqref{flatspaceMellin} as long as we can trust the saddle point approximation.

As for the contact diagram itself, we can reproduce all the computations of subsection \ref{subsec:contact} in Mellin space. For example, for $n = 4$ and all equal $\Delta_i = \Delta$ we can solve the Mellin saddle point equations to find:
\be
\sigma^*_{12} = \sigma^*_{34} = \frac{\Delta \sqrt{P_{12} P_{34}}/R}{\sqrt{P_{12} P_{34}} + \sqrt{P_{13}P_{24}} + \sqrt{P_{14} P_{23}}}
\ee
and the obvious permutations. With some work, the saddle point approximation can then be shown to yield
\be
\begin{split}
G_c(P_{ij})& \overset{\mathrm{R \to \infty}}{\longrightarrow}  
\frac{2\pi^2 e^{ 2 \D \log(\D) - 2\D}}{\D^2}
\frac{ \left(\sqrt{P_{12} P_{34}} + \sqrt{P_{13}P_{24}} + \sqrt{P_{14} P_{23}} \right)^{-2\D + 3/2} }{(P_{12}P_{13} P_{14} P_{23} P_{24} P_{34})^{1/4}} M_c(R \sigma^*_{ij})
\end{split}
\ee
which reduces to precisely the same expression as equation \eqref{Eqn_flat space limit G_c} and so the position space and Mellin space analyses of this diagram are in complete agreement.

\subsection{The steepest descent contour}
In our previous analysis we determined that the Mellin space prescription \eqref{flatspaceMellin} follows from our conjectures if we use a saddle point approximation for the Mellin integration variables. What remains to be checked is where the saddle point analysis can be trusted. In section \ref{sec:examples} we found issues with the position-space analysis because the steepest descent contour for a bulk-bulk propagator may not pass inbetween the poles at $c = \pm \Delta$. As we discuss below, the same can happen in Mellin space where the steepest descent contour for the Mellin variables may lie on the wrong side of poles in the Mellin amplitude itself. We will see that the additional contribution from these poles is the Mellin space analogue of the AdS Landau diagram contributions discussed in section \ref{sec:landaudiagrams}.

First consider the poles in the Gamma functions that are part of the definition of the Mellin amplitude given in equation \eqref{mellindefinition}. In our analysis these disappeared when we used the Stirling approximation, so really we should verify if none of the $\sigma_{ij}$ is real and negative so this approximation is trustworthy. In terms of the Mandelstam invariants $s_{ij} \colonequals - (k_i + k_j)^2$ we find that
\be
\sigma_{ij}^* = \frac{1}{2 \sum_k m_k} ((m_i + m_j)^2 - s_{ij})
\ee
We see $\sigma_{ij}^*$ becomes real and negative precisely when the corresponding Mandelstam invariant lies above the two-particle threshold. We can therefore use the Mellin saddle point as long as we are on the principal sheet for all the Mandelstam invariants and stay at least an infinitesimal amount away from the physical values.\footnote{This analysis also highlights a potentially important difference between the Mellin space prescription and our amplitude conjecture: the former only works on the principal sheet but the latter has a chance of giving the right answer also on the other sheets that can be reached by passing through the multi-particle cuts. It would be interesting to explore this further.}

\begin{figure}[t!]
\begin{subfigure}[t]{.5\textwidth}
\centering
\includegraphics[width=9cm]{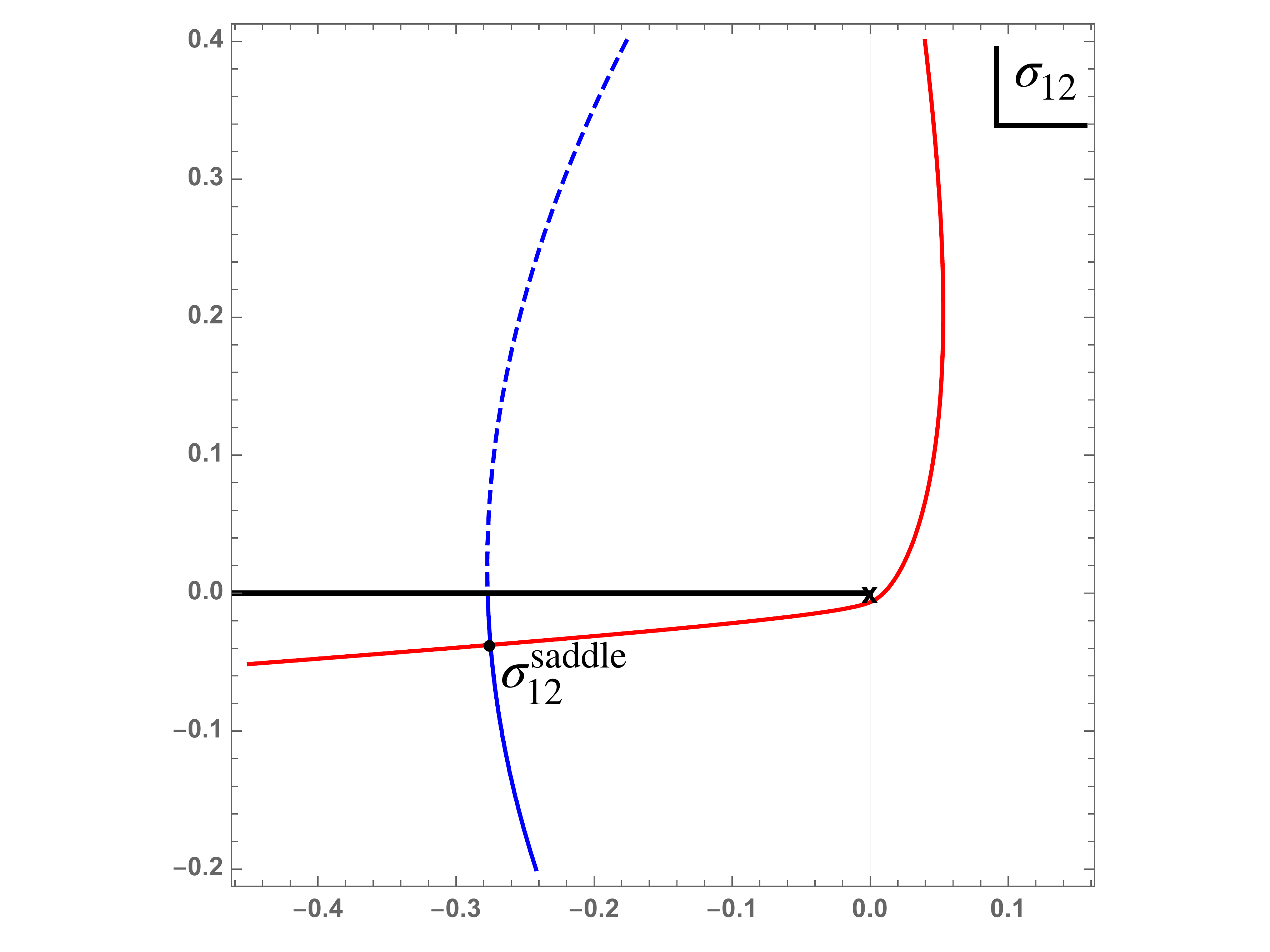}
\caption{$s=6.2+0.3i$}
\end{subfigure}
\begin{subfigure}[t]{.5\textwidth}
\centering
\includegraphics[width=9cm]{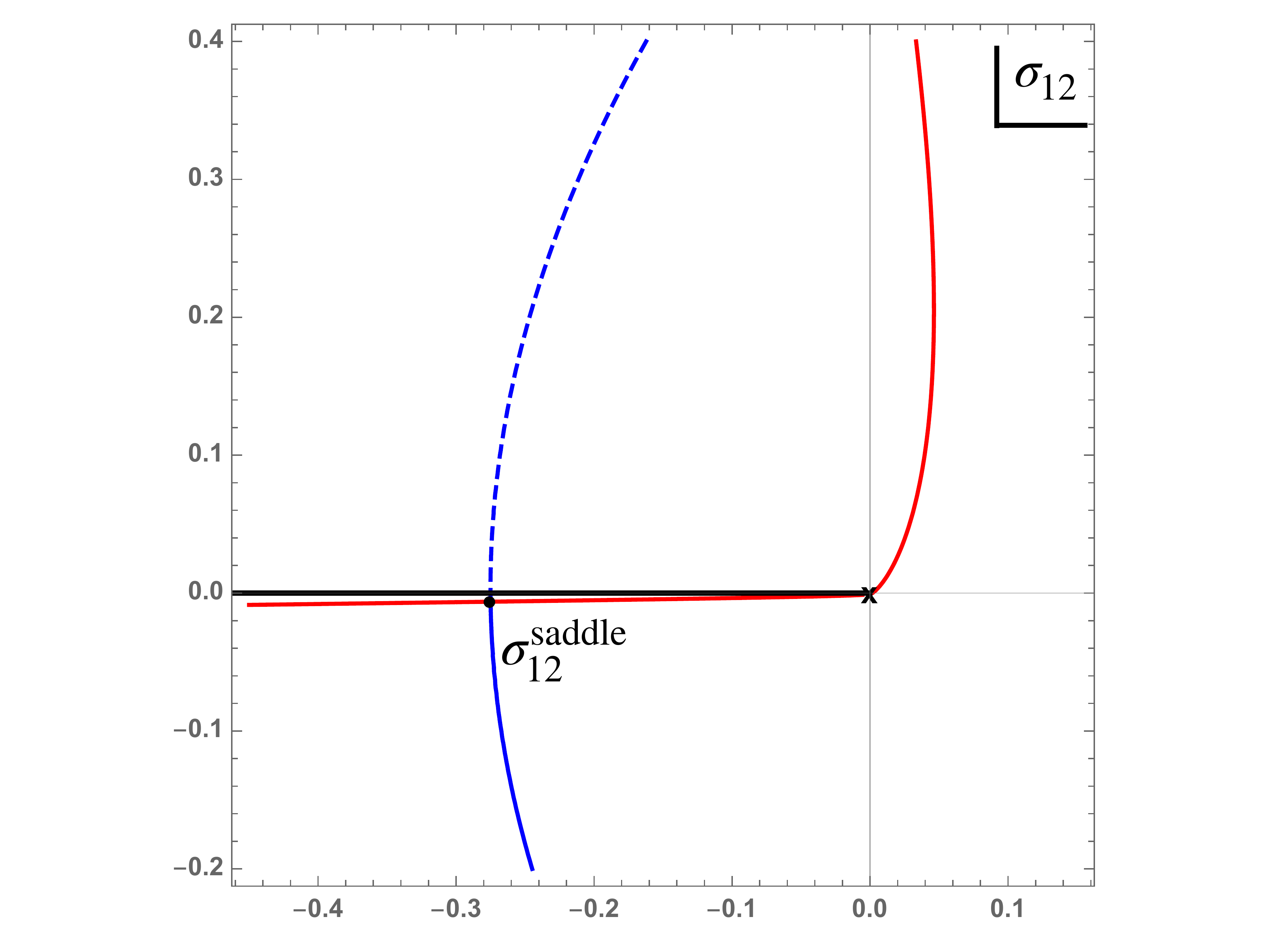}
\caption{$s=6.2+0.05i$}
\end{subfigure}
\caption{
Steepest descent (red curve) and ascent (blue curve) contours of $\sigma_{12}$ integral for $s=6.2+0.3i$ (left) or $s=6.2+0.05i$ (right). The black dot is the saddle point, and the cross indicates the starting poles of the semi-infinite sequences of poles in $\Gamma^2(R \s_{12})$. The other poles are represented by the black solid line. As $s$ approaches the physical region, the steepest descent contour gets close to the first $\Gamma$-function pole, but does not cross it.}
\label{Fig_Steepest descent of contact diagram}
\end{figure}

To illustrate this phenomenon we consider a four-point function of identical operators, see also figure \ref{Fig_Steepest descent of contact diagram}. After resolving the constraints on the Mellin variables we can write that
\be
G(P_{ij}) = (P_{13} P_{24})^{-\D} \int \frac{d \gamma_{12} d\gamma_{14}}{(2\pi i)^2} M(\g_{12},\g_{14}) \Gamma^2(\gamma_{12}) \Gamma^2(\gamma_{14}) \Gamma^2(\D - \gamma_{12} - \gamma_{14}) u^{-\gamma_{12}} v^{-\gamma_{14}}
\ee
The Gamma functions provide exponential damping for large imaginary values of their arguments, but this can be offset by extra phases introduced by the rotation of $u$ and $v$ in the complex plane. Starting from real and positive values of these cross ratios, we can rotate them by at most $2 \pi$ before the exponential damping is overcome and the Mellin representation becomes invalid. In more detail we can say that the Mellin representation works for $\sqrt{u}$, $\sqrt{v}$ and $\sqrt{u/v}$ away from the negative real axis. In terms of the Mandelstam invariants this means that
\be \label{mellinvaliditymandelstam}
- \pi < \text{arg}\left( \frac{s - 4m^2}{\tilde u - 4m^2} \right) < \pi, \qquad- \pi < \text{arg}\left( \frac{t - 4m^2}{\tilde u - 4m^2} \right) < \pi, \qquad- \pi < \text{arg}\left( \frac{s - 4m^2}{t - 4m^2} \right) < \pi. 
\ee
We see that for physical values of the Mandelstam variables (in either of the three channels) we are just at the boundary of the range of validity of the Mellin space representation and we need to use a small $i \epsilon$ prescription.\footnote{In practice we can probably send $\epsilon$ to zero as $R$ goes to infinity. We have not worked out this scaling in detail.}

Figure \ref{Fig_Steepest descent of contact diagram} shows the steepest descent contour in the $\sigma_{12}$ plane for values of Mandelstam $s$ approaching the physical threshold. We remark that in order to construct this figure we already performed the $\gamma_{14}$ integral, after which the location of the saddle point and the steepest descent contour for large $R$ depend only on Mandelstam $s$. More precisely, if we do the $\gamma_{14}$ integral by saddle point approximation then 
\begin{multline}
\label{eq:Wsigma12only}
G(P_{ij})|_{\text{cons}} \overset{\mathrm{R \to \infty}}{\longrightarrow}  \int \frac{d \sigma_{12}}{2\pi i} (\ldots) M(R\s_{12},R \s^*_{14}) \\ \times \exp\left( 2 R \left( \sigma_{12} \log(\sigma_{12}) + (m-\sigma_{12})\log(m - \sigma_{12}) + \sigma_{12} \log\left( \frac{s + 4m^2}{s-4m^2} \right)\right) \right)
\end{multline}
where we have also made the substitution $\gamma_{12} =  R\sigma_{12}$ and expressed the cross ratios in terms of the Mandelstam invariants. In deriving this expression we assumed that the $P_{i}$ were chosen to lie on the support of the momentum-conserving delta function. The ellipsis refer to subleading terms in $1/R$ which do not affect the location of the steepest descent contour in figure \ref{Fig_Steepest descent of contact diagram} (but which are essential for recovering the delta function as described above).


A second potential issue is due to the existence of poles in the Mellin amplitude itself. These are present for non-contact Witten diagrams in AdS and therefore we need to understand how they affect the saddle point and the steepest descent contour analysis. Unfortunately at present we lack the technology to analyze the most general Witten diagram in the most general kinematical setup. We will therefore focus on the scalar exchange diagram as a case study.

\subsection{The exchange diagram revisited}
\label{subsec:mellinexchange}
Let us again consider the scalar exchange diagram, already discussed in section \ref{subsec:exchange}, involving four external operators of dimension $\Delta$ with an exchange of a scalar operator with dimension $\D_b$. The corresponding Mellin amplitude is \cite{joaomellin}:
\begin{align}
\mathcal N^{-1} M(\gamma_{12})= - R^{5-d} \sum_{k=0}^{\infty} 
\frac{R_k}{2 \D -2\gamma_{12} - \Delta_b - 2k}, 
\end{align}
with
\begin{align}
R _ { k } = \frac { \Gamma ( \frac { 2 \D + \Delta_b - d } { 2 } )^2} { 2 \Gamma ( \frac { 4 \Delta - d } { 2 } ) } \frac { ( 1 + \frac { \Delta_b - 2\Delta } { 2 } )^2 _ { k } } { k ! \Gamma ( \Delta_b - \frac { d } { 2 } + 1 + k ) }.
\end{align}
For the flat-space limit we recall the analysis in \cite{QFTinAdS}: set $k = R \kappa$ to find that, at large $R$,
\be \label{eq:Rkasympt}
R_{k} \overset{\mathrm{R \to \infty}}{\longrightarrow} \frac{1}{4 m R^2} \sqrt{\frac{64 R m^3}{\pi (4 m^2-m_b^2)^2}} \exp\left( - \frac{64 R m^3}{(4m^2 - m_b^2)^2} \left( \kappa - (2m - m_b)^2/8m\right)^2 + \ldots \right)
\ee
The sum over $k$ therefore localizes at $\kappa^* = (2m - m_b)^2/8m$ and the remaining Gaussian sum (over $k$, not $\kappa$) gives a factor $(4 m R)^{-1}$ so we find that
\be
R^{d-5} \mathcal N^{-1} M(R\sigma_{12}^*) \overset{\mathrm{R \to \infty}}{\longrightarrow} - \frac{1}{4 m R^2(2 m - 2 \sigma^*_{12} - m_b - 2 \kappa^*)} = - \frac{1}{ R^2} \frac{1}{s - m_b^2}
\ee
as expected from the flat-space formula in Mellin space \cite{QFTinAdS}.

\begin{figure}[t!]
\hspace*{-.3in}
\begin{subfigure}[t]{.5\textwidth}
\centering
\includegraphics[width=9cm]{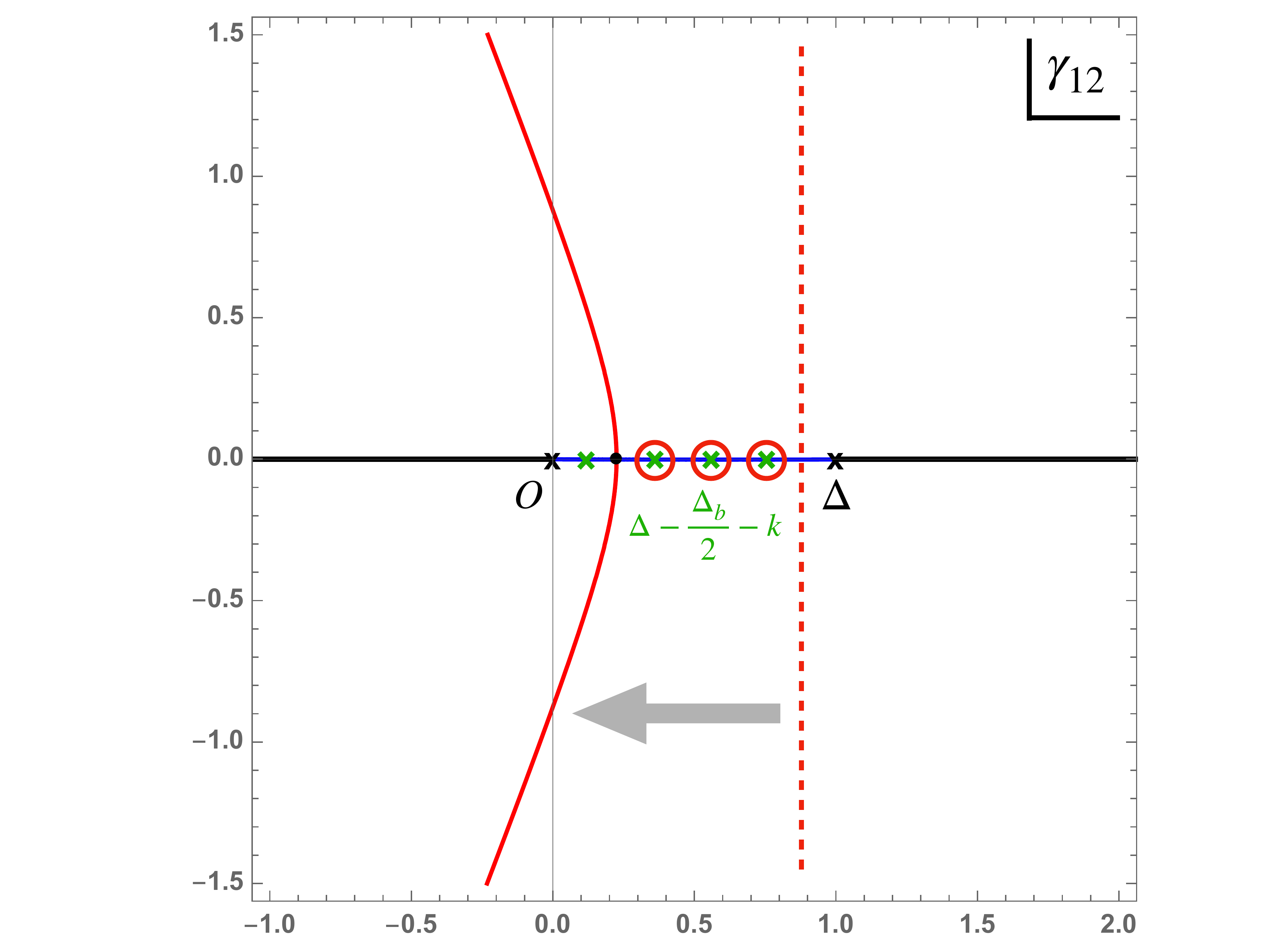}
\caption{$s=2.2$}
\end{subfigure}
\begin{subfigure}[t]{.5\textwidth}
\centering
\includegraphics[width=9cm]{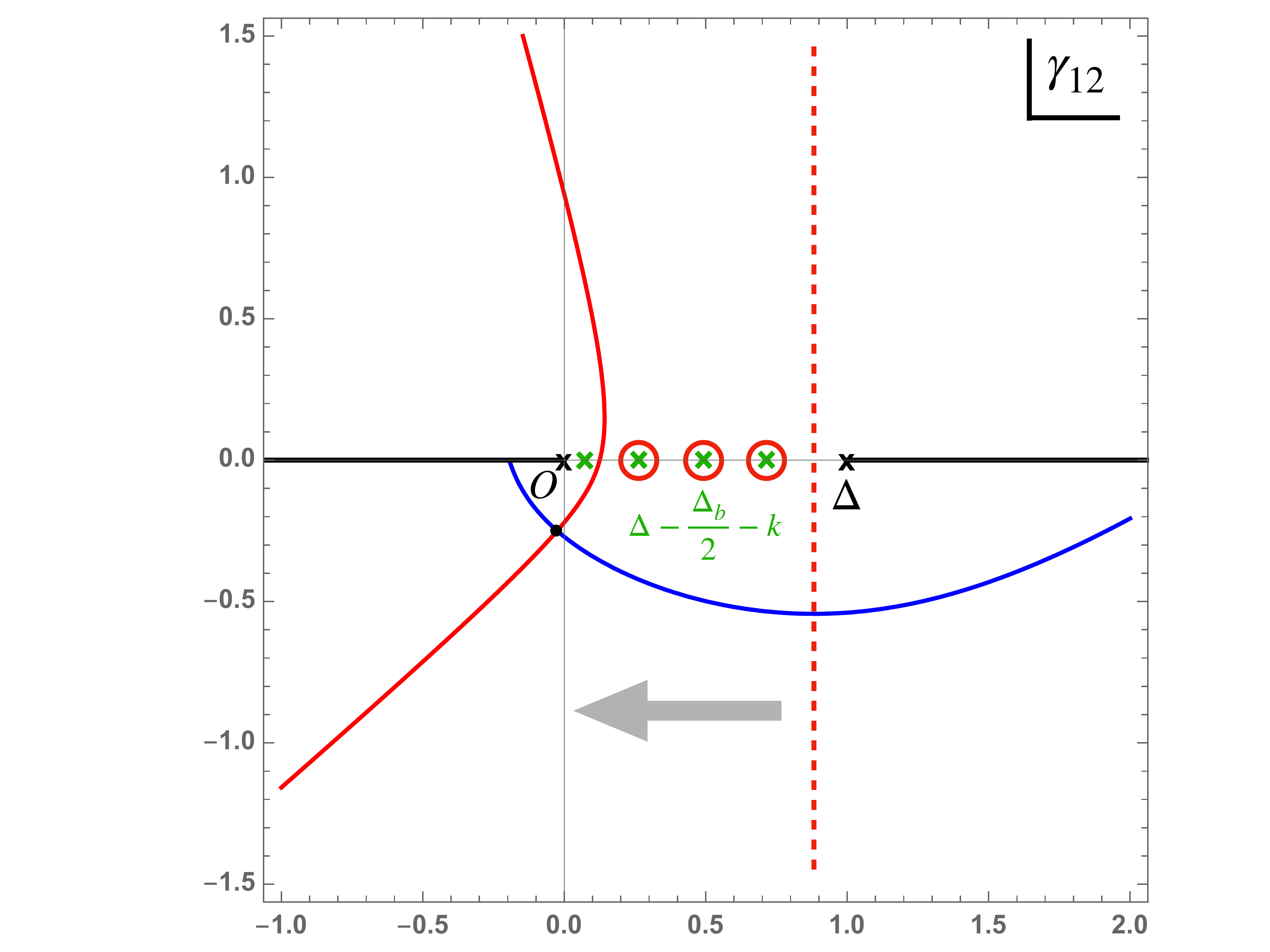}
\caption{$s=4.2+2i$}
\end{subfigure}
\caption{
Steepest descent (red curve) and ascent (blue curve) contours of scalar exchange diagram with $m=1$, $s=2.2$ (left) or $s=4.2+2i$ (right). The black dot is the saddle point. The red dashed line indicates the original integral contour. The green crosses are poles of the Mellin amplitude and those with a red circle are picked up during contour deformation. The black crosses are the starting poles of $\Gamma$-functions and the rest are represented by black solid lines.}
\label{Fig_Steepest descent of scalar exchange diagram}
\end{figure}

Now we can analyze the steepest descent contour. Since the Mellin amplitude for the exchange diagram only depends on $\gamma_{12}$, we can safely do the $\gamma_{14}$ integral by a saddle point approximation and start our analysis from \eqref{eq:Wsigma12only}, the important part of which is:
\be
\label{eq:Wsigma12only2}
\begin{split}
G(P_{ij})|_{\text{cons}} &\overset{\mathrm{R \to \infty}}{\longrightarrow}  \int \frac{d \sigma_{12}}{2\pi i} (\ldots) M(R\s_{12},R \s^*_{14}) \exp\left( R \,\phi\left(m,\frac{4m^2 + s}{4m^2 -s},\sigma_{12}\right)\right)\\
\phi(\a,\b,\sigma) &\colonequals 2 \sigma \log(\sigma) + 2(\a-\sigma)\log(\a - \sigma) + 2 \sigma \log\left( \b \right)
\end{split}
\ee
We will once more assume that we are on the support of the momentum-conserving delta function. The main idea is illustrated in figure \ref{Fig_Steepest descent of scalar exchange diagram}, where deforming the original integration contour to the steepest descent contour may yield a number of contributions from the Mellin poles. Of course, if Mellin poles get picked up then we still have to check whether their contribution is leading or subleading at large $R$. So we naturally arrive at three regions: in region I no Mellin poles are picked up, in region II Mellin poles are picked up but their contribution is subleading, and finally in region III Mellin poles are picked up and leading. Only in region III does the flat space limit not work.

\begin{figure}[t]
\begin{center}
\includegraphics[width=10cm]{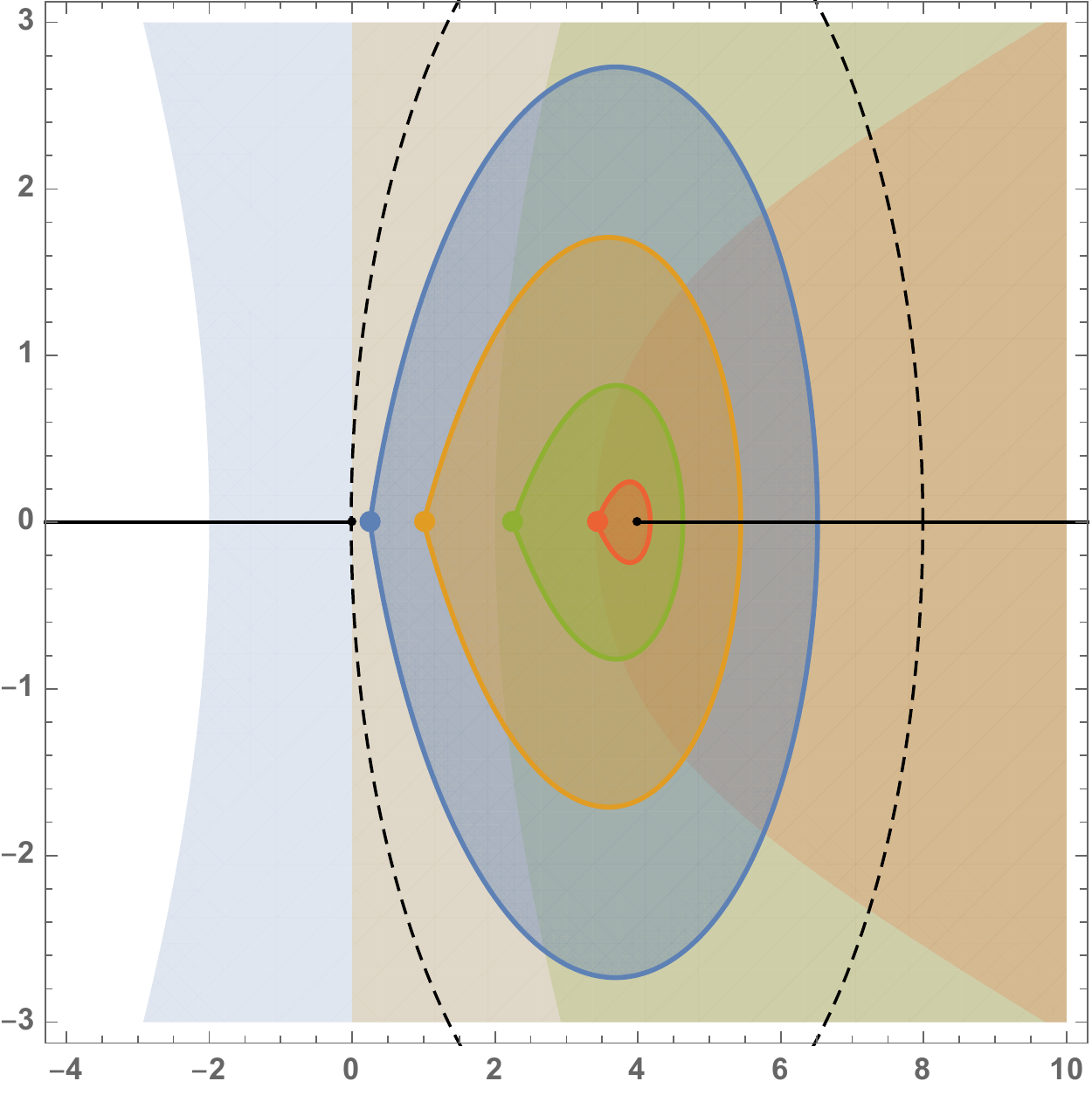}
\caption{\label{fig_regionsmellin}Light shading: regions in the complex $s$ plane where we pick up poles in the Mellin amplitude. Dark shading: regions where these poles dominate and our flat-space limit diverges. We have set $m=1$ and the blue, orange, green and red domains respectively correspond to $m_b = 0.5, 1, 1.5, 1.85$. We have also highlighted the pole at $s = m_b^2$ for each color as well as the cut at $s = 4$. Although this is not entirely obvious from the plot, the blue region extends rightward to include the orange, green and red regions and similarly for the other colors.}
\end{center}
\end{figure}

Finding these three regions is a technical exercise that involves finding the stationary phase contour in \eqref{eq:Wsigma12only2} and then estimating the contribution of any poles that get picked up in regions II and III. The details are left to appendix \ref{app:exchangemellin} and the result is sketched in figure \ref{fig_regionsmellin}. This figure should of course be compared with figure \ref{fig_regions3} in section \ref{sec:examples}. Clearly the main results in each figure, namely the region III `blobs' where the flat-space limit diverges, are exactly the same. The analyses were also remarkably similar: in both cases there were `anomalous' contributions from either the pole at $c = \Delta_b$ or from the poles in the Mellin amplitude. Nevertheless the regions of type II are different: sometimes Mellin poles get picked up whereas the pole at $c = \Delta_b$ does not. This is not a contradiction: outside the blobs these contributions vanish anyway, so the discrepancy just shows that not all zeroes are created equally.

\subsection{A bound on anomalous thresholds?}
Compared to scattering amplitudes, conformal correlation functions have an extra feature that we have not exploited so far: a conformal block decomposition. Since a conformal block for an operator of dimension $\Delta_b$ and spin $\ell$ corresponds to a semi-infinite sequences of pole in the Mellin amplitude starting at $\Delta_b - \ell$, we are in the unique position that we know the singularities in the Mellin amplitudes if we know the spectrum of the theory (and assume that the correlation function can be represented as a Mellin amplitude). In other words, unlike scattering amplitudes Mellin amplitudes cannot have anomalous thresholds like the one discussed for the triangle diagram: if the first physical state, so the first non-trivial conformal block, corresponds to an operator with twist $\tau_b$ then the Mellin amplitude should have poles only for $\gamma_{12} \leq \Delta - \tau_b/2$. As examples we note that for the exchange diagram $\tau_b = m_b R$ whereas for the triangle diagram discussed in section \ref{subsec:triangle} $\tau_b = 2 \mu R$ with $\mu$ the mass of the internal particle.

In the flat-space limit we are supposed to evaluate the Mellin amplitude at
\be
\gamma_{12} = \frac{\D}{8 m^2}(4 m^2 - s)
\ee
and therefore the Mellin poles can only interfere with the flat-space limit if
\be \label{sineq}
s > 4 m (\tau_b/R - m)
\ee
on the real axis. For complex $s$ the Mellin conjecture \eqref{flatspaceMellin} would predict that there are no other singularities in the complex $s$ plane as long as \eqref{mellinvaliditymandelstam} is obeyed; for our amplitude conjecture we in addition need to restrict ourselves to $s$ in region I so no Mellin poles are picked up. (This region is easily found using the results in appendix \ref{app:exchangemellin}.) Either way, we see that the absence of anomalous thresholds in the Mellin amplitude leads to a region in the Mandelstam plane where Landau singularities should never appear, no matter how complicated the diagram.

Could the inequality be a universal threshold for Landau singularities? As a zeroth order check we have plotted the inequality \eqref{sineq} on the right in figure \ref{Fig_Triangle_numerics} and indeed find that it is obeyed by the anomalous triangle threshold. Of course, to obtain a general result we need to demand that equation \eqref{sineq} holds also for $t$ and $u$ with $\tau_b$ replaced with the first physical operator in the corresponding conformal block channel, and there may be additional subtleties if the Mellin amplitude depends non-trivially on both the Mandelstam variables. We should also note that this putative bound hinges on the validity of either the Mellin or the amplitude conjecture, but if it holds then it is a result that applies purely to flat-space scattering amplitudes. It would be interesting to investigate this further, either with QFT in AdS methods or perhaps even without reference to AdS.


\section{S-matrices from conformal block expansions}
\label{sec:blocks}
Up to now we have motivated and examined our conjecture that CFT correlators become S-matrices in the flat space limit from a perturbative standpoint, examining specific diagrams for weakly coupled QFTs in AdS space. The goal of this section is to offer a complementary perspective by providing a non-perturbative argument for the validity of our central claim, in the case of 2-to-2 scattering. In that case we can explicitly use the OPE to express amplitudes in terms of the CFT data, and this will allow us to show that CFT correlators do indeed, under certain assumptions, become objects which obey the expected unitarity conditions for scattering amplitudes.

For simplicity we will consider identical scalar particles of mass $m$ and we will mostly focus on physical kinematics, i.e. $s>4 m^2$ and real scattering angle. We will be able to see explicitly under which circumstances our conjectures have a chance of being valid, and under which it certainly fails. 

\subsection{Preliminaries}

For the purposes of this section, we set:
\bea
\langle \underline{\tilde k}_3,\underline{\tilde k}_4| S |\underline{k}_1,\underline{k}_2\rangle=i (2\pi)^{d+1} \delta^{(d+1)}(k_1+k_2+k_3+k_4) \mathcal S(\tilde k_3,\tilde k_4;k_1,k_2)
\eea
Note that $\mathcal S(k_i)$ contains both connected and disconnected contributions, which may seem a bit unusual. In terms of $\mathcal S(k_i)$ our claim is:
\bea
 \mathcal S(\tilde k_3,\tilde k_4;k_1,k_2)=\lim_{R\to \infty} \frac{\langle \mathcal O(\tilde n_3)\mathcal O(\tilde n_4)\mathcal O(n_1) \mathcal O(n_2)\rangle}{G_c(\tilde n_3,\tilde n_4,n_1,n_2)} \bigg|_{\mbox{\tiny S-matrix, cons}} \label{eq:claim}
\eea
where $G_c(x_1,\ldots,x_4)$ is the contact diagram in AdS. Implicit in the equation is that $\Delta \to m R$ with the physical mass $m$ fixed, which henceforth we set to unity. We can think of the above formula as something that can be done for any family of CFT correlators that depends on the parameter $R$ such that the scaling dimensions grow with $R$. Below we will show that under certain assumptions the resulting $\mathcal S(k_i)$ will be finite and obey the expected unitarity conditions for an elastic amplitude.

By conformal invariance we can of course write:
\bea
\langle \cO(x_1)\cO(x_2)\cO(x_3)\cO(x_4)\rangle=\frac{\mathcal G_{\Df}(z,\bar z)}{x_{13}^{2\Df} x_{24}^{2\Df}}\,,\qquad
G_c(x_1,x_2,x_3,x_4)=\frac{\mathcal D_{\Df}(z,\bar z)}{x_{13}^{2\Df} x_{24}^{2\Df}}
\eea
and with the conformal block decomposition:
\bea
\mathcal G_{\Df}(z,\bar z)=\sum_{\Delta,\ell} a_{\Delta,\ell} \frac{G_{\Delta,\ell}(z,\bar z)}{(z \bar z)^{\Df}}
\eea
with $a_{\Delta,\ell}=\lambda_{\mathcal O_{\Delta,\ell}}^2$ the squared OPE coefficient, $\phi\times \phi\sim \sum_{\Delta} \lambda_{\mathcal O_{\Delta,\ell}} \mathcal O_{\Delta,\ell}$.

As reviewed in section \ref{Section_Erasing the circle}, the conformal cross ratios $u$ and $v$ are often parametrized in terms of the Dolan-Osborn variables $z$ and $\bar z$ or in terms of the radial coordinates $\rho$ and $\bar \rho$. In the following we will work with $r$ and $\eta$ which are related to the $\rho$ and $\bar \rho$ variables via:
\be
r(z,\bar z)=\sqrt{\rho} \sqrt{\bar \rho},\qquad \eta=\frac 12\left(\sqrt{\frac{\rho}{\bar \rho}}+\sqrt{\frac{\bar \rho}{ \rho}}\right)\,.
\ee
Equation \eqref{eq:conformalMandelstam} is easily adapted to find the map between $(r,\eta)$ and the Mandelstam variable $s$ and the scattering angle\footnote{Recall that Mandelstam $t=-\frac{1}{2} (s-4)\left(1-\cos(\theta)\right)$.} $\theta$:
\be
s = 4(1-z_{\mbox{\tiny eff}}),\qquad \qquad \eta =-\cos(\theta),
\ee
where the introduction of
\be
z_{\mbox{\tiny eff}} \colonequals \frac{4r}{(1+r)^2}
\ee
will be useful to simplify the notation below. The technical advantage of working with $r,\eta$ is that we may reach the physical kinematics scattering region, which is $s - i \epsilon > 4$ and real $\theta$, from the Euclidean section without worrying about branch cuts. Indeed, if we start from the Euclidean region where $r \geq 0$ and $\eta \in [-1,1]$ then we can reach the physical $s$-channel scattering region by taking $r$ negative with a small negative imaginary part. At the same time $\eta$ is taken to $-\eta$ while always remaining real.

A particular family of correlators that will be important for us is that of generalized free fields, which describe free massive particles propagating in $AdS$ space. For this theory the correlator of the elementary fields takes the form:
\bea
\label{gffcorr}
\langle \phi(x_1) \phi(x_2)\phi(x_3) \phi(x_4)\rangle &=\frac{1}{x_{12}^{2\Df} x_{34}^{2\Df}} + \frac{1}{x_{13}^{2\Df} x_{24}^{2\Df}} + \frac{1}{x_{14}^{2\Df} x_{23}^{2\Df}} \\
&=\,\frac{1}{x_{13}^{2\Df} x_{24}^{2\Df}} \left( \frac{1}{(z\bar z)^\Df}+ \sum_{\Delta,\ell} a_{n,\ell} \frac{G_{\Delta_{n,\ell},\ell}(z,\bar z)}{(z \bar z)^{\Df}} \right)\,.
\eea
where $\Delta_{n,\ell}=2\Df+2n+\ell$ corresponds to the scaling dimension of double twist operators of the form $\phi\partial^{\ell} \Box^{n} \phi$, and the OPE coefficients are given by:
\bea
a_{n,\ell}=\frac{(d-2)_\ell}{(d/2-1)_\ell}\,\frac{2 \left[(\Df-d/2+1)_n (\Df)_{\ell+n}\right]^2}{\ell!n! (d/2+\ell)_n (n+2\Df-d+1)_n (\ell+2n+2\Df-1)_\ell (\ell+n+2\Df-d/2)_n}.
\eea
Below it will be useful to define an OPE density which is defined for continuous values of $\Delta$:
\bea
a_{\Delta,\ell}^{\mbox{\tiny cont}}\equiv a_{n,\ell}\bigg|_{n=\frac{\Delta-2\Df-\ell}2}\,.
\eea

Note that in the flat-space limit the generalized free correlator should reduce to trivial scattering. For the last two terms in equation \eqref{gffcorr} this is indeed the case, since they are just products of two-point functions between one `in' and one `out' particle and the analysis of the two-point functions in section \eqref{subsec:twopt} is directly applicable. The first term corresponds to a Witten diagram where the two `in' particles and the two `out' particles are contracted. The corresponding Feynman diagram is certainly not part of a scattering amplitude, and we will analyze what remains of this contribution below.

\subsection{Conformal block expansion at large \texorpdfstring{$\Df$}{Delta}}
After these preliminary remarks, we begin by examining the flat space limit. In the limit of large scaling dimensions $\Delta$ the conformal block reduces to \cite{Kos:2014bka}:
\bea
G_{\Delta,\ell}\underset{\Delta\to\infty}{\sim} N_G \frac{r^{\Delta}}{(1-r^2)^\nu}\,\frac{C_\ell^{(\nu)}(\eta)}{\sqrt{(1+r^2)^2-4 r^2 \eta^2}}\,, \qquad N_G=\frac{\ell!}{(2\nu)_\ell}\, 4^\Delta\,.
\eea
with $\nu = (d-2)/2$. The second ingredient in our formula for $\mathcal S(k_i)$ is the contact term. In section \ref{subsec:contact} we determined that at large $\Df$ and in Euclidean kinematics we have:
\bea
\mathcal D_{\Df}(r,\eta)=\frac{1}{N_D}\left(\frac{1+ r^2 + 2 r \eta}{(1+r)^2}\right)^{2\Df} \frac{(1+r)^3}{\sqrt{r (1+r^2)^2-4 r^3 \eta^2} }\,.
\eea
where\footnote{In this section we work with unit normalized operators which means that this normalization factor is a little different from equation \eqref{Eqn_flat space limit G_c}.}
\bea
N_D=2^{\frac{7+d}2}\, \Df^{5\frac{d-1}2} \pi^{\frac{d-1}2}\,.
\eea

To proceed we need to make some simplifying assumptions consistent with the existence of a dual QFT description in flat space in the large radius limit. Let us assume that in the particular family of correlators under consideration, the spectrum of states consists of the identity, a finite set of scalar ``bound states'' with $1\ll\Delta<2 \Df$ and infinite towers of states with spins $\ell\geq 0$ and with $\Delta\geq2\Df+\ell$. The latter will become the multiparticle continuum in the large $R$ limit. Note that these assumptions are satisfied for instance for the generalized free field considered above. Physically we are stating that a hypothetical holographic flat-space QFT description should contain only massive states, and that we are examining the scattering of the lightest particle (otherwise multiparticle states would appear below $2\Df$). There are further assumptions that must be made on the OPE coefficients, as we shall see shortly.

We can now use the expressions above to write:
\begin{multline}
\lim_{\Df\to \infty} \frac{\mathcal G_{\Df}(r,\eta)}{\mathcal D_{\Df}(r,\eta)}=\lim_{\Df\to \infty}\left(\frac{\sqrt{r}}{(1+r)^3(1-r^2)^\nu}\right)\\
\left[I(r,\eta)+\sum_{0<\Delta_b<2\Df} \mathcal N_{\Delta_b}^{\mbox{\tiny bound}} \left(\frac{a_{\Delta_b,0}}{a_{\Delta_b}^{\mbox{\tiny bound}}}\right) \frac{r^{\Delta_b}}{z_{\mbox{\tiny eff}}^{2\Df}}
+\sum_{\substack{\Delta>2\Df+\ell\\\ell=0,2,\ldots}} \mathcal N_{\Delta}^{\mbox{\tiny cont}} \left(\frac{a_{\Delta,\ell}}{a_{\Delta,\ell}^{\mbox{\tiny cont}}}\right) \frac{r^{\Delta}}{z_{\mbox{\tiny eff}}^{2\Df}}\,  C_{\ell}^{(\nu)}(\eta)\right]
\end{multline}
where $I(r,\eta)$ is the identity block contribution,
\bea
I(r,\eta)\equiv N_D \frac{\sqrt{(1+r^2)^2-4 r^2 \eta^2}}{z_{\mbox{\tiny eff}}^{2\Df}} (1-r^2)^\nu
\eea
and we have defined the parameters
\bea
\mathcal N_{\Delta}^{\mbox{\tiny cont}}=N_G\,N_D\, a_{\Delta,\ell}^{\mbox{\tiny cont}}\,,\qquad\,\mathcal N_{\Delta_b}^{\mbox{\tiny bound}}=N_G\,N_D\, a_{\Delta_b}^{\mbox{\tiny bound}}\,.
\eea
We will give $a_{\Delta}^{\mbox{\tiny bound}}$ below. 

Our expression contains three kinds of contributions: the identity, ``bound states'' and the ``continuum'', i.e. states above the twist gap $\tau>2
\Df$. We will examine below each contribution in turn, but the logic is as follows. The third contribution is the most interesting one. We will see that for physical kinematics i.e. $s>4m^2$ and under certain assumptions on the OPE coefficients it is finite and becomes the partial wave decomposition of the S-matrix, with an expression for the spin-$\ell$ phase shifts in terms of the CFT data. But before we see this we must take care of the first two contributions, which will turn out to be either zero or divergent depending on the choice of kinematical region.

\subsubsection{Identity and bound states}
We begin with the contribution of the identity. As stated above, from an S-matrix perspective it corresponds to an unphysical diagram that connects the particles in the `in' states and the particles in the `out' states with nothing propagating in between. It might therefore be reasonable to subtract it by hand. Alternatively, we see that in the large $\Df$ limit the factor $z_{\mbox{\tiny eff}}^{-2 \Df}$ dominates and as long as
\bea
|z_{\mbox{\tiny eff}}|>1 \Leftrightarrow |s-4|>4\,,
\eea
the unwanted contribution vanishes automatically. In particular for physical kinematics this requires $s>8$. On the other hand, the factor diverges as long as $|z_{\mbox{\tiny eff}}|<1$.

Now for the bound states. In this case we need to know something about the behaviour of their OPE coefficients in the large $\Df$ limit. We will assume that for each such state we have
\bea
g_{\Delta_b}^2\equiv \lim_{R\to \infty} \frac{a_{\Delta_b,0}}{a_{\Delta_b}^{\mbox{\tiny bound}}}<\infty\, \qquad (\Delta_b<2\Df)
\eea
where:
\bea
a_{\Delta_b}^{\mbox{\tiny bound}}:=
\frac{\pi^d \Gamma\left(\frac{2\Df+\Delta_b-d}2\right)^2}{4 \Df^{d-5}}\, \frac{\mathcal C_{\Df}^2 \mathcal C_{\Delta_b} \Gamma\left(\frac{\Delta_b}2\right)^4 \Gamma\left(\frac {2\Df-\Delta_b}2\right)^2}{\Gamma(\Df)^4 \Gamma(\Delta_b)^2}
\eea
We have defined $g_{\Delta}^2$ in this way since it becomes a physical coupling in the flat-space limit for a QFT in AdS space, as discussed in \cite{QFTinAdS}. There is also some evidence for the validity of this bound from the numerical conformal bootstrap \cite{QFTinAdS,Paulos:2017fhb} and a proof \cite{Mazac:2018mdx} for the special case $d=1$.

Going to physical kinematics means we take $r < 0$, $z_{\mbox{\tiny eff}} < 0$ and changing the sign of $\eta$. With the appropriate $i \epsilon$ insertions this yields:
\bea
\frac{r^{\Delta_b}}{z_{\mbox{\tiny eff}}^{2\Df}}\to \frac{(-r)^{\Delta_b}}{(-z_{\mbox{\tiny eff}})^{2\Df}}\,e^{-i \pi(\Delta-2\Df)}\,, \qquad \eta \to -\eta\,.
\eea
Note that the change in sign of $\eta$ is immaterial in our case, because all states appearing in the OPE must have even spin and so $\eta$ always appears squared.

In the flat space limit we set $s_b\equiv \Delta_b^2/ \Df^2$ and find
\bea
\mathcal N_{\Delta_b}^{\mbox{\tiny bound}} \left(\frac{a_{\Delta_b,0}}{a_{\Delta_b}^{\mbox{\tiny bound}}}\right) \frac{(-r)^{\Delta_b}}{(-z_{\mbox{\tiny eff}})^{2\Df}}\underset{\Df\to \infty}{\sim} \Df^{2d-\frac 32} g^2_{\Delta_b}\,R(s,s_b)\times E(s,s_b)^{\Df}
\eea
where $R(s,s_b)$ is independent of $\Df$ and
\bea \label{Eblock}
E(s,s_b):=\left(\frac{4-s_b}{s-4}\right)^2\left[\frac{(\sqrt{s}-2)(2+\sqrt{s_b})}{(\sqrt{s}+2)(2-\sqrt{s_b})}\right]^{\sqrt{s_b}}
\eea
The exponential factor implies that the contribution of the bound states is either zero or infinite. In fact, the expression for a single `bound state' conformal block is exactly the same as the contribution of the pole at $c = \Delta_b$ in the exchange diagram discussion of section \ref{subsec:exchange}, as follows from the discussion in appendix \ref{ap:exchange}. (If it diverges then it also agrees with the contribution from the Mellin poles discussed in section \ref{subsec:mellinexchange}.) Therefore, as can also be gleaned from the expression \eqref{Eblock} itself, the region where the contribution from a bound state diverges is the same as already shown in figures \ref{fig_regions3} and \ref{fig_regionsmellin}. The worst case scenario corresponds to $s_b=0$, i.e. massless bound states, for which $|E(s,s_b)|<1$ if $|s-4|>4$, i.e. the same result as for the identity. As we increase $s_b$ divergences are avoided in an increasingly wider region.

\subsection{The phase shift formula}

Now let us look at those states which lie above the twist gap $\tau>2\Df$. We would like to commute the large $\Df$ limit with the sum over blocks. For general kinematics we cannot do this: each term in the sum will diverge. This does not necessarily mean that our formula for $\mathcal S(k_i)$ is wrong, but merely that we cannot commute the OPE and flat space limits. However, let us again restrict to  physical kinematics. As explained above, after continuation we get the sum
\bea
i \left(\frac{\sqrt{-r}}{(1+r)^3(1-r^2)^\nu}\right) \sum_{\substack{\Delta>2\Df+\ell\\\ell=0,2,\ldots}} \mathcal N_{\Delta}^{\mbox{\tiny cont}} \left(\frac{a_{\Delta,\ell}}{a_{\Delta,\ell}^{\mbox{\tiny cont}}}\right) \frac{(-r)^{\Delta}}{(-z_{\mbox{\tiny eff}})^{2\Df}} e^{-i\pi(\Delta-2\Df)}\,  C_{\ell}^{(\nu)}(\eta) \label{eq:sumtemp}
\eea
Notice that here it is important that we perform the continuation, and in particular take $r$ to be arbitrarily close to the negative real axis, {\em before} taking the large radius limit.

The S-matrix is usually expressed in terms of polynomials related to Gegenbauer polynomials in the following way:
\bea
P_\ell^{(d)}(\eta):=p_{d,\ell}\, C_\ell^{(\nu)}(\eta)\,,\qquad p_{d,\ell}=(d-2+2\ell) 2^{2d-3}\pi^{\nu}\,\Gamma(\nu)]\,.
\eea
We now notice that
\bea
\left(\frac{\sqrt{-r}}{(1+r)^3(1-r^2)^\nu}\right)\frac{1}{p_{d,\ell}}\,\mathcal N_{\Delta}\frac{(-r)^{\Delta}}{(-z_{\mbox{\tiny eff}})^{2\Df}}\underset{R\to \infty}{\sim} \frac{4\sqrt{s}}{(s-4)^{\nu}}\, \left(\frac{e^{-\frac{(\Delta-2\Df\sqrt{1-z_{\mbox{\tiny eff}}})^2}{2 \Df(-z_{\mbox{\tiny eff}})}}}{\sqrt{2\pi \Df(-z_{\mbox{\tiny eff}})}}\right)\,, \qquad \Delta>2\Df
\eea
We see that for real and negative $z_{\mbox{\tiny eff}}$ this is bounded and exponentially suppressed except in a region centered at
\bea
\left(\frac{\Delta}{2\Df}\right)^2\sim 1-z_{\mbox{\tiny eff}}=\frac s4
\eea
and with half-width of order $\sqrt{2\Df(-z_{\mbox{\tiny eff}})}$. Hence, the sum over states effectively receives only contributions from that region as long as the OPE ratios which appear in \reef{eq:sumtemp}  are suitably bounded.

Let us now restrict ourselves to a kinematic region where both identity and bound states do not contribute. Putting all the ingredients together we find
\bea
\mathcal S(k_i)&=
\frac{2i \sqrt{s}}{(s-4)^{\nu}}\,\sum_{\ell=0,2,\ldots}\, e^{2i \delta_\ell(s)} P_\ell^{(d)}(\eta)\\
e^{2i \delta_\ell(s)}&:=\lim_{\Df\to \infty} \sum_{\Delta>2\Delta_\cO+\ell} 2  \left(\frac{a_{\Delta,\ell}}{a_{\Delta,\ell}^{\mbox{\tiny cont}}}\right) e^{-i\pi (\Delta-2\Df)} \left(\frac{e^{-\frac{(\Delta-2\Df\sqrt{1-z_{\mbox{\tiny eff}}})^2}{2 \Df(-z_{\mbox{\tiny eff}})}}}{\sqrt{2\pi \Df(-z_{\mbox{\tiny eff}})}}\right)\label{eq:ourphaseshift}
\eea
This is the main result of this section. It tells us that in the physical region the conformal block expansion computes an object which at least kinematically takes the same form as the partial wave decomposition of an S-matrix, with spin-$\ell$ phase shifts computable in terms of the CFT data. Note that the derivation of this formula required restricting ourselves to the region $s>8$. However, this is a shortcoming of our original conjecture relating the S-matrix to the correlator, and not of the formulae above which are expected to be valid for the full range of physical kinematics. For instance, for the GFF correlator our formula gives:
\bea
e^{2i \delta_{\ell}(s)}=\lim_{\Df\to \infty} \sum_{n=0}^{\infty} 2 \left(\frac{e^{-\frac{(\Delta_{n,\ell}-2\Df\sqrt{1-z_{\mbox{\tiny eff}}})^2}{2 \Df(-z_{\mbox{\tiny eff}})}}}{\sqrt{2\pi \Df(-z_{\mbox{\tiny eff}})}}\right)=1\,.
\eea
as it should be, independently of the range of $s$.

Arguably one of the most important properties of the S-matrix is that it should satisfy unitarity, which in the current context states that:
\bea
|e^{2i \delta_l(s)}|\leq 1\,.
\eea
This is not automatic from our formula, and to show it will require us to make one last but crucial assumption:
\bea
\lim_{\Df\to \infty}\sum_{|\Delta-E \Df|<\Df^{\alpha}} \left(\frac{a_{\Delta,\ell}}{a_{\Delta,\ell}^{\mbox{\tiny cont}}}\right)=\Df^{\alpha},\qquad \mbox{with $E>2$ fixed, and for some $\alpha\in[0,\frac 12)$}\,.\label{eq:opeassumption}
\eea
Physically we are demanding that in the flat space limit, the average OPE density per unit size bin in scaling dimension space matches that of a generalized free field. Note that the requirement of $\alpha<1/2$ is such that the averaging must be apparent on a scale which is smaller than the scale of variation of the gaussian, namely $O(\sqrt{\Df})$.
We can think of this assumption as a natural condition for a family of CFTs to have a well-defined flat-space limit. It would be interesting to explore whether this is really an assumption or if it can be proved as a general property of CFT correlators. Using this assumption it is not hard to show that firstly the sum over states indeed localizes in a region around $\Delta\sim \sqrt{s} \Df$ of width $\sim \sqrt{\Df}$ and secondly that the all-important unitarity condition on the S-matrix actually holds. 

The attentive reader may have noticed that in fact, for both these statements to be true it would actually be sufficient that the equal sign in \eqref{eq:opeassumption} was demoted to a less-than sign. She may have also noticed that a similarly looking formula for the phase shift was given already in \cite{QFTinAdS}. These two observations are in fact related: equality is needed in order for our formula to match the one given in \cite{QFTinAdS}, as we show in detail in appendix \ref{app:phaseshift}. A second argument for the equality sign to hold in equation \eqref{eq:opeassumption} is that it leads to a nice relation between the imaginary part of the connected amplitude and the double discontinuity of the CFT correlator as defined in \cite{Caron-Huot:2017vep}, as we now discuss.

Consider the connected part of the S-matrix, which arises by subtracting the full GFF solution from the correlator,
\bea
\mathcal T(s,\eta)=\frac{2\sqrt{s}}{(s-4)^{\nu}}\,\sum_{\ell=0,2,\ldots}\, i(e^{2i \delta_\ell(s)}-1) P_\ell^{(d)}(\eta)
\eea
Thanks to our assumption \reef{eq:opeassumption} we may write
\bea
e^{2i \delta_\ell(s)}-1=\lim_{\Df\to \infty} \sum_{\Delta>2\Delta_\cO+\ell} 2  \left(\frac{a_{\Delta,\ell}}{a_{\Delta,\ell}^{\mbox{\tiny cont}}}\right) (e^{-i\pi (\Delta-2\Df)}-1) \left(\frac{e^{-\frac{(\Delta-2\Df\sqrt{1-z_{\mbox{\tiny eff}}})^2}{2 \Df(-z_{\mbox{\tiny eff}})}}}{\sqrt{2\pi \Df(-z_{\mbox{\tiny eff}})}}\right)
\eea
where the equality sign in that equation was crucial to move the subtracted `1' inside the sum. This now means that the imaginary part of the amplitude will involve
\bea
1-\mbox{Re}\, e^{2i\delta_\ell(s)}=\lim_{\Df\to \infty} \sum_{\Delta>2\Delta_\cO+\ell} 2  \left(\frac{a_{\Delta,\ell}}{a_{\Delta,\ell}^{\mbox{\tiny cont}}}\right) \sin\left[\frac{\pi(\Delta-2\Df)}2\right]^2\,\left(\frac{e^{-\frac{(\Delta-2\Df\sqrt{1-z_{\mbox{\tiny eff}}})^2}{2 \Df(-z_{\mbox{\tiny eff}})}}}{\sqrt{2\pi \Df(-z_{\mbox{\tiny eff}})}}\right).
\eea
and hence
\bea
\mbox{Im}\, \mathcal T(s,\eta)\,=\lim_{\Df\to \infty} \frac{\mbox{dDisc}\, \mathcal G(r,\eta)}{\mathcal D(r,\eta)}\,.
\eea
where we may define here
\bea
\mbox{dDisc}\, \mathcal G(r,\eta)=
\lim_{\epsilon\to 0^+} \frac 14\left[ 2\,\mathcal G(-r,-\eta)-\mathcal G(r+i\epsilon,\eta)-\mathcal G(r-i\epsilon,\eta)\right]\,, \quad r<0\,.
\eea
This relation is the precise sense that the double discontinuity of a correlator captures the imaginary part of a scattering amplitude. Notice that we have derived the above equation for physical kinematics, since there we could use the conformal block decomposition, but for the Lorentzian inversion formula of \cite{Caron-Huot:2017vep} or the dispersion relation of \cite{Carmi:2019cub} the double discontinuity is integrated over a region in cross ratio space that does not reduce to a physical kinematics in the flat-space limit. The precise flat-space limit of these equations will be discussed in future work.


\section{Conclusions}

In this work we presented two related conjectures concerning the flat-space limit of the correlators on the conformal boundary for a gapped QFT in AdS. For a given Witten diagram we have seen that all the propagators and vertices reduce to their flat-space counterparts, and our S-matrix conjecture is the natural one that `erases the circle' and reduces it to a Feynman diagram contributing to an S-matrix element in the flat-space limit. In contrast with many other conjectures, the flat-space limit can be taken directly in position space and no integral transform is necessary. On the other hand, for certain choices of the external momenta the interactions may be spread out over distances comparable to the AdS scale and therefore the conjectures do not always work.

Earlier recipes for extracting amplitudes from correlators include one based on Mellin space and a phase shift formula, both presented in \cite{QFTinAdS}. In regions where our conjectures work we can recover these recipes, as we showed in detail in section \ref{sec:mellin} and \ref{sec:blocks}, but we view our conjectures as more general: they also work for functions that are not representable in Mellin space (with disconnected correlators being the easiest example) and they are not restricted to elastic scattering like the phase shift formula. An important qualification is that, within the domain given in equation \eqref{mellinvaliditymandelstam}, the Mellin-space prescription seems to work for any correlator that can be written in Mellin space and does not suffer from the same divergences as our conjectures.

In section \ref{sec:examples} we discussed how the conjectures work for simple Witten diagrams but fail for some values of the Mandelstam $s$ for the exchange diagram. This led us to introduce the general notion of Landau diagrams in AdS in section \ref{sec:landaudiagrams}. As in flat space, they correspond to classical particles propagating over long distances and interacting in a momentum-conserving fashion; unlike in flat space, the momentum conservation equations can always be solved and an AdS Landau diagram exists for all values of the external momenta. They can however only be important if the real part of the corresponding on-shell `action' is positive, which indicates a failure of the conjectures because the flat-space limit of the correlator diverges.

The most important open question concerns the range of external momenta where the limit used in our conjectures is actually finite, and the assumptions needed to show this. Notice that in all the examples we considered a divergence in the flat-space limit only occurred within a subregion of
\be
|s - 4 m^2| \leq 4m^2\,,
\ee
and similarly for $t$ and $u$. This is in good agreement with the physical picture given in section \ref{subsec:capsandsubleties}: close operator insertions in the Euclidean cap generate less energetic particles, and the Euclidean geodesic connecting them dominates over the Lorentzian one. Furthermore, for functions that can be represented in Mellin space we have also seen that singularities are also unlikely to arise if $s$ is real and
\be
s < 4 m (\mu - m)
\ee
and again similarly for $t$ and $u$, and with $\mu$ the energy of the first state in the corresponding channel. Could these inequalities hold more generally? Are there more refined inequalities to be found? We can try to answer these questions either at a perturbative or at a non-perturbative level.

Perturbatively there are important open questions concerning the structure of the most general AdS Landau diagram and the region  where it diverges. We also need to better understand the AdS version of the well-known intricacies of flat-space Landau diagrams as described for example in \cite{eden2002analytic} and references therein. For example, we can extend the numerical investigation of the triangle diagram in section \ref{subsec:triangle} to include the box and the acnode diagrams which are known to have a richer set of anomalous thresholds. Similarly we should investigate singularities on other Riemann sheets, which generally have complex displacement parameters $\alpha$, as well as understand the AdS analogue of `second type' singularities that correspond to pinch configurations with infinite momenta. Furthermore, recall that flat-space Landau diagrams can only be drawn for specific configurations of the external momenta whereas AdS Landau diagrams can always be drawn, as discussed in section \ref{sec:landaudiagrams}. Correspondingly, the flat-space Landau diagrams capture the threshold where (anomalous) singularities can appear but do not capture the branch cut that is often attached to such thresholds. It would be interesting to see if the AdS Landau diagrams can do better and whether branch cuts in the flat-space amplitude can always be taken to lie in the `blobs' where the AdS Landau diagrams diverge.

In this work we have not yet made enough use of the conformal block decomposition of conformal correlation functions. As is by now well-known, this property implies all sorts of wonderful boundedness and analyticity properties of the correlation functions themselves. Can we use them to infer boundeness\footnote{There are many interesting and well-known consequences of analyticity and unitarity including elastic unitariy as was recently explored in \cite{Correia:2020xtr}. For all of these it would be very interesting to find their AdS ancestors. As an example we can mention the high-energy behavior, which was done in \cite{Dodelson:2019ddi,Haldar:2019prg}.} (unitarity) and analyticity of the flat-space scattering amplitudes? Abstractly, consider a one-parameter family of consistent conformal correlation functions and make a few natural assumptions concerning its spectrum as the parameter $R \to \infty$. When does such a family limit to a consistent S-matrix? And supposing that it does, what are the analyticity properties of the resulting scattering amplitudes? To answer these questions it is essential to obtain a better handle on the possible behavior of correlation functions and their OPE coefficients in the flat-space limit. The analysis in section \ref{sec:blocks} provided a first step in this direction. In particular we showed that, for a four-point function of identical operators the unitarity condition can be rather precisely tied to the OPE coefficient density. On the other hand,  additional tools are necessary to conclude anything about the analyticity of the resulting amplitude. We expect to soon report some further progress in this direction.

We should note that in all of the above the divergences in the flat-space limit were considered a `given' and that we only attempted to avoid them by choosing the external momenta appropriately. However one could also try to \emph{subtract} all the AdS Landau diagrams by hand so the flat-space limit is everywhere finite, thereby improving the region of validity of our original conjectures. A possible way to do this is suggested by our computation in section \ref{sec:blocks}: from the conformal block perspective one could naturally subtract crossing-symmetric sums of AdS exchange diagrams for every block below the two-particle threshold, the effect of which can then be reinstated by adding simple poles to the resulting S-matrix after taking the flat-space limit. It would be interesting to find out whether this procedure can also be done if the number of blocks below threshold becomes unbounded as $R \to \infty$, which should be the case if we do not scatter the lightest particle in the theory.

Finally it would be interesting to generalize our formalism to scattering amplitudes of massless particles, including photons and gravitons. One important point which is not fully understood is how infrared divergences in such theories  arise when taking the flat-space limit. Landau diagrams in AdS introduced in this paper might provide a useful tool for addressing this question. Another related question is to develop a detailed understanding of the soft theorem in flat space \cite{Weinberg:1965nx,He:2014laa,Strominger:2017zoo} (see also recent discussions in \cite{Hijano:2020szl} on the soft therem and the flat-space limit of AdS/CFT). We hope our position-space approach will prove useful for this purpose since it would allow us to directly analyze the flat-space limit of the Ward identity of the boundary correlators. Of course the ultimate goal would be to understand  quantum gravity in flat space by taking the flat-space limit of AdS/CFT, but also on this front we still have a long journey ahead of us.

\section*{Acknowledgements}We thank Martin Kruczenski and Jo\~ao Penedones for useful discussions and suggestions. We are also grateful to the co-organizers and participants of the ``Workshop on S-matrix bootstrap'' at ICTP-SAIFR for providing a stimulating environment where part of this work could be completed. The work of SK is supported by DOE grant number DE-SC0009988. BvR and XZ are supported by the Simons Foundation grant \#488659 (Simons Collaboration on the non-perturbative bootstrap).

\appendix


\section{Analytic continuation in cross ratio space}
\label{Analytic continuation subsec}
It is worthwhile to see what the analytic continuation to the S-matrix configuration becomes in terms of the familiar cross-ratios $(u,v)$ and $(z,\bar{z})$ for four-point conformal correlation functions. In spherical boundary coordinates they are
\be
\begin{aligned}
u &=z\bar{z}= \frac{(1 - \hat n_1 \cdot \hat n_2)(1 - \hat n_3 \cdot \hat n_4)}{(1 - \hat n_1 \cdot \hat n_3)(1 - \hat n_2 \cdot \hat n_4)}, \\
v &=(1-z)(1-\bar{z}) =\frac{(1 - \hat n_1 \cdot \hat n_4)(1 - \hat n_2 \cdot \hat n_3)}{(1 - \hat n_1 \cdot \hat n_3)(1 - \hat n_2 \cdot \hat n_4)} \,.
\end{aligned}
\ee
For the analytic continuation to the S-matrix configuration it will be necessary to view two cross ratios as independent complex variables.

\begin{figure}[t]
\centering
\begin{subfigure}{.5\textwidth}
\centering
\includegraphics[width=8.5cm]{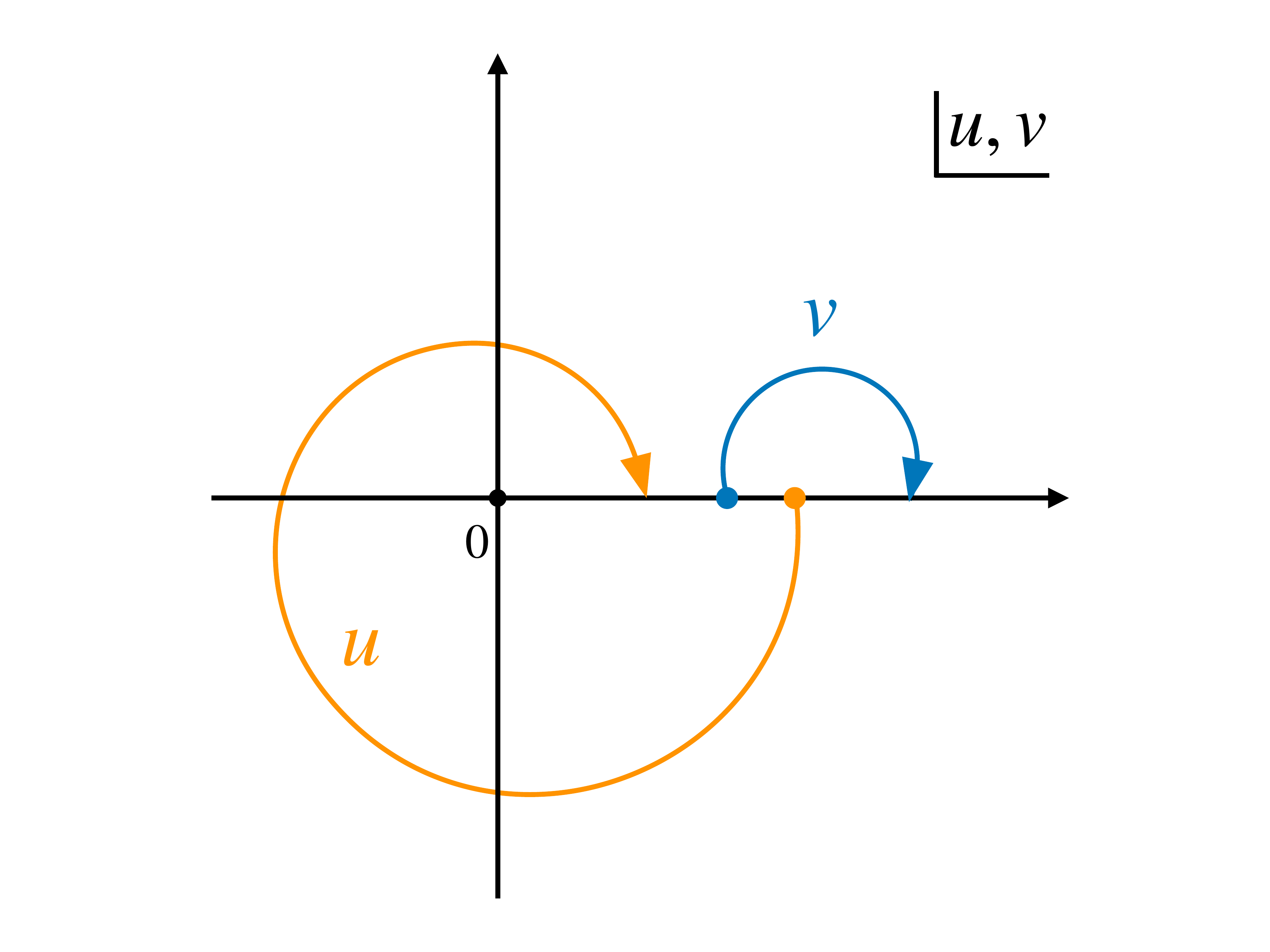}
\caption{Analytic continuation of $u$ and $v$}
\label{Fig_uv_continuation}
\end{subfigure}%
\begin{subfigure}{.5\textwidth}
\centering
\includegraphics[width=8.5cm]{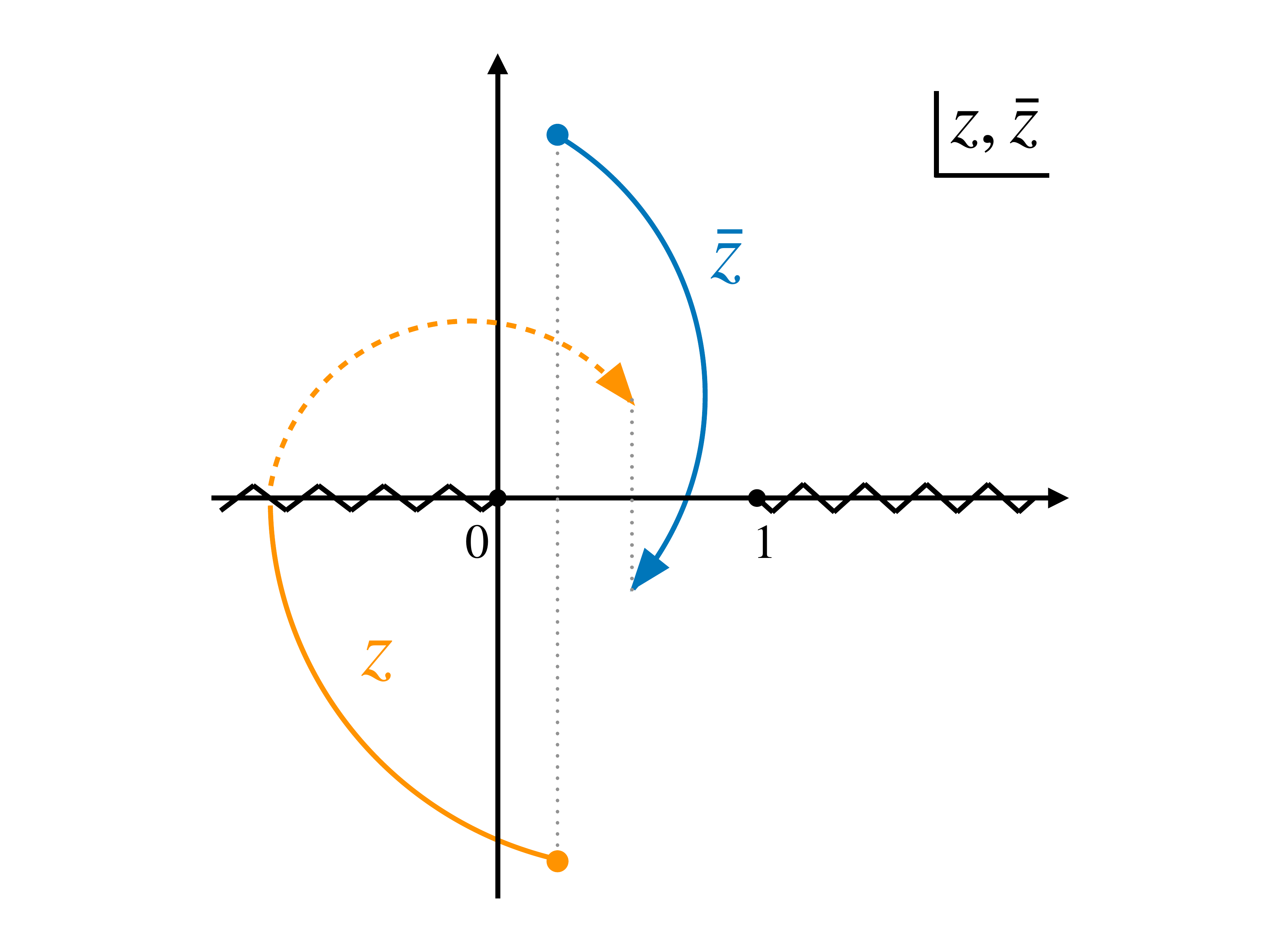}
\caption{Analytic continuation of $z$ and $\bar{z}$}
\label{Fig_z_zbar_continuation}
\end{subfigure}
\caption{Analytic continuation to $s$-physical $S$-matrix configuration. Left: $u$ takes a full clockwise turn around 0 while $v$ does not. Right: the orange dashed curve indicates that $z$ moves on the second Riemann sheet. Black dotted lines indicate that $z$ and $\bar{z}$ are complex conjugate of each other at the starting point (corresponding to a Euclidean configuration) and the end point (corresponding to an S-matrix configuration).}
\end{figure}

We will consider the analytic continuation to a scattering amplitude where particles 1 and 2 are incoming and particles 3 and 4 are outgoing. Therefore, upon substitution of
\be
(n^0, \underline{n}) = - (k^0, - i \underline{k}) / m\,,
\ee
we will take $k^0_1$ and $k^0_2$ positive and $k^0_3$ and $k^0_4$ negative. We will assume that we are on the support of the momentum conserving delta function $\sum_i k_i^\mu = 0$, where the remaining kinematical Lorentz-invariant degrees of freedom are captured in terms of the Mandelstam invariants\footnote{Notice that the Mandelstam $t$ variable is defined following CFT conventions where the `$t$-channel' is traditionally the one where operators 1 and 4 are fused together.}
\be
s = - (k_1 + k_2)^2, \qquad t = - (k_1 + k_4)^2, \qquad \tilde u = - (k_1 + k_3)^2 = \sum_i m_i^2 - s - t
\ee
In terms of which we find that
\be
u = \frac{\left(s-(m_1+m_2)^2\right) \left(s-(m_3+m_4)^2\right)}{\left(\tilde u-(m_1+m_3)^2\right) \left(\tilde u-(m_2+m_4)^2\right)}, \qquad v = \frac{\left(t - (m_2+m_3)^2\right) \left(t - (m_1+m_4)^2\right)}{\left(\tilde u-(m_1+m_3)^2\right) \left(\tilde u-(m_2+m_4)^2\right)}
\ee
We can now analytically continue from the Euclidean region, where the Mandelstam invariants $s$, $t$ and $\tilde u$ are all real and positive, to the physical S-matrix region for the 12 $\to$ 34 process, where $s \geq \text{max}\left((m_1 + m_2)^2, (m_3+m_4)^2\right) + i \epsilon$ and $t$ and $\tilde u$ such that the scattering angle is real. The corresponding continuation in the cross-ratios $u$ and $v$ is shown in figure \ref{Fig_uv_continuation}. We see that $u$ makes a clockwise turn around the origin and $v$ remains real. The corresponding continuation in terms of the $z$ and $\zb$ variables is shown in figure \ref{Fig_z_zbar_continuation}: we start from a Euclidean configuration with $\zb = z^*$, take one variable (which we take to be $z$) in a clockwise fashion through the branch cut on the negative real axis, and end on an `S-matrix' configuration where again $\zb = z^*$ but on another sheet.

In the equal mass case we can write that
\be
\sqrt{u} = \frac{s - 4 m^2}{\tilde u - 4 m^2}, \qquad \qquad \sqrt{v} = \frac{t - 4 m^2}{\tilde u - 4m^2}
\ee
In this formulation the Euclidean configuration corresponds to the principal branch of the square roots. We find that restricting to physical Lorentzian momenta implies that $\sqrt{u} \leq 0$ and $\sqrt{v} \geq 0$. In terms of the scattering angle $\theta$, defined as
\be
t = \frac{1}{2} (4m^2 - s) (1 - \cos(\theta)),\qquad \tilde u = \frac{1}{2} (4m^2 - s)(1 + \cos(\theta))
\ee 
we find a particularly simple form in the $\rho$ variables, namely
\be
\rho = \frac{\sqrt{s}-2 m}{\sqrt{s} + 2 m}e^{i (\theta - 2\pi)},\qquad \rhob = \frac{\sqrt{s}-2 m}{\sqrt{s} + 2m}e^{-i \theta}
\ee
So the physical parameters $s$ and $\theta$ simply correspond to the modulus and argument of the $\rho$ variables. Notice that it is again understood that $\rho$, like $z$ above, is evaluated on the second sheet obtained by circling around zero in a clockwise fashion in order to ensure that $\sqrt{u} \leq 0$.

\begin{figure}[ht!]
\centering
\includegraphics[width=0.9\linewidth]{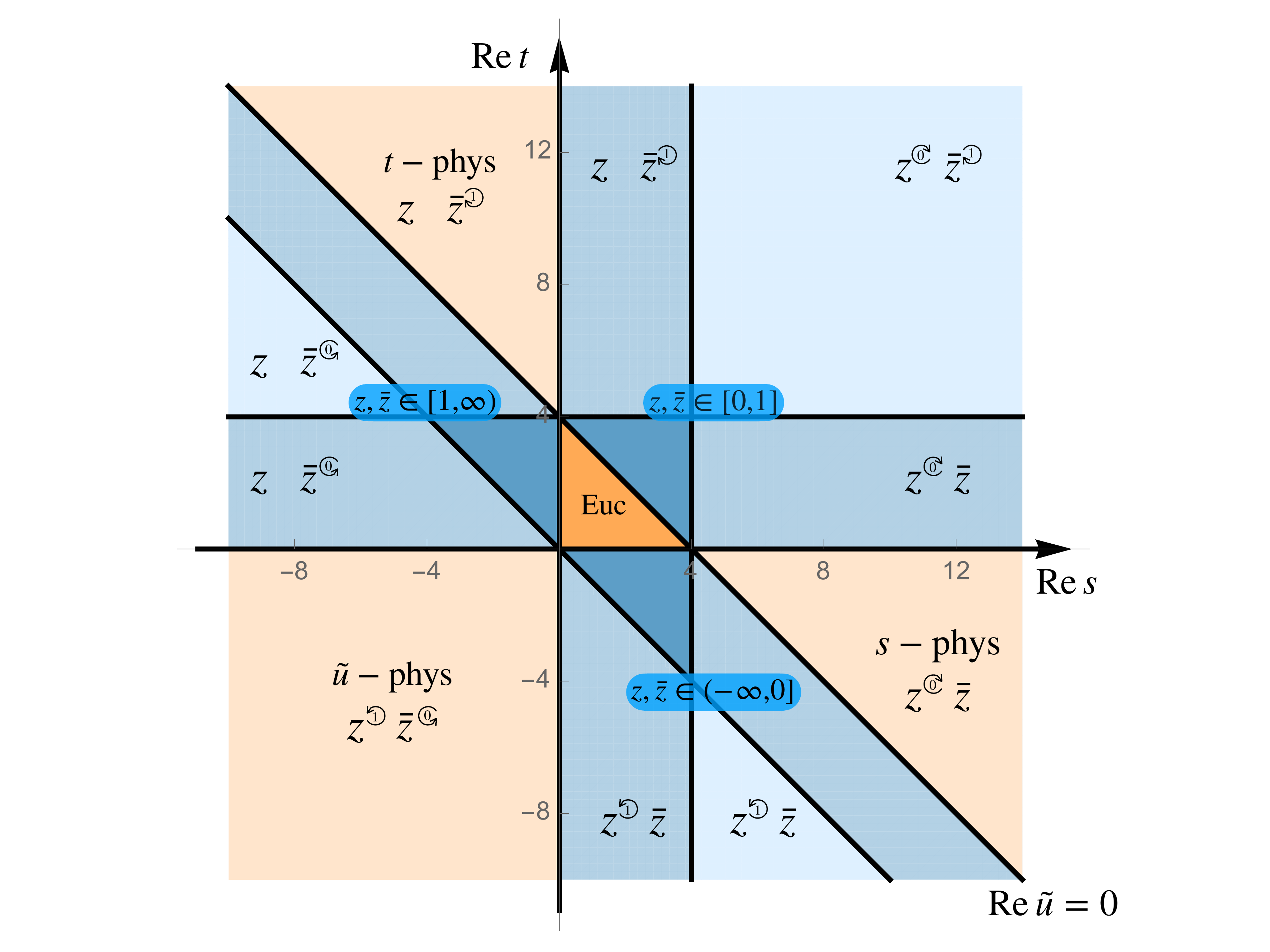}
\caption{\label{Fig_Analytic region for real s and t}Analytic continuation of cross ratios on the real $s-t$ plane. The orange triangle is the Euclidean region and where all analytic continuations start from. The larger triangle (including the Euclidean region) is the region where $u$ and $v$ stay in their principal branch. It is also the so-called Mandelstam triangle where all Mandelstam variables are below their two-particle threshold. The lighter orange regions correspond to $s/t/\tilde{u}$-physical region, respectively. All the blue regions are Lorentzian regions where $z$ and $\bar{z}$ are real and independent, and lie within the indicated intervals. The round arrows (notice the different directions) indicate how the cross ratios $z,\bar{z}$ should be analytically continued through the $(-\infty,0]$ or $[1,\infty)$ branch cut in the complex plane. In unlabelled regions $z,\bar{z}$ stay in the principal branch.}
\end{figure}

So far we have focused on $s$-channel physics, with operators 1 and 2 moved to an `in' configuration and particles 3 and 4 to an `out' configuration. Of course we could have continued the positions differently in order to reach the physical regions of the $t$ and the $u$ channel which would involve different analytic continuations of the cross ratios. We can also consider the `Euclidean' configuration where all Mandelstam invariants lie between $0$ and $4m^2$ -- in this configuration $\sqrt{u}$ and $\sqrt{v}$ are both real and positive, and the $\rho$ variable does not need to be analytically continued around zero. Figure \ref{Fig_Analytic region for real s and t} shows the various continuations that are necessary to reach these regions.\footnote{Note that the continuations of $z,\bar{z}$ are somewhat arbitrary because those of Mandelstam variables only depend on products of $z,\bar{z}$ (for example see \eqref{eq:conformalMandelstam}). In addition, the symmetry $z\leftrightarrow\bar{z}$ adds even more arbitrariness. In particular, to reach a given point on $s-t$ plane, a phase rotation in $z(\bar{z})$ can be traded with one in $\bar{z}(z)$ provided the phases are the same. Therefore, depending on different conventions the reader may find different continuation prescriptions for $z,\bar{z}$, but the one for $u,v$ should have no ambiguity.} For each of the physical regions the indicated continuation lands us above the cut in the corresponding Mandelstam variable.

\subsection{Different kinematic limits}

\begin{figure}[!ht]
\centering
\includegraphics[width=0.9\linewidth]{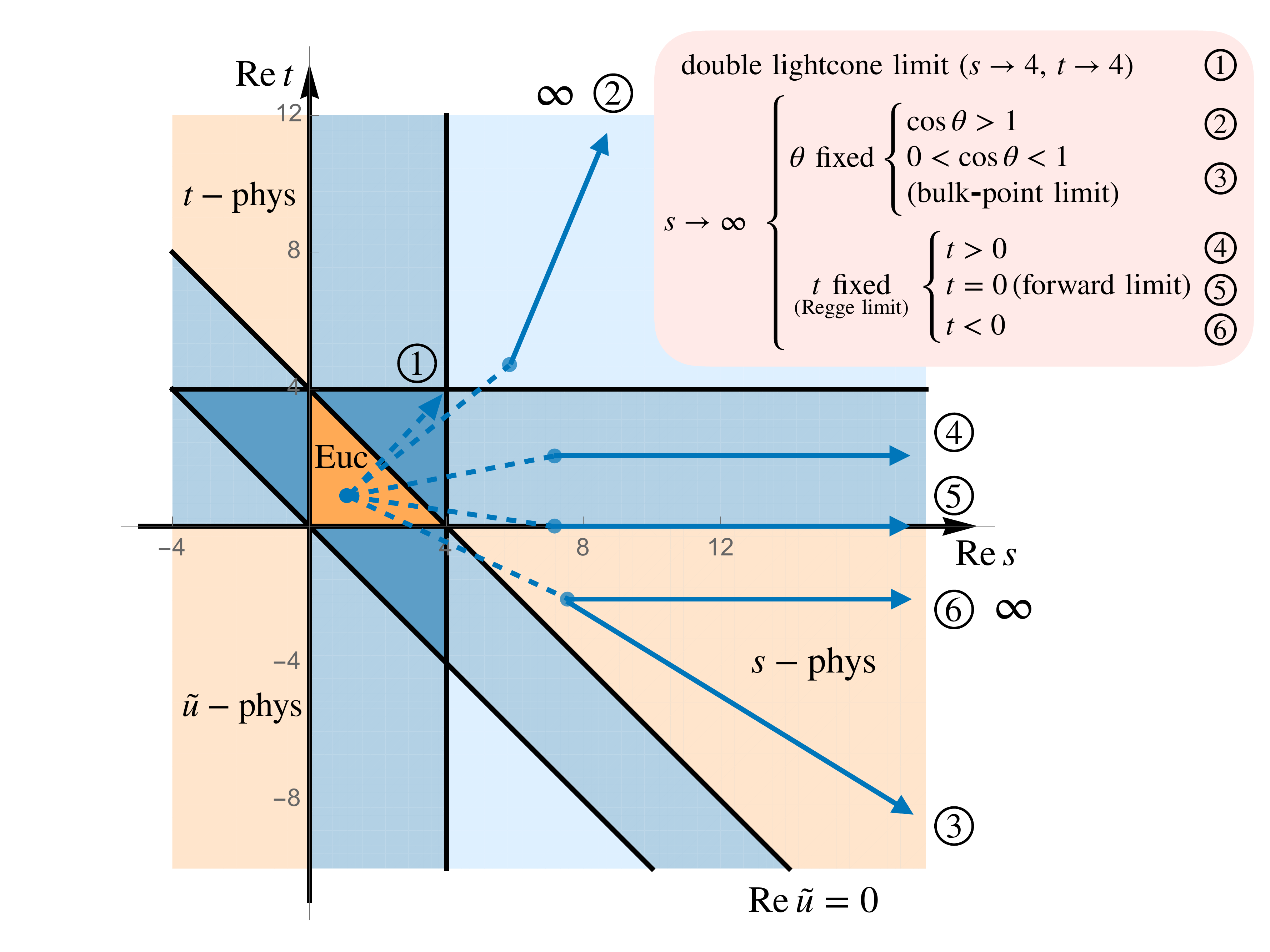}
\caption{Different kinematic limits in real $s-t$ plane. The dark blue dashed lines indicate analytic continuation of $s,\ t$ into their complex planes, respectively. Details of analytic continuation are explained in the text. The dark blue arrows lie on the real $s-t$ plane. The orange triangle is the Euclidean region, and within it lies the starting point of all limits and their corresponding analytic continuation. }
\label{Fig_different_kinematic_lim_s_t_plane}
\end{figure}

\begin{figure}[!ht]
\renewcommand\thesubfigure{\protect\circled{\arabic{subfigure}}}
\centering
\begin{subfigure}[b]{.45\linewidth}
\centering
\includegraphics[width=\linewidth]{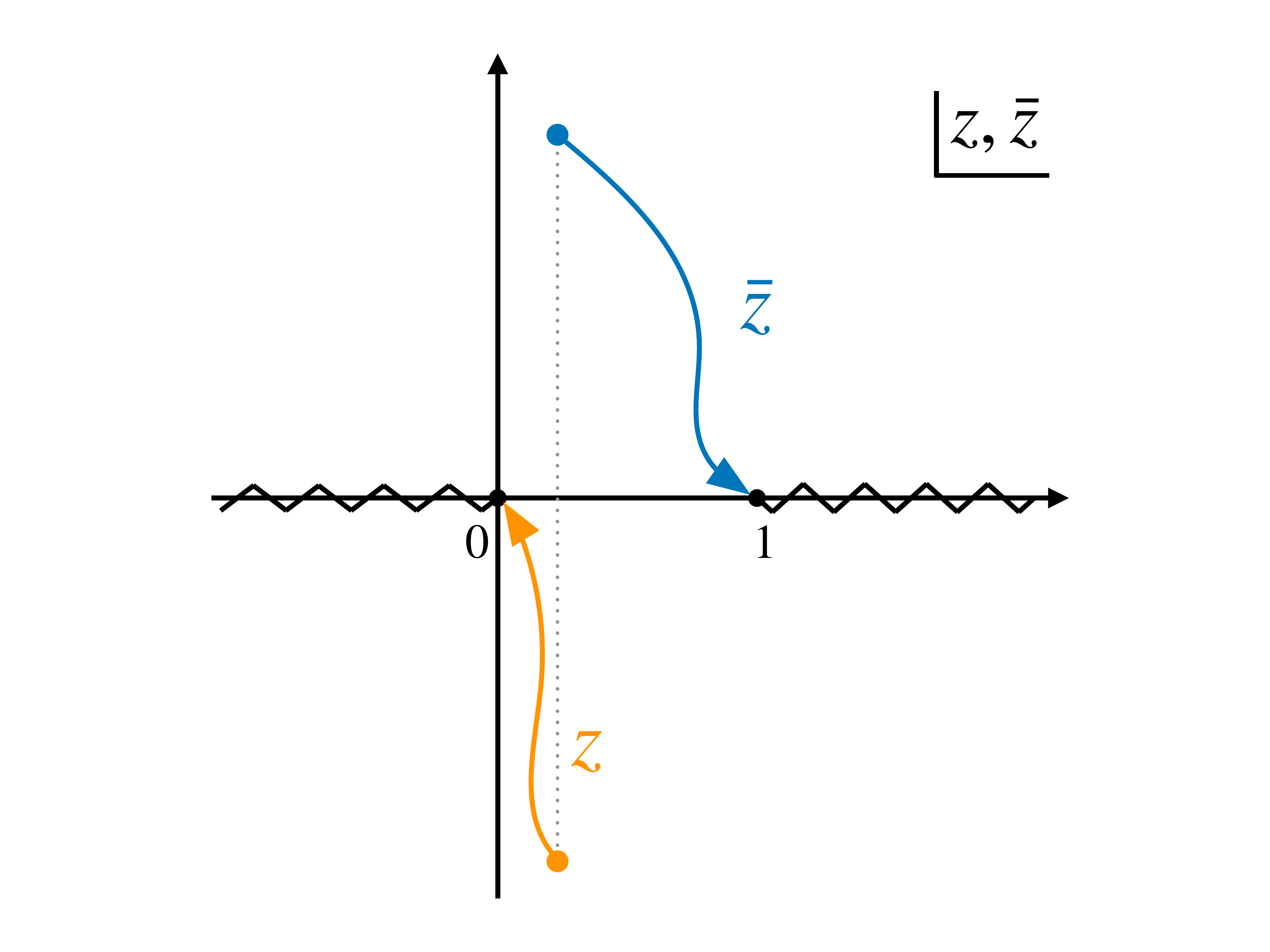}
\caption{$s\to4, t\to4$}
\end{subfigure}%
\begin{subfigure}[b]{.45\linewidth}
\centering
\includegraphics[width=\linewidth]{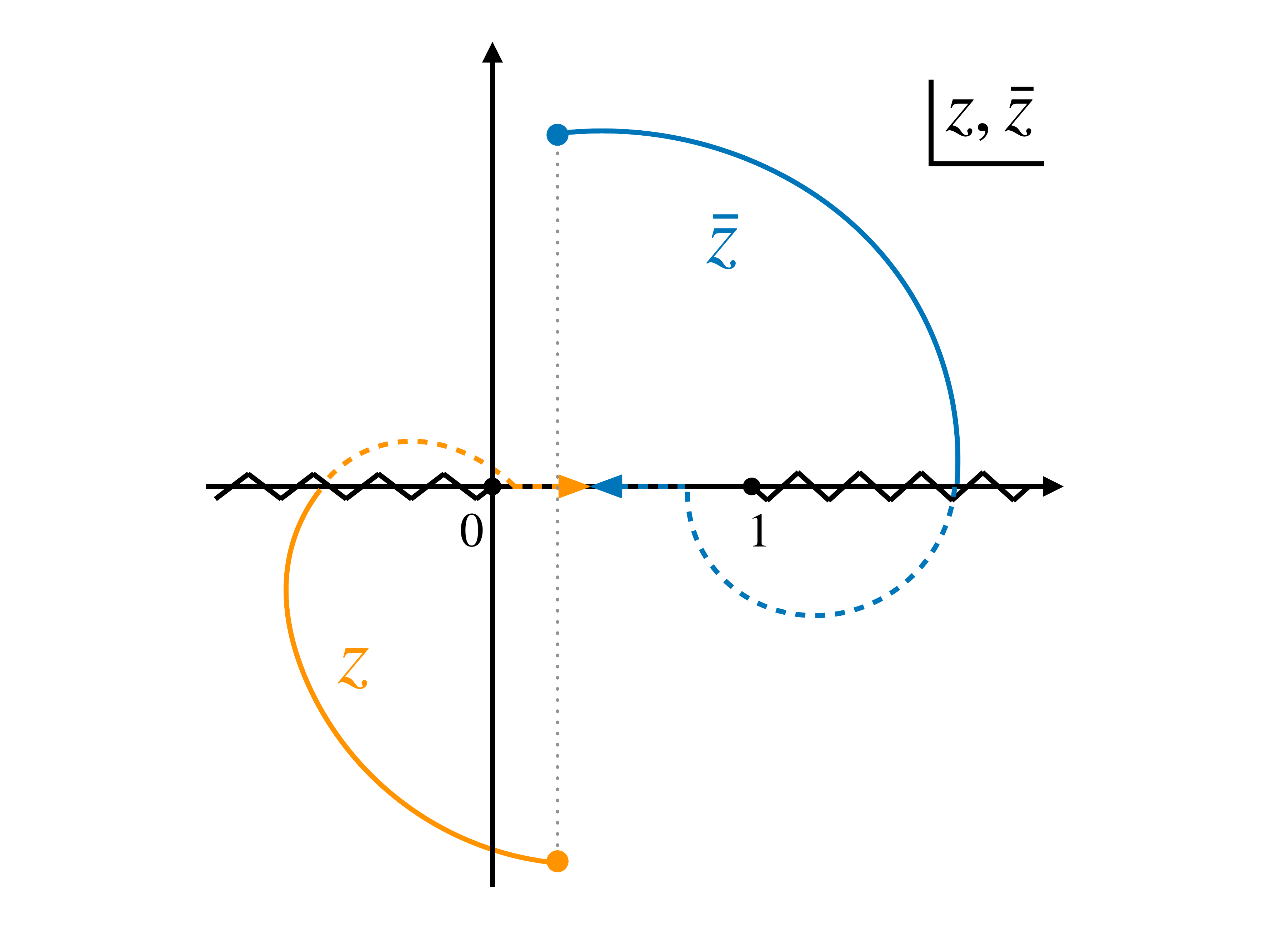}
\caption{$s\to\infty,\ \cos\theta>1$}
\end{subfigure}%

\begin{subfigure}[b]{.45\linewidth}
\centering
\includegraphics[width=\linewidth]{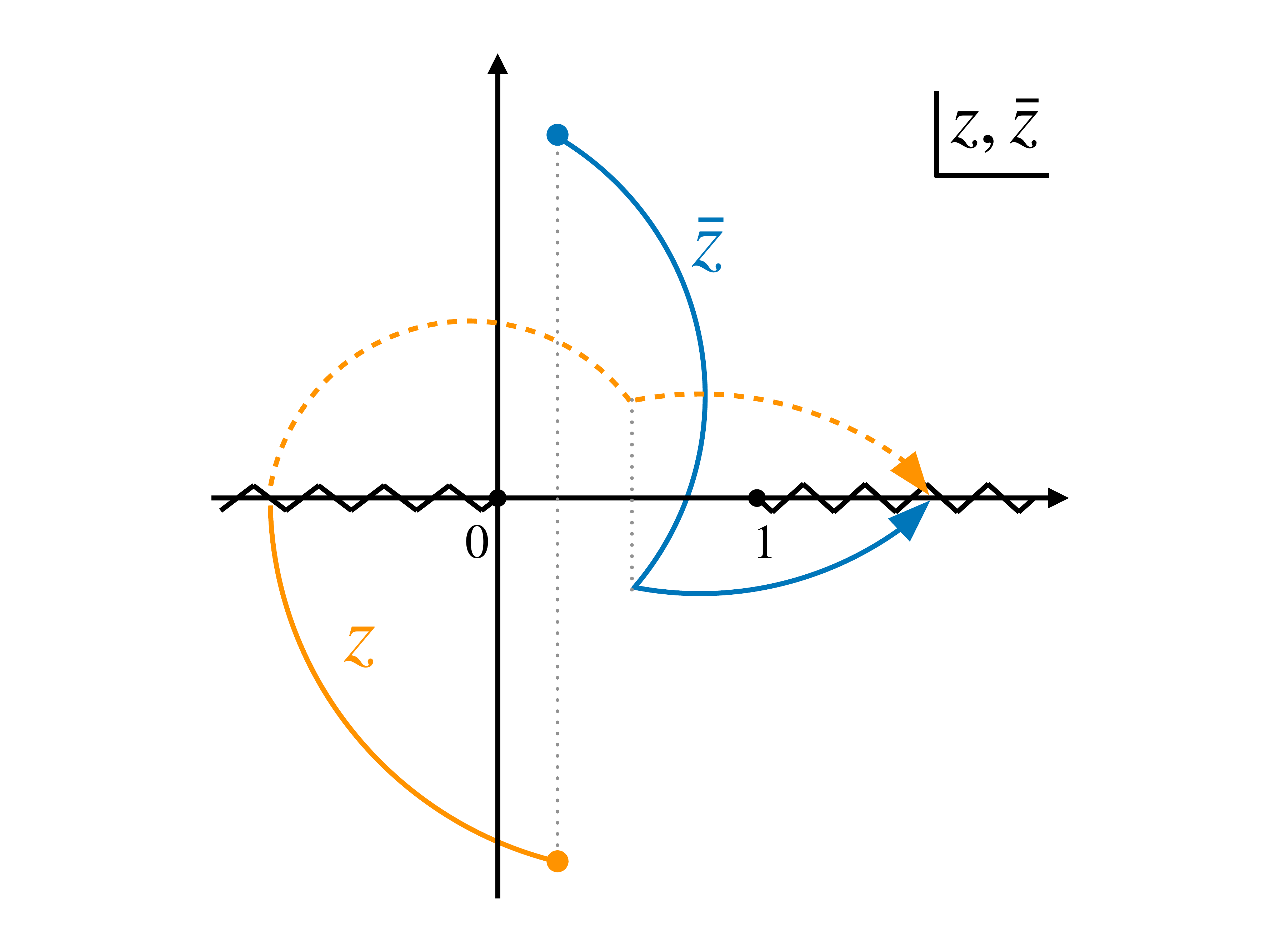}
\caption{$s\to\infty,\ 0<\cos\theta<1$}
\end{subfigure}
\begin{subfigure}[b]{.45\linewidth}
\centering
\includegraphics[width=\linewidth]{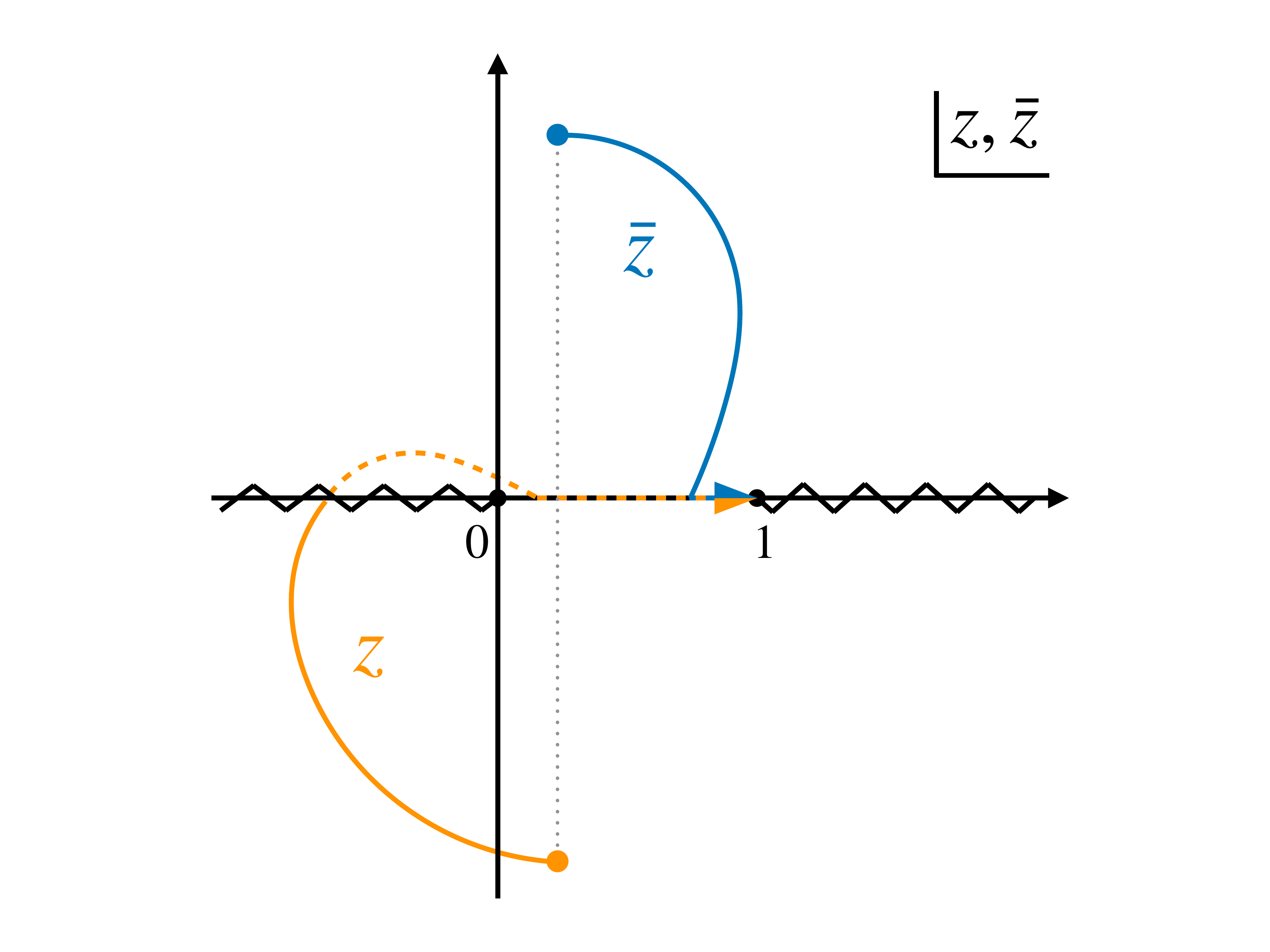}
\caption{$s\to\infty,\ t>0$}
\end{subfigure}

\begin{subfigure}[b]{.45\linewidth}
\centering
\includegraphics[width=\linewidth]{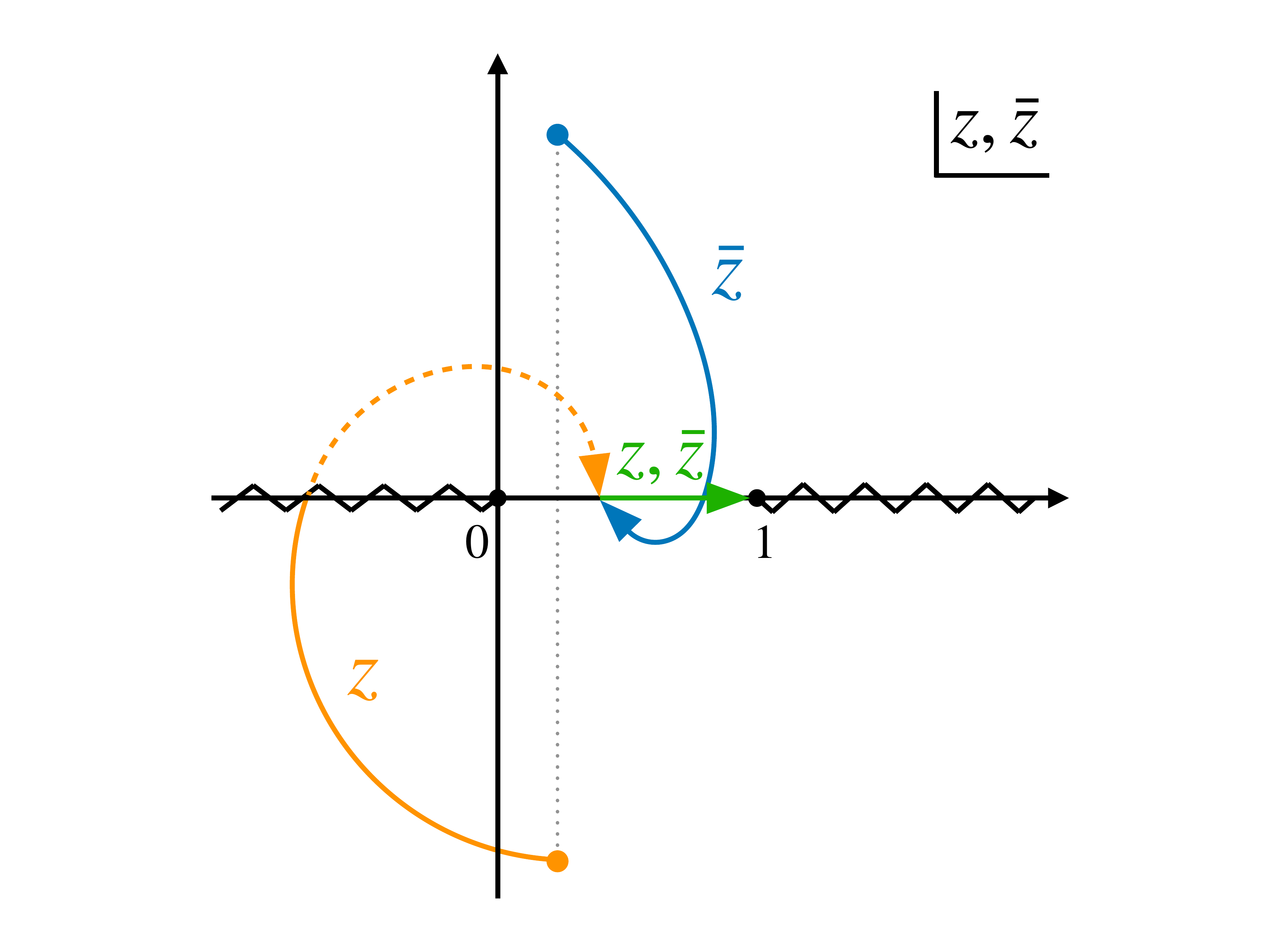}
\caption{$s\to\infty,\ t=0$}
\end{subfigure}%
\begin{subfigure}[b]{.45\linewidth}
\centering
\includegraphics[width=\linewidth]{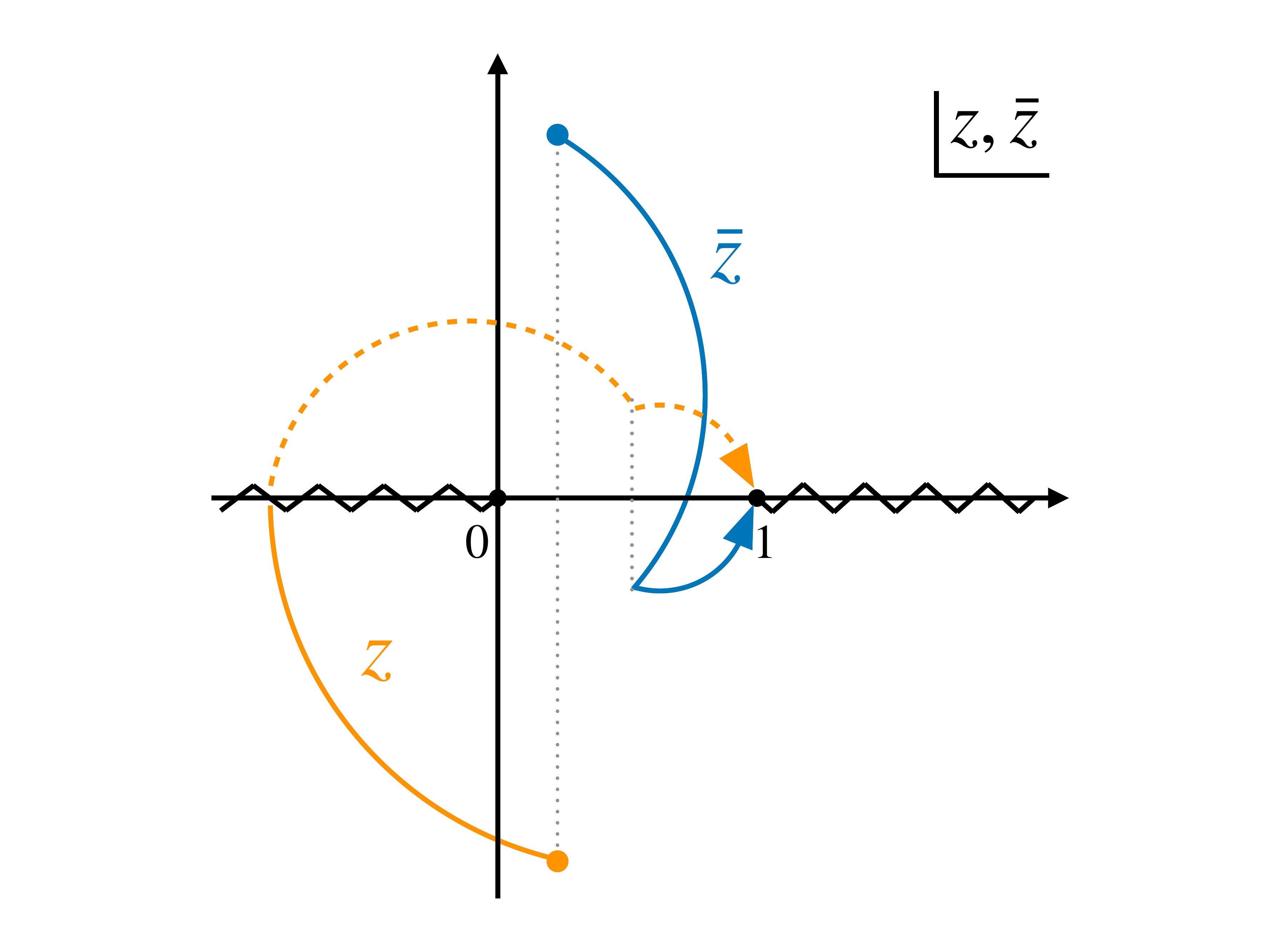}
\caption{$s\to\infty,\ t<0$}
\end{subfigure}%
\caption{Different kinematic limits in $z,\ \bar{z}$. The dashed curves indicate that the cross ratios are in the second sheet. The grey dotted lines indicate that $z$ and $\bar{z}$ are complex conjugate of each other.}
\label{Fig_different_kinematic_lim_z_zb}
\end{figure}

To gain a bit more insight we will now explore various kinematic limits and relate each of them between Mandelstam invariants $s,\ t$ and cross ratios $z,\ \bar{z}$. The limits under consideration are a double lightcone limit and five $s\to\infty$ limits:
\begin{align}
\begin{split}
\begin{cases}
&\text{double lightcone limit:}\  
z\to0,\ \bar{z}\to1
\\
&s\to\infty:\
\begin{cases}
\theta\ \mathrm{fixed}
&\begin{cases}
\cos(\theta)>1
\\
0<\cos(\theta)<1\,\text{(bulk-point\ limit)}
\end{cases}
\\[14pt]
\underset{\text{(Regge limit)}}{t\ \text{fixed}}
&\begin{cases}
t>0
\\
t=0\,\text{(forward limit)}
\\
t<0
\end{cases}
\end{cases}
\end{cases}
\end{split}
\end{align}
where we recall that
\begin{align}
t= \frac{1}{2} (4m^2-s)(1-\cos(\theta)) .
\end{align}
In the Mandelstam $s$ and $t$ plane these different limits are shown in figure \ref{Fig_different_kinematic_lim_s_t_plane} and in the complex $z$ and $\bar z$ plane they are shown in figure \ref{Fig_different_kinematic_lim_z_zb}. Note that the starting point for each limit always lies inside the Euclidean region and the endpoints of the limits are always outside. The $i \epsilon$ prescription we adopted for these continuations is as follows: if the endpoint of $s/t$ is greater than 4, then keep $\im(s/t)$ \emph{non-negative}; if the endpoint of $s/t$ is smaller than 0, then keep $\im(s/t)$ \emph{non-positive}; otherwise the continuation can be arbitrary. 

Let us offer a few comments on these limits. First of all, we find that the double lightcone limit where $z \to 0$ and $\bar z \to 1$ corresponds to $s \to 4m^2$ and $t \to 4m^2$. (We are not too concerned about the precise order in which these limits are taken here.) At this point the two lightcones in the position space correlation function intersect, and similarly the $s$-channel and $t$-channel cuts intersect in the amplitude. In both cases the region \emph{beyond} this point is a bit mysterious. For example, as shown in figure \ref{Fig_different_kinematic_lim_z_zb}, limit \circled{2} requires an analytic continuation of $z$ around 0 and $\bar z$ around 1, and so we are beyond the radius of convergence of any conformal block decomposition. It would be interesting to see if some theory can be developed in order to better understand this region.

Secondly let us consider the fixed angle high-energy limit \circled{3}. In cross ratio space this corresponds to the familiar bulk point limit. This is known to be the region in cross ratio space where the flat-space scattering amplitude can be obtained from the conformal correlation function in the case of \emph{massless} external particles, which is of course important in the context of AdS/CFT. In our case we recover this limit at high energies where in some approximate sense the masses of the external particles no longer matter. We have however not checked in detail that the bulk point limit prescription is exactly reproduced here.

Lastly there are the various Regge limits\footnote{Note that here we discuss the Regge limits in the $s$-channel, as opposed to $t$-channel one in section \ref{sec:kinematic limits}. The two limits are simply related by $s\leftrightarrow t, u\leftrightarrow v$.} \circled{4}, \circled{5} and \circled{6} which correspond to large $s$ and fixed $t$. In plot \circled{4}, $z$ and $\bar{z}$ approach 1 in a way such that
\begin{align}
\frac{1-z}{1-\bar{z}}\overset{s\to\infty}{=}\left(\frac{\sqrt{t}+2}{\sqrt{t}-2}\right)^2.
\end{align}
In plot \circled{5}, $z=\bar{z}$ when approaching 1, whereas in plot \circled{6}, $z$ and $\bar{z}$ are complex conjugate of each other up to a phase $e^{2\pi i}$.

As discussed in the main text, these limits exactly correspond to the Regge limit of conformal correlation functions. In this case only the $z$ variable is continued around $0$ and so we are at the boundary of the radius of convergence of the $s$ channel conformal block decomposition.


\section{Steepest descent contours for the exchange diagram in Mellin space}
\label{app:exchangemellin}

In this appendix we analyze the location of the steepest descent contour for the exchange diagram in Mellin space and determine the contribution of any Mellin poles that are picked up in deforming the original integration contour to the steepest descent contour.

First we have to decide which poles are picked up. This is determined by the crossing point of the steepest descent contour with the real axis. With the integrand as in \eqref{eq:Wsigma12only2} the saddle point is located at
\be
\sigma_{12}^* = \frac{4m^2-s}{8 m}
\ee
and at this point
\be
\phi\left(m,\frac{4m^2 + s}{4m^2 -s},\sigma_{12}^*\right) = - 2m \log\left(\frac{8m}{4m^2 + s}\right)
\ee
so we see that 
\be
\text{Im}\left[\phi\left(m,\frac{4m^2 + s}{4m^2 -s},\sigma_{12}^*\right)\right] = 2m\, \text{arg}\left(4m^2 + s\right)
\ee
so the real $\sigma_{12}$ axis is crossed at
\be
\bar \sigma_{12}(s) \colonequals \frac{m\, \text{arg}[4m^2 + s]}{\text{arg}[4 m^2 + s] - \text{arg}[4 m^2 - s]}
\ee
where we used that the steepest descent contour crossed the real axis for $0 < \sigma_{12} < m$ as explained above. (As a consistency check we recover $\sigma_{12}^*$ from $\bar \sigma_{12}(s)$ for $s$ real and Euclidean via a limiting procedure.) We pick up Mellin poles if:
\be
\bar \sigma_{12}(s) < m - m_b/2\,.
\ee
This gives rise to a region in the complex $s$ plane that is shown as the lightly shaded region in figure \ref{fig_regionsmellin}. Notice that it in particular includes the entire physical line $s > 4 m^2$ and the pole at $s = m_b^2$. A little further experimentation shows that we do not pick up any Mellin poles for $\text{Re}(s) < 0$ as long as $m_b > m$, and also that for fixed $\text{Re}(s) > 0$ the function $\bar \sigma_{12}(s)$ is increasing with $|\text{Im}(s)|$, so if we do not pick up poles for a given $s$ with $\text{Re}(s) > 0$ then we will also not pick up poles for any $s$ with greater imaginary part. Altogether this means that the region where we pick up poles is within the region defined by $\text{Re}(s) > 4m (m_b - m)$ for $m_b > m$.

The picked up Mellin poles give rise to a sum which we should compare to the order of magnitude of the Mellin saddle point at $\sigma_{12}^*$. In equations, we can say that we are safe if
\be
\sum_{k = 0}^{\bar k(s)} R_k \exp\left( R \,\phi\left(m,\frac{4m^2 + s}{4m^2 -s},m - m_b/2 - k/R\right) - R \, \phi\left(m,\frac{4m^2 + s}{4m^2 -s},\s_{12}^*\right)\right) \overset{\mathrm{R \to \infty}}{\longrightarrow} 0
\ee
with $\bar k(s) = R (m - m_b/2 - \bar \sigma_{12}(s))$. We will henceforth assume $\bar k(s) > 0$ otherwise there is nothing to estimate. To analyze the $k$ sum we will ignore the phase factor and instead investigate the slightly larger expression
\be \label{mellinabstes}
\sum_{k = 0}^{\bar k(s)} R_k \exp\left( R \,\phi\left(m,\frac{|4m^2 + s|}{|4m^2 -s|},m - m_b/2 - k/R\right) - R \, \phi\left(m,\frac{4m^2 + s}{4m^2 -s},\s_{12}^*\right)\right) \overset{\mathrm{R \to \infty}}{\longrightarrow} 0\,.
\ee
We now substitute the large $R$ expression for the $R_k$ and can attempt to find a saddle point approximation for the sum. Notice that the location of the saddle point has changed compared to the discussion around equation \eqref{eq:Rkasympt} because of the extra $k$-dependent factor. We find two stationary points at
\be
k_\pm^*(s)/R =  \frac{m_b}{8} \left( -4 \pm \frac{|4m^2 + s|}{m \sqrt{\text{Re}(s)}} \right)
\ee
If $\text{Re}(s) > 0$ then exactly one of the saddle points is real and positive. The value of the summand at the saddle point is
\be \label{mellinabstestsaddle}
\begin{split}
\exp\Bigg(R\, &(2m + m_b) \left( \log(2m + m_b) - \frac{1}{2}\log\left(|4m^2 + s| + 4m\sqrt{\text{Re(s)}}\right)\right) \\
+ R \, & (2m - m_b) \left( \log(2m - m_b) - \frac{1}{2}\log\left(|4m^2 + s| - 4m\sqrt{\text{Re(s)}}\right)\right)\Bigg)
\end{split}
\ee
up to an unimportant prefactor.

We can now distinguish several possibilities:
\begin{itemize}
\item If $0 < \bar k(s)$ but $\text{Re}(s) < 0$ then the stationary points in the sum are complex. We checked numerically that the summand is monotonically increasing and so it is bounded from above by its value at $\bar k(s)$. This can only happen when $m_b < m$.
\item If $\text{Re(s)} > 0$ and $0 < k^*_+(s) < \bar k(s)$ then the saddle point approximation should work well for the sum.
\item If $\text{Re}(s) > 0$ and $0 < \bar k(s) < k_+^*(s)$ then we can again take the value at $\bar k(s)$ as an upper bound for the sum.
\end{itemize}
Each of these cases corresponds to a domain in the complex $s$ plane (which all lie within region I as defined around figure \ref{fig_regionsmellin} because we assume $\bar k(s) > 0$).  We have numerically checked that the summand in \eqref{mellinabstes} in the first and third possibility always goes to zero in the large $R$ limit, so the problematic region is the subregion of the second possibility where \eqref{mellinabstestsaddle} does not vanish at large $R$. This is now easily plotted numerically, leading to the domains shaded in figure \ref{fig_regionsmellin}.


\section{Verification of momentum conserving delta function}
\label{app:delta function}

In this appendix we show that in the flat-space limit, the normalized, analytically continued four-point contact diagram $G_c(P_i)$ equals the momentum conserving delta function,
\begin{align}
Z^2 G_c(P_i)|_{\text{S-matrix}}
\overset{R\to\infty}{\xrightarrow{\hspace*{0.6cm}}}
i(2\pi)^{d+1} \delta^{(d+1)}( k_1 + k_2 + k_3 + k_4 ),
\end{align}
where the normalization factor is defined in \eqref{Eqn_normalization factor Z}:
\begin{align}
Z= \frac{2^{2\D}R^{d-1}}{\CC_\D^2}.
\end{align}
Our starting point is \eqref{Eqn_flat space limit G_c}, which is reproduced here:
\be
G_c(P_i) \overset{R\to\infty}{\xrightarrow{\hspace*{0.6cm}}} 
R^{-d + 3} 2^{2 \D - d/2 - 6} \pi^{-3d/2 + 1/2} \Delta^{3d/2 - 9/2}
\frac{\left( \sqrt{P_{12} P_{34}} + \sqrt{P_{13}P_{24}} + \sqrt{P_{14}P_{23}} \right)^{-2\D + 3/2}}{\left(P_{12}P_{13}P_{14}P_{23}P_{24}P_{34}\right)^{1/4}}.
\ee
In Euclidean signature the $P_{ij}$ are all real and positive and so the amplitude in the large $\Delta$ limit gets support only around $P_{ij} = 0$. This changes if we analytically continue to the S-matrix configuration \eqref{momentaSmatrix} as our conjecture dictates. Consider again the case where particles 1 and 2 are the `in' particles and 3 and 4 the `out' particles. We are then instructed to take $P_{12}$ and $P_{34}$ negative with a small negative imaginary part. We may write:
\begin{multline}
\label{wittenlimit}
G_c(P_{ij})|_{\text{S-matrix}}  \overset{\mathrm{R \to \infty}}{\longrightarrow} 
i   R^{-d + 3} 2^{2 \D - d/2 - 6} \pi^{-3d/2 + 1/2} \Delta^{3d/2 - 9/2} 
\left.\frac{ \left(f(P_{ij})\right)^{-2\D + 3/2} }{(P_{12}P_{13} P_{14} P_{23} P_{24} P_{34})^{1/4}} \right|_{\text{S-matrix}}
\end{multline}
where we have introduced
\begin{align}
f(P_{ij})\coloneqq - \sqrt{P_{12} P_{34}} + \sqrt{P_{13}P_{24}} + \sqrt{P_{14} P_{23}}.
\end{align}

Our first benefit is the factor of $i$, which matches \eqref{MomentumConservation}. Taking into account the wave function factors $\sqrt{Z}$ we realize that the large $\D$ limit of the normalized contact diagram should, after substitution of $P_{ij} = 2( 1 + m^{-2} k_i \cdot \eta \cdot k_j )$, become equal to the momentum conserving delta function:
\begin{align}
\lim_{R \to \infty}  i Z^2
R^{-d+3} 2^{2 \D - d/2 - 6} \pi^{-3d/2 + 1/2} \Delta^{3d/2 - 9/2}
&\frac{ \left(f(P_{ij})\right)^{-2\D + 3/2} }{(P_{12}P_{13} P_{14} P_{23} P_{24} P_{34})^{1/4}}
\nonumber \\
&\qquad
\overset{?}{=}
i(2\pi)^{d+1} \delta^{(d+1)}( k_1 + k_2 + k_3 + k_4 )
\end{align}

A complication in the verification of this claim is that both sides are meant to be understood as functions of \emph{on-shell} momenta, so the independent variables on either side are really the spatial components $\underline k_i$ of the four momenta.

Our first claim is that the contact diagram is maximized on the support of the delta function. At large $\Delta$ it is the function $f(P_{ij})$ that provides exponential damping and it is maximized whenever
\begin{align}
\frac{\partial{\log(f(P_{ij}))}}{\partial{\underline k_3^i}}= \frac{\partial{\log(f(P_{ij}))}}{\partial{\underline k_4^i}}=0
\end{align}
which is solved on the momentum conserving configuration. More generally the form of $f(P_{ij})$ leads us to believe that there should be a Minkowskian version of Ptolemy's inequality that rigorously proves the maximization of $f(P_{ij})$ uniquely and exactly when momentum conservation is obeyed, but we have not tried to obtain it. At the saddle point we easily find that
\begin{align}
f(P_{ij})|_{\text{saddle point}} = 8,
\end{align} 
independently of the values of the momenta themselves.

Next we need to check the volume of the bump centered at the momentum conserving configuration, so whether the integrations over both sides of \eqref{Eqn_delta function claim} agree. To integrate, we parametrize $P_i$, or equivalently $k_i^\mu$, using rapidity and spherical coordinates:
\begin{align}
\begin{gathered}
P_i = 
(1, \mp\cosh(\theta_i), i \sinh(\theta_i) \hat{n}_i^{(d)}) 
\qquad\ \,
-/+:\text{in/out-going}
\\
k_i^\mu = (\pm m\cosh(\theta_i), m\sinh(\theta_i) \hat{n}_i^{(d)})
\qquad
+/-:\text{in/out-going}
\\
\hat{n}_i^{(d)} = (\cos(\phi_{i,1}), \sin(\phi_{i,1})\, \hat{n}_i^{(d-1)})
\end{gathered}
\label{Eqn_integral coordinates}
\end{align}
Integrating over the right-hand side of \eqref{Eqn_delta function claim} gives
\begin{align}
\begin{split}
&\int d|\underline k_3| |\underline{k}_3|^{d-1} d^d\underline{k}_4\ 
\left[ i(2\pi)^{d+1}\delta^{(d+1)}(k_1+k_2+k_3+k_4) \right]\\
&= i (2\pi)^{d+1}\left(\frac{E_{CM}^2}{4}-m^2\right)^{d/2-1}\frac{E_{CM}}{4}\\
&= i (2\pi)^{d+1}\frac{m^{d-1}}{2}\cosh (\theta_1) \sinh^{d-2}(\theta_1).
\label{Eqn_4pt delta function integral}
\end{split}
\end{align}
This function of $\theta_1$ needs to be matched on the left-hand side. To perform the integrals we again resort to the saddle point approximation. Going to the center of mass frame of ingoing particles 1 and 2, the momentum-conserving saddle point now reads
\begin{align}
\begin{split}
\theta_3^*= \theta_4^* = \theta_1, \qquad
\hat{n}_4^{(d)*} = -\hat{n}_3^{(d)}.
\end{split}
\end{align}
We also need to calculate the determinant of the matrix of second-order derivatives of $\log(f(P_{ij}))$. Let us denote this as Det$_d$ for the $d$-dimensional case. It turns out that using our coordinates the determinant at the saddle point can be easily calculated for any dimension. To be precise, we set the ordering of the coordinates as
\begin{align}
\{\theta_3, \theta_4, \phi_{4,1}, \ldots, \phi_{4,d-1}\}
\end{align}
and then the result reads
\begin{align}
\left.\frac{\mathrm{Det}_d}{\mathrm{Det}_2}\right|_{\text{saddle point}} =  \left(\frac{\sinh^2(\theta_1)}{8}\right)^{d-2}
\sin^{2d-4}(\phi_{3,1}) \sin^{2d-6}(\phi_{3,2}) \ldots \sin^{2}(\phi_{3,d-2})
\label{Eqn_d dimensional 2nd order determinant}
\end{align}
where Det$_2$ satisfies
\begin{align}
\left.\sqrt{\frac{1}{\text{Det}_2}} \frac{ \left(f(P_{ij})\right)^{3/2} }{(P_{12}P_{13} P_{14} P_{23} P_{24} P_{34})^{1/4}} \right|_{\text{saddle point}}
=
\frac{32}{\cosh(\theta_1)\sinh^2(\theta_1)}.
\end{align}

The simplicity of this calculation follows from the fact that additional matrix elements in higher dimensional cases compared to the two-dimensional case are all zeros except for the diagonal terms. To check this, we need to calculate
\begin{align}
\begin{split}
&\frac{\partial^2\log(f(P_{ij}))}{\partial\theta_k \partial \phi_{4,l}},
\qquad\qquad
k=3,4,\quad 2\leq l \leq d-1\\
&\frac{\partial^2\log(f(P_{ij})}{\partial\phi_{4,l} \partial \phi_{4,m}},
\qquad\qquad\ \,
1\leq l \leq m \leq d-1
\end{split}
\label{2nd-order partial deriv}
\end{align}
A useful observation is that when $l\geq2$, $\partial f(P_{ij})/\partial \phi_{4,l}$ is only proportional to $\partial P_{34}/\partial \phi_{4,l}$ and it vanishes at the saddle point. Explicity, we have
\begin{align}
&P_{34}=
2(1-\cosh(\theta_3)\cosh(\theta_4)+\sinh(\theta_3)\sinh(\theta_4)\hat{n}_3\cdot\hat{n}_4)
\\
&\left.\frac{\partial P_{34}}{\partial \phi_{4,l}}\right|_\text{saddle point}
=
\left.2\sinh(\theta_3)\sinh(\theta_4) 
\frac{\partial\left(\hat{n}_3^{(d)}\cdot\hat{n}_4^{(d)}\right)}{\partial \phi_{4,l}}\right|_\text{saddle point}=0
\end{align}
where the last equal sign is because at the saddle point $\partial \hat n_3^{(d)}/\partial \phi_{4,l}$ is orthogonal to $\hat n_4^{(d)}$ and vice versa. It follows similarly that all the second-order partial derivatives in \eqref{2nd-order partial deriv} are zero due to orthogonality, except for one term, which is 
\begin{align}
\left.\frac{\partial^2 f(P_{34})}{\partial \phi_{4,l}^2}\right|_{\text{saddle point}}=
2 \sinh^2(\theta_1) \sin(\phi_{3,1})^2 \ldots \sin(\phi_{3,l-1})^2\,.
\end{align}
Altogether we find that the integral over the left-hand side becomes\footnote{The product of sine functions in \eqref{Eqn_d dimensional 2nd order determinant} precisely cancels against the spherical integration measure of $\hat n_4^{(d)}$.}
\begin{align}
\begin{split}
&\int d|\underline k_3| |\underline{k}_3|^{d-1} d^d\underline{k}_4\ 
\left[ 
i Z^2
R^{-d+3} 2^{2 \D - d/2 - 6} \pi^{-3d/2 + 1/2} \Delta^{3d/2 - 9/2}
\frac{ \left(f(P_{ij})\right)^{-2\D + 3/2} }
{(P_{12}P_{13} P_{14} P_{23} P_{24} P_{34})^{1/4}}
\right]\\
&\ \,\quad\simeq
i Z^2
R^{-d+3} 2^{2 \D - d/2 - 6} \pi^{-3d/2 + 1/2} \Delta^{3d/2 - 9/2}
(m\cosh(\theta_1))^2 (m\sinh(\theta_1))^{2d-2} \\
&\qquad\quad \times
\left(\frac{\pi}{\D}\right)^{(d+1)/2}
\left(\frac{\sinh^2(\theta_1)}{8}\right)^{-(d-2)/2}
\frac{8^{-2\D + 5/3} }{\cosh(\theta_1)\sinh^2(\theta_1)}
\\
&\quad\overset{R\to\infty}{=}i (2\pi)^{d+1}\frac{m^{d-1}}{2}\cosh (\theta_1) \sinh^{d-2}(\theta_1)
\end{split}
\end{align}
which exactly agrees with \eqref{Eqn_4pt delta function integral}. We have therefore shown that our S-matrix conjecture is also correct for the four-point contact Witten diagram of identical operators in any spacetime dimension.

\section{Saddle-point analysis of the exchange diagram\label{ap:exchange}}
In this appendix, we explain details of the saddle-point computation of the exchange diagram discussed in subsection \ref{subsec:exchange}. For this purpose, we first integrate out $X$, $Y$ and $Q$ in \eqref{eq:contourintegralGe} to get
\begin{align}\label{eq:apsaddleint}
G_e (P_i)=&\frac{R^{-d+5}}{32\pi^{h}(P_{12})^{\Delta}(P_{34})^{\Delta}}\int^{i\infty}_{-i\infty}\frac{dc}{2\pi i}\frac{1}{c^2-(\Delta_b-h)^2}\left(\frac{\Gamma \left(\Delta-\frac{h}{2}+\frac{c}{2}\right)\Gamma \left(\Delta-\frac{h}{2}-\frac{c}{2}\right)}{2\pi^{h}(\Gamma (\Delta-h+1))^2}\right)^2 \nonumber\\
&\times \left[{\sf k}_{c}G_{h+c,0}(\rho,\bar{\rho})+{\sf k}_{-c}G_{h-c,0}(\rho,\bar{\rho})\right]\,,
\end{align}
where $G_{\Delta,l}$ is the conformal block and ${\sf k}_{a}$ is defined by
\be
{\sf k}_{a}\colonequals \frac{\left(\Gamma (\frac{h}{2}+\frac{a}{2})\right)^4}{\Gamma (h+c)\Gamma(c)}\,.
\ee
In the flat-space limit ($\Delta \sim c\gg 1$), \eqref{eq:apsaddleint} can be approximated by
\be
G_e (P_i)\sim \frac{R^{-d+5}}{32\pi^{h}(P_{12})^{\Delta}(P_{34})^{\Delta}}\int^{i\infty}_{-i\infty}\frac{dc}{2\pi i}\frac{\mathcal{N}}{c^2-\Delta_b^2}\left(c^{h-1}e^{g(c)}+(-c)^{h-1}e^{g(-c)}\right)\,,
\ee
with
\be
\begin{aligned}
\mathcal{N}=&\frac{2\pi^{1-2h} \Delta^{4h-2}(\rho\bar{\rho})^{h/2}}{\sqrt{(1-\rho^2)(1-\bar{\rho}^2)}(\Delta^2-\frac{c^2}{4})^{h+1}(1-\rho\bar{\rho})^{h-1}}\,,\\
g(x)=&-4\Delta\log (\Delta)+2(\Delta +\tfrac{x}{2})\log (\Delta+\tfrac{x}{2}) +2(\Delta -\tfrac{x}{2})\log (\Delta-\tfrac{x}{2})+\frac{x}{2}\log \rho\bar{\rho}\,,
\end{aligned}
\ee
where $\rho$ and $\bar{\rho}$ are the radial coordinates. 
We can then perform the saddle-point analysis for the $c$-integral. The two exponential factors $e^{g(c)}$ and $e^{g(-c)}$ give two different saddle points but their contributions turn out to be identical. Including the one-loop fluctuations around the saddle-point we get the result quoted in the main text
\be
G_{c}(P_i) \sim \left.G_{c}(P_i)\right|_{R\to \infty}\times \frac{R^2}{\Delta_b^2-c^2}\,.
\ee

On the other hand, the contributions from the poles can be determined by evaluating $e^{g(c)}$ and $e^{g(-c)}$ at the positions of the poles $c=\pm\Delta_b$. For the $c=\Delta_b$ pole, $e^{g(-c)}$ can be neglected and the dominant contribution at large $R$ is given by a product of $e^{g(\Delta_b)}$ and $(P_{12}P_{34})^{\Delta}$. This leads to the formula \eqref{eq:polesaddlevalue} in the main text.


\section{Comparison with the phase shift formula of \texorpdfstring{\cite{QFTinAdS}}{[1]}}
\label{app:phaseshift}
In reference \cite{QFTinAdS}, following a different derivation, a formula for the spin-$\ell$ phase shifts was proposed for the scattering of two identical particles which in our notation can be written as
\bea
e^{2i\delta_\ell(s)}&=\lim_{\Df\to \infty} N_\ell(s)^{-1}\sum_{|\Delta-\sqrt{s}\Df| <\Df^{\alpha}} w(\Delta)^2 a_{\Delta,\ell} e^{i\pi(\Delta-2\Df)}\,,\\
N_\ell(s)&=\sum_{|\Delta-\sqrt{s}\Df| <\Df^{\alpha}} w(\Delta)^2 a_{\Delta,\ell}\,,
\eea
where the constraint on $\alpha>0$ was not examined very carefully, other than it should be smaller than one. The explicit form of the weight $w(\Delta)$ will not be required here. It is sufficient to say that its dependence on $\Delta$ is such that $w(\Delta)^2 a_{\Delta,\ell}^{\mbox{\tiny cont}}$ does not vary exponentially with $\Delta$ in the limit of large $\Df$. This implies that for the GFF theory we have
\bea
N_\ell^{\mbox{\tiny gff}}(s)\underset{\Df\to \infty}{=} \Df^{\alpha} \left[w(\Delta)^2 a_{\Delta,\ell}^{\mbox{\tiny cont}}\right]\bigg |_{\Delta=\sqrt{s}\Df}
\eea
Let us now assume that $N_\ell(s)$ is universal for CFTs which describe QFTs in AdS, and in particular equal to the GFF result above. This assumption was originally made in \cite{QFTinAdS} in order to match the above phase shift formula with the prescription for the S-matrix based on the Mellin amplitude in that same reference, which is our equation \eqref{flatspaceMellin}. We now point out that using the result above this is nothing but the condition
\bea
\sum_{|\Delta-\sqrt{s}\Df| <\Df^{\alpha}} \left(\frac{a_{\Delta,\ell}}{a_{\Delta,\ell}^{\mbox{\tiny cont}}}\right)= \Df^{\alpha}
\eea
i.e.~the same as our assumption \reef{eq:opeassumption}. This also leads to
\bea
e^{2i\delta_\ell(s)}=\lim_{\Df\to \infty} \Df^{-\alpha} \sum_{|\Delta-\sqrt{s}\Df| <\Df^{\alpha}}\left(\frac{a_{\Delta,\ell}}{a_{\Delta,\ell}^{\mbox{\tiny cont}}}\right) e^{i\pi(\Delta-2\Df)}\,.
\eea
Comparing this to our own expression \reef{eq:ourphaseshift} we see that they differ only in the way the averaging in the sum over states is being done: with a gaussian in our case and with a rectangular ``window'' in the above. Both averaging kernels tend to delta functions in the large $\Df$ limit. As long as both limits exist they must describe the same limiting function. The main difference is that the Gaussian prescribes that the sum over states must be done over a precise region of width $O(\sqrt{\Delta})$, whereas the exact width of the averaging window was not as precisely determined in the above.

\bibliography{biblio}

\providecommand{\href}[2]{#2}\begingroup\raggedright\begin{thebibliography}{10}

\bibitem{QFTinAdS}
M.~F. Paulos, J.~Penedones, J.~Toledo, B.~C. van Rees, and P.~Vieira, ``{The
  S-matrix bootstrap. Part I: QFT in AdS},''
  \href{http://dx.doi.org/10.1007/JHEP11(2017)133}{{\em JHEP} {\bfseries 11}
  (2017) 133},
\href{http://arxiv.org/abs/1607.06109}{{\ttfamily arXiv:1607.06109 [hep-th]}}.

\bibitem{Polchinski:1999ry}
J.~Polchinski, ``{S matrices from AdS space-time},''
  \href{http://arxiv.org/abs/hep-th/9901076}{{\ttfamily arXiv:hep-th/9901076}}.

\bibitem{Giddings:1999jq}
S.~B. Giddings, ``{Flat space scattering and bulk locality in the AdS / CFT
  correspondence},'' \href{http://dx.doi.org/10.1103/PhysRevD.61.106008}{{\em
  Phys. Rev. D} {\bfseries 61} (2000) 106008},
  \href{http://arxiv.org/abs/hep-th/9907129}{{\ttfamily arXiv:hep-th/9907129}}.

\bibitem{Gary:2009ae}
M.~Gary, S.~B. Giddings, and J.~Penedones, ``{Local bulk S-matrix elements and
  CFT singularities},''
  \href{http://dx.doi.org/10.1103/PhysRevD.80.085005}{{\em Phys. Rev. D}
  {\bfseries 80} (2009) 085005},
  \href{http://arxiv.org/abs/0903.4437}{{\ttfamily arXiv:0903.4437 [hep-th]}}.

\bibitem{Okuda:2010ym}
T.~Okuda and J.~Penedones, ``{String scattering in flat space and a scaling
  limit of Yang-Mills correlators},''
  \href{http://dx.doi.org/10.1103/PhysRevD.83.086001}{{\em Phys. Rev. D}
  {\bfseries 83} (2011) 086001},
  \href{http://arxiv.org/abs/1002.2641}{{\ttfamily arXiv:1002.2641 [hep-th]}}.

\bibitem{joaomellin}
J.~Penedones, ``{Writing CFT correlation functions as AdS scattering
  amplitudes},'' \href{http://dx.doi.org/10.1007/JHEP03(2011)025}{{\em JHEP}
  {\bfseries 03} (2011) 025},
\href{http://arxiv.org/abs/1011.1485}{{\ttfamily arXiv:1011.1485 [hep-th]}}.

\bibitem{Goncalves:2014ffa}
V.~Goncalves, ``{Four point function of $\mathcal{N}=4$ stress-tensor multiplet
  at strong coupling},'' \href{http://dx.doi.org/10.1007/JHEP04(2015)150}{{\em
  JHEP} {\bfseries 04} (2015) 150},
  \href{http://arxiv.org/abs/1411.1675}{{\ttfamily arXiv:1411.1675 [hep-th]}}.

\bibitem{Paulos:2016but}
M.~F. Paulos, J.~Penedones, J.~Toledo, B.~C. van Rees, and P.~Vieira, ``{The
  S-matrix bootstrap II: two dimensional amplitudes},''
  \href{http://dx.doi.org/10.1007/JHEP11(2017)143}{{\em JHEP} {\bfseries 11}
  (2017) 143}, \href{http://arxiv.org/abs/1607.06110}{{\ttfamily
  arXiv:1607.06110 [hep-th]}}.

\bibitem{Paulos:2017fhb}
M.~F. Paulos, J.~Penedones, J.~Toledo, B.~C. van Rees, and P.~Vieira, ``{The
  S-matrix bootstrap. Part III: higher dimensional amplitudes},''
  \href{http://dx.doi.org/10.1007/JHEP12(2019)040}{{\em JHEP} {\bfseries 12}
  (2019) 040}, \href{http://arxiv.org/abs/1708.06765}{{\ttfamily
  arXiv:1708.06765 [hep-th]}}.

\bibitem{Homrich:2019cbt}
A.~Homrich, J.~Penedones, J.~Toledo, B.~C. van Rees, and P.~Vieira, ``{The
  S-matrix Bootstrap IV: Multiple Amplitudes},''
  \href{http://dx.doi.org/10.1007/JHEP11(2019)076}{{\em JHEP} {\bfseries 11}
  (2019) 076}, \href{http://arxiv.org/abs/1905.06905}{{\ttfamily
  arXiv:1905.06905 [hep-th]}}.

\bibitem{Aharony:2012jf}
O.~Aharony, M.~Berkooz, D.~Tong, and S.~Yankielowicz, ``{Confinement in Anti-de
  Sitter Space},'' \href{http://dx.doi.org/10.1007/JHEP02(2013)076}{{\em JHEP}
  {\bfseries 02} (2013) 076}, \href{http://arxiv.org/abs/1210.5195}{{\ttfamily
  arXiv:1210.5195 [hep-th]}}.

\bibitem{Aharony:2015zea}
O.~Aharony, M.~Berkooz, and S.-J. Rey, ``{Rigid holography and six-dimensional
  $\mathcal{N}=(2,0)$ theories on AdS$_{5}\times S^{1}$},''
  \href{http://dx.doi.org/10.1007/JHEP03(2015)121}{{\em JHEP} {\bfseries 03}
  (2015) 121}, \href{http://arxiv.org/abs/1501.02904}{{\ttfamily
  arXiv:1501.02904 [hep-th]}}.

\bibitem{Carmi:2018qzm}
D.~Carmi, L.~Di~Pietro, and S.~Komatsu, ``{A Study of Quantum Field Theories in
  AdS at Finite Coupling},''
  \href{http://dx.doi.org/10.1007/JHEP01(2019)200}{{\em JHEP} {\bfseries 01}
  (2019) 200}, \href{http://arxiv.org/abs/1810.04185}{{\ttfamily
  arXiv:1810.04185 [hep-th]}}.

\bibitem{Giombi:2020rmc}
S.~Giombi and H.~Khanchandani, ``{CFT in AdS and boundary RG flows},''
  \href{http://arxiv.org/abs/2007.04955}{{\ttfamily arXiv:2007.04955
  [hep-th]}}.

\bibitem{Giombi:2017cqn}
S.~Giombi, R.~Roiban, and A.~A. Tseytlin, ``{Half-BPS Wilson loop and
  AdS$_2$/CFT$_1$},''
  \href{http://dx.doi.org/10.1016/j.nuclphysb.2017.07.004}{{\em Nucl. Phys. B}
  {\bfseries 922} (2017) 499--527},
  \href{http://arxiv.org/abs/1706.00756}{{\ttfamily arXiv:1706.00756
  [hep-th]}}.

\bibitem{Beccaria:2019stp}
M.~Beccaria and A.~A. Tseytlin, ``{On boundary correlators in Liouville theory
  on AdS$_{2}$},'' \href{http://dx.doi.org/10.1007/JHEP07(2019)008}{{\em JHEP}
  {\bfseries 07} (2019) 008}, \href{http://arxiv.org/abs/1904.12753}{{\ttfamily
  arXiv:1904.12753 [hep-th]}}.

\bibitem{Hijano:2019qmi}
E.~Hijano, ``{Flat space physics from AdS/CFT},''
  \href{http://dx.doi.org/10.1007/JHEP07(2019)132}{{\em JHEP} {\bfseries 07}
  (2019) 132},
\href{http://arxiv.org/abs/1905.02729}{{\ttfamily arXiv:1905.02729 [hep-th]}}.

\bibitem{TTBar}
S.~Dubovsky, V.~Gorbenko, and M.~Mirbabayi, ``{Asymptotic fragility, near
  AdS$_{2}$ holography and $ T\overline{T} $},''
  \href{http://dx.doi.org/10.1007/JHEP09(2017)136}{{\em JHEP} {\bfseries 09}
  (2017) 136},
\href{http://arxiv.org/abs/1706.06604}{{\ttfamily arXiv:1706.06604 [hep-th]}}.

\bibitem{Penedones:2019tng}
J.~Penedones, J.~A. Silva, and A.~Zhiboedov, ``{Nonperturbative Mellin
  Amplitudes: Existence, Properties, Applications},''
  \href{http://arxiv.org/abs/1912.11100}{{\ttfamily arXiv:1912.11100
  [hep-th]}}.

\bibitem{Maldacena:2015iua}
J.~Maldacena, D.~Simmons-Duffin, and A.~Zhiboedov, ``{Looking for a bulk
  point},'' \href{http://dx.doi.org/10.1007/JHEP01(2017)013}{{\em JHEP}
  {\bfseries 01} (2017) 013}, \href{http://arxiv.org/abs/1509.03612}{{\ttfamily
  arXiv:1509.03612 [hep-th]}}.

\bibitem{Skenderis:2002wp}
K.~Skenderis, ``{Lecture notes on holographic renormalization},''
  \href{http://dx.doi.org/10.1088/0264-9381/19/22/306}{{\em Class. Quant.
  Grav.} {\bfseries 19} (2002) 5849--5876},
\href{http://arxiv.org/abs/hep-th/0209067}{{\ttfamily arXiv:hep-th/0209067
  [hep-th]}}.

\bibitem{Beccaria:2019dws}
M.~Beccaria, S.~Giombi, and A.~A. Tseytlin, ``{Correlators on
  non-supersymmetric Wilson line in $ \mathcal{N}=4 $ SYM and
  AdS$_{2}$/CFT$_{1}$},'' \href{http://dx.doi.org/10.1007/JHEP05(2019)122}{{\em
  JHEP} {\bfseries 05} (2019) 122},
  \href{http://arxiv.org/abs/1903.04365}{{\ttfamily arXiv:1903.04365
  [hep-th]}}.

\bibitem{Beccaria:2019ibr}
M.~Beccaria and G.~Landolfi, ``{Toda theory in AdS$_{2}$ and $\mathcal
  WA_{n}$-algebra structure of boundary correlators},''
  \href{http://dx.doi.org/10.1007/JHEP10(2019)003}{{\em JHEP} {\bfseries 10}
  (2019) 003}, \href{http://arxiv.org/abs/1906.06485}{{\ttfamily
  arXiv:1906.06485 [hep-th]}}.

\bibitem{Beccaria:2019mev}
M.~Beccaria, H.~Jiang, and A.~A. Tseytlin, ``{Non-abelian Toda theory on
  AdS$_2$ and AdS$_2$/CFT$_2^{1/2}$ duality},''
  \href{http://dx.doi.org/10.1007/s13130-019-11219-y}{{\em JHEP} {\bfseries 09}
  (2019) 036}, \href{http://arxiv.org/abs/1907.01357}{{\ttfamily
  arXiv:1907.01357 [hep-th]}}.

\bibitem{Beccaria:2019dju}
M.~Beccaria, H.~Jiang, and A.~A. Tseytlin, ``{Supersymmetric Liouville theory
  in AdS$_{2}$ and AdS/CFT},''
  \href{http://dx.doi.org/10.1007/JHEP11(2019)051}{{\em JHEP} {\bfseries 11}
  (2019) 051}, \href{http://arxiv.org/abs/1909.10255}{{\ttfamily
  arXiv:1909.10255 [hep-th]}}.

\bibitem{Beccaria:2020qtk}
M.~Beccaria, H.~Jiang, and A.~A. Tseytlin, ``{Boundary correlators in WZW model
  on AdS$_{2}$},'' \href{http://dx.doi.org/10.1007/JHEP05(2020)099}{{\em JHEP}
  {\bfseries 05} (2020) 099}, \href{http://arxiv.org/abs/2001.11269}{{\ttfamily
  arXiv:2001.11269 [hep-th]}}.

\bibitem{Skenderis:2008dh}
K.~Skenderis and B.~C. van Rees, ``{Real-time gauge/gravity duality},''
  \href{http://dx.doi.org/10.1103/PhysRevLett.101.081601}{{\em Phys. Rev.
  Lett.} {\bfseries 101} (2008) 081601},
\href{http://arxiv.org/abs/0805.0150}{{\ttfamily arXiv:0805.0150 [hep-th]}}.

\bibitem{Skenderis:2008dg}
K.~Skenderis and B.~C. van Rees, ``{Real-time gauge/gravity duality:
  Prescription, Renormalization and Examples},''
  \href{http://dx.doi.org/10.1088/1126-6708/2009/05/085}{{\em JHEP} {\bfseries
  05} (2009) 085},
\href{http://arxiv.org/abs/0812.2909}{{\ttfamily arXiv:0812.2909 [hep-th]}}.

\bibitem{Dolan_Osborn}
F.~A. Dolan and H.~Osborn, ``{Conformal four point functions and the operator
  product expansion},''
  \href{http://dx.doi.org/10.1016/S0550-3213(01)00013-X}{{\em Nucl. Phys.}
  {\bfseries B599} (2001) 459--496},
\href{http://arxiv.org/abs/hep-th/0011040}{{\ttfamily arXiv:hep-th/0011040
  [hep-th]}}.

\bibitem{Dolan:2003hv}
F.~Dolan and H.~Osborn, ``{Conformal partial waves and the operator product
  expansion},'' \href{http://dx.doi.org/10.1016/j.nuclphysb.2003.11.016}{{\em
  Nucl. Phys. B} {\bfseries 678} (2004) 491--507},
  \href{http://arxiv.org/abs/hep-th/0309180}{{\ttfamily arXiv:hep-th/0309180}}.

\bibitem{Hogervorst:2013sma}
M.~Hogervorst and S.~Rychkov, ``{Radial Coordinates for Conformal Blocks},''
  \href{http://dx.doi.org/10.1103/PhysRevD.87.106004}{{\em Phys. Rev.}
  {\bfseries D87} (2013) 106004},
\href{http://arxiv.org/abs/1303.1111}{{\ttfamily arXiv:1303.1111 [hep-th]}}.

\bibitem{Aprile:2020luw}
F.~Aprile and P.~Vieira, ``{Large $p$ explorations. From SUGRA to big STRINGS
  in Mellin space},'' \href{http://arxiv.org/abs/2007.09176}{{\ttfamily
  arXiv:2007.09176 [hep-th]}}.

\bibitem{Cornalba:2006xk}
L.~Cornalba, M.~S. Costa, J.~Penedones, and R.~Schiappa, ``{Eikonal
  Approximation in AdS/CFT: From Shock Waves to Four-Point Functions},''
  \href{http://dx.doi.org/10.1088/1126-6708/2007/08/019}{{\em JHEP} {\bfseries
  08} (2007) 019}, \href{http://arxiv.org/abs/hep-th/0611122}{{\ttfamily
  arXiv:hep-th/0611122}}.

\bibitem{Cornalba:2007fs}
L.~Cornalba, ``{Eikonal methods in AdS/CFT: Regge theory and multi-reggeon
  exchange},'' \href{http://arxiv.org/abs/0710.5480}{{\ttfamily arXiv:0710.5480
  [hep-th]}}.

\bibitem{Costa:2012cb}
M.~S. Costa, V.~Goncalves, and J.~Penedones, ``{Conformal Regge theory},''
  \href{http://dx.doi.org/10.1007/JHEP12(2012)091}{{\em JHEP} {\bfseries 12}
  (2012) 091}, \href{http://arxiv.org/abs/1209.4355}{{\ttfamily arXiv:1209.4355
  [hep-th]}}.

\bibitem{Gross:1987ar}
D.~J. Gross and P.~F. Mende, ``{String Theory Beyond the Planck Scale},''
  \href{http://dx.doi.org/10.1016/0550-3213(88)90390-2}{{\em Nucl. Phys. B}
  {\bfseries 303} (1988) 407--454}.

\bibitem{Joao_string_scattering}
T.~Okuda and J.~Penedones, ``{String scattering in flat space and a scaling
  limit of Yang-Mills correlators},''
  \href{http://dx.doi.org/10.1103/PhysRevD.83.086001}{{\em Phys. Rev.}
  {\bfseries D83} (2011) 086001},
\href{http://arxiv.org/abs/1002.2641}{{\ttfamily arXiv:1002.2641 [hep-th]}}.

\bibitem{Minahan:2012fh}
J.~A. Minahan, ``{Holographic three-point functions for short operators},''
  \href{http://dx.doi.org/10.1007/JHEP07(2012)187}{{\em JHEP} {\bfseries 07}
  (2012) 187}, \href{http://arxiv.org/abs/1206.3129}{{\ttfamily arXiv:1206.3129
  [hep-th]}}.

\bibitem{Mack:2009mi}
G.~Mack, ``{D-independent representation of Conformal Field Theories in D
  dimensions via transformation to auxiliary Dual Resonance Models. Scalar
  amplitudes},''
\href{http://arxiv.org/abs/0907.2407}{{\ttfamily arXiv:0907.2407 [hep-th]}}.

\bibitem{GeodesicWittenDiagram}
E.~Hijano, P.~Kraus, E.~Perlmutter, and R.~Snively, ``{Witten Diagrams
  Revisited: The AdS Geometry of Conformal Blocks},''
  \href{http://dx.doi.org/10.1007/JHEP01(2016)146}{{\em JHEP} {\bfseries 01}
  (2016) 146},
\href{http://arxiv.org/abs/1508.00501}{{\ttfamily arXiv:1508.00501 [hep-th]}}.

\bibitem{Landau:1959fi}
L.~Landau, ``{On analytic properties of vertex parts in quantum field
  theory},'' \href{http://dx.doi.org/10.1016/B978-0-08-010586-4.50103-6}{{\em
  Nucl. Phys.} {\bfseries 13} no.~1, (1960) 181--192}.

\bibitem{Coleman:1965xm}
S.~Coleman and R.~Norton, ``{Singularities in the physical region},''
  \href{http://dx.doi.org/10.1007/BF02750472}{{\em Nuovo Cim.} {\bfseries 38}
  (1965) 438--442}.

\bibitem{bjorken1965relativistic}
J.~D. Bjorken and S.~D. Drell, {\em Relativistic quantum fields}.
\newblock McGraw-Hill, 1965.

\bibitem{eden2002analytic}
R.~J. Eden, P.~V. Landshoff, D.~I. Olive, and J.~C. Polkinghorne, {\em The
  analytic S-matrix}.
\newblock Cambridge University Press, 2002.

\bibitem{sterman1993introduction}
G.~Sterman, {\em An introduction to quantum field theory}.
\newblock Cambridge university press, 1993.

\bibitem{Fitzpatrick:2011ia}
A.~Fitzpatrick, J.~Kaplan, J.~Penedones, S.~Raju, and B.~C. van Rees, ``{A
  Natural Language for AdS/CFT Correlators},''
  \href{http://dx.doi.org/10.1007/JHEP11(2011)095}{{\em JHEP} {\bfseries 11}
  (2011) 095}, \href{http://arxiv.org/abs/1107.1499}{{\ttfamily arXiv:1107.1499
  [hep-th]}}.

\bibitem{Paulos:2011ie}
M.~F. Paulos, ``{Towards Feynman rules for Mellin amplitudes},''
  \href{http://dx.doi.org/10.1007/JHEP10(2011)074}{{\em JHEP} {\bfseries 10}
  (2011) 074}, \href{http://arxiv.org/abs/1107.1504}{{\ttfamily arXiv:1107.1504
  [hep-th]}}.

\bibitem{Kos:2014bka}
F.~Kos, D.~Poland, and D.~Simmons-Duffin, ``{Bootstrapping Mixed Correlators in
  the 3D Ising Model},'' \href{http://dx.doi.org/10.1007/JHEP11(2014)109}{{\em
  JHEP} {\bfseries 11} (2014) 109},
  \href{http://arxiv.org/abs/1406.4858}{{\ttfamily arXiv:1406.4858 [hep-th]}}.

\bibitem{Mazac:2018mdx}
D.~Mazac and M.~F. Paulos, ``{The analytic functional bootstrap. Part I: 1D
  CFTs and 2D S-matrices},''
  \href{http://dx.doi.org/10.1007/JHEP02(2019)162}{{\em JHEP} {\bfseries 02}
  (2019) 162}, \href{http://arxiv.org/abs/1803.10233}{{\ttfamily
  arXiv:1803.10233 [hep-th]}}.

\bibitem{Caron-Huot:2017vep}
S.~Caron-Huot, ``{Analyticity in Spin in Conformal Theories},''
  \href{http://dx.doi.org/10.1007/JHEP09(2017)078}{{\em JHEP} {\bfseries 09}
  (2017) 078}, \href{http://arxiv.org/abs/1703.00278}{{\ttfamily
  arXiv:1703.00278 [hep-th]}}.

\bibitem{Carmi:2019cub}
D.~Carmi and S.~Caron-Huot, ``{A Conformal Dispersion Relation: Correlations
  from Absorption},'' \href{http://arxiv.org/abs/1910.12123}{{\ttfamily
  arXiv:1910.12123 [hep-th]}}.

\bibitem{Correia:2020xtr}
M.~Correia, A.~Sever, and A.~Zhiboedov, ``{An Analytical Toolkit for the
  S-matrix Bootstrap},'' \href{http://arxiv.org/abs/2006.08221}{{\ttfamily
  arXiv:2006.08221 [hep-th]}}.

\bibitem{Dodelson:2019ddi}
M.~Dodelson and H.~Ooguri, ``{High-energy behavior of Mellin amplitudes},''
  \href{http://dx.doi.org/10.1103/PhysRevD.101.066008}{{\em Phys. Rev. D}
  {\bfseries 101} no.~6, (2020) 066008},
  \href{http://arxiv.org/abs/1911.05274}{{\ttfamily arXiv:1911.05274
  [hep-th]}}.

\bibitem{Haldar:2019prg}
P.~Haldar and A.~Sinha, ``{Froissart bound for/from CFT Mellin amplitudes},''
  \href{http://dx.doi.org/10.21468/SciPostPhys.8.6.095}{{\em SciPost Phys.}
  {\bfseries 8} (2020) 095}, \href{http://arxiv.org/abs/1911.05974}{{\ttfamily
  arXiv:1911.05974 [hep-th]}}.

\bibitem{Weinberg:1965nx}
S.~Weinberg, ``{Infrared photons and gravitons},''
  \href{http://dx.doi.org/10.1103/PhysRev.140.B516}{{\em Phys. Rev.} {\bfseries
  140} (1965) B516--B524}.

\bibitem{He:2014laa}
T.~He, V.~Lysov, P.~Mitra, and A.~Strominger, ``{BMS supertranslations and
  Weinberg's soft graviton theorem},''
  \href{http://dx.doi.org/10.1007/JHEP05(2015)151}{{\em JHEP} {\bfseries 05}
  (2015) 151}, \href{http://arxiv.org/abs/1401.7026}{{\ttfamily arXiv:1401.7026
  [hep-th]}}.

\bibitem{Strominger:2017zoo}
A.~Strominger, ``{Lectures on the Infrared Structure of Gravity and Gauge
  Theory},'' \href{http://arxiv.org/abs/1703.05448}{{\ttfamily arXiv:1703.05448
  [hep-th]}}.

\bibitem{Hijano:2020szl}
E.~Hijano and D.~Neuenfeld, ``{Soft photon theorems from CFT Ward identites in
  the flat limit of AdS/CFT},''
  \href{http://arxiv.org/abs/2005.03667}{{\ttfamily arXiv:2005.03667
  [hep-th]}}.

\end{thebibliography}\endgroup
\bibliographystyle{utphys}

\end{document}